\documentclass[12pt]{iopart}
\usepackage{iopams}  
\usepackage{cite}

\def\be{\begin{equation}}
\def\ee{\end{equation}}
\def\beq{\begin{eqnarray}}
\def\eeq{\end{eqnarray}}

\def\inbar{\vrule height1.5ex width.4pt depth0pt}
\def\Cop{\relax\,\hbox{$\inbar\kern-.3em{\rm C}$}}
\def\Zop{\mathbb{Z}}
\def\Rop{\relax{\rm I\kern-.18em R}}
\def\Nop{\relax{\rm I\kern-.18em N}}

\def\A{{\cal A}}

\def\C{{\cal C}}

\def\F{{\cal F}}

\def\H{{\cal H}}
\def\K{{\cal K}}
\def\sL{{\cal L}}
\def\M{{\cal M}}
\def\Nu{{\cal N}}
\def\O{{\cal O}}

\def\V{{\cal V}}
\def\W{{\cal W}}

\def\bz{{\bar{z}}}
\def\bw{{\bar{w}}}

\def\Fbar{\overline{\F}}

\def\psbar{\overline{\psi}}

\def\hF{\widehat\F}

\def\bF{{\pmb{{\cal F}}}}
\def\bH{{\pmb{{\cal H}}}}

\def\z{\zeta}
\def\rs{{\mathbb P}}
\def\nox{{\scriptstyle{\times \atop \times}}}

\def\half{{1\over 2}}
\def\thalf{{3\over 2}}

\def\ie{{\it i.e.}}

\def\SL{\hbox{SL}}

\def\bbbone {{\mathchoice {{\rm 1\mskip-4mu l}} {{\rm 1\mskip-4mu l}}
{{\rm 1\mskip-4.5mu l}} {{\rm 1\mskip-5mu l}}}}

\begin{document}
\bibliographystyle{paper2}

\jl{19}

\begin{flushright}
  DAMTP-1999-143
\end{flushright}

\review[Conformal Field Theory]{An Introduction to Conformal Field Theory}

\author{Matthias R Gaberdiel\footnote{Email: {\tt
M.R.Gaberdiel@damtp.cam.ac.uk}}}

\address{Department of Applied Mathematics and Theoretical Physics,
Silver Street, Cambridge, CB3 9EW, UK and}
\address{Fitzwilliam College, Cambridge, CB3 0DG, UK}

\begin{abstract}
A comprehensive introduction to two-dimensional conformal field theory
is given.
\end{abstract}


\submitted

\maketitle

\section{Introduction}

Conformal field theories have been at the centre of much attention
during the last fifteen years since they are relevant for at least
three different areas of modern theoretical physics: conformal field
theories provide toy models for genuinely interacting quantum field
theories, they describe two-dimensional critical phenomena, and they
play a central r\^ole in string theory, at present the most promising 
candidate for a unifying theory of all forces. Conformal field theories
have also had a major impact on various aspects of modern mathematics,
in particular the theory of vertex operator algebras and Borcherds
algebras, finite groups, number theory and low-dimensional topology.
\smallskip

{}From an abstract point of view, conformal field theories are
Euclidean quantum field theories that are characterised by the
property that their symmetry group contains, in addition to the
Euclidean symmetries, local conformal transformations, \ie\
transformations that preserve angles but not lengths. The local
conformal symmetry is of special importance in two dimensions
since the corresponding symmetry algebra is infinite-dimensional in
this case. As a consequence, two-dimensional conformal field theories
have an infinite number of conserved quantities, and are 
completely solvable by symmetry considerations alone. 

As a bona fide quantum field theory, the requirement of conformal
invariance is very restrictive. In particular, since the theory is
scale invariant, all particle-like excitations of the theory are
necessarily massless. This might be seen as a strong argument against
any possible physical relevance of such theories. However, all
particles of any (two-dimensional) quantum field theory are
approximately massless in the limit of high energy, and many
structural features of quantum field theories are believed to be
unchanged in this approximation. Furthermore, it is possible to
analyse deformations of conformal field theories that describe
integrable massive models \cite{Zam88,ZamZam79}. Finally, it might
be hoped that a good mathematical understanding of interactions
in any model theory  should have implications for realistic theories.  
\smallskip

The more recent interest in conformal field theories has different
origins. In the description of statistical mechanics in terms of
Euclidean quantum field theories, conformal field theories describe
systems at the critical point, where the correlation length
diverges. One simple system where this occurs is the so-called 
{\em Ising model}. This model is formulated in terms of a
two-dimensional lattice whose lattice sites represent atoms of an
(infinite) two-dimensional crystal. Each atom is taken to have a spin
variable $\sigma_i$ that can take the values $\pm 1$, and the
magnetic energy of the system is the sum over pairs of adjacent 
atoms  
\be
E = \sum_{(ij)} \sigma_i \sigma_j \,.
\ee
If we consider the system at a finite temperature $T$, the thermal
average $\langle \cdots \rangle$ behaves as 
\be\label{Issing}
\langle \sigma_i \sigma_j \rangle - 
\langle \sigma_i \rangle \cdot \langle \sigma_j \rangle
\sim \exp \left(-{|i-j| \over \xi} \right) \,,
\ee
where $|i-j| \gg 1$ and $\xi$ is the so-called {\em correlation
length} that is a function of the temperature $T$. Observable
(magnetic) properties can be derived from such correlation functions,
and are therefore directly affected by the actual value of $\xi$. 

The system possesses a {\em critical temperature}, at which the
correlation length $\xi$ diverges, and the exponential decay in
(\ref{Issing}) is replaced by a power law. The continuum theory that
describes the correlation functions for distances that are large
compared to the lattice spacing is then scale invariant. Every
scale-invariant two-dimensional local quantum field theory is 
actually conformally invariant \cite{Pol70}, and the critical point of 
the Ising model is therefore described by a conformal field
theory \cite{BPZ84}. (The conformal field theory in question will be
briefly described at the end of section~4.)

The Ising model is only a rather rough approximation to the actual
physical system. However, the continuum theory at the critical point 
--- and in particular the different {\em critical exponents} that 
describe the power law behaviour of the correlation functions at the
critical point --- are believed to be fairly insensitive to the
details of the chosen model; this is the idea of {\em universality}. 
Thus conformal field theory is a very important method in the study
of critical systems.
\smallskip

The second main area in which conformal field theory has played a
major r\^ole is {\em string theory} \cite{GSW,Pol98}. String
theory is a generalised quantum field theory in which the basic
objects are not point particles (as in ordinary quantum field theory)
but one dimensional strings. These strings can either form closed
loops ({\em closed string theory}), or they can have two end-points,
in which case the theory is called {\em open string theory}. Strings
interact by joining together and splitting into two; compared to the
interaction of point particles where two particles come arbitrarily
close together, the interaction of strings is more spread out, and
thus many divergencies of ordinary quantum field theory are absent. 

Unlike point particles, a string has internal degrees of freedom that
describe the different ways in which it can vibrate in the ambient
space-time. These different vibrational modes are interpreted as the
`particles' of the theory --- in particular, the whole particle
spectrum of the theory is determined in terms of one fundamental
object. The vibrations of the string are most easily described from the 
point of view of the so-called {\em world-sheet}, the two-dimensional
surface that the string sweeps out as it propagates through
space-time; in fact, as a theory on the world-sheet the vibrations of
the string are described by a conformal field theory. 

In closed string theory, the oscillations of the string can be
decomposed into two waves which move in opposite directions around the
loop. These two waves are essentially independent of each other, and
the theory therefore factorises into two so-called {\em chiral
conformal field theories}. Many properties of the local theory can be
studied separately for the two chiral theories, and we shall therefore
mainly analyse the chiral theory in this article. The main advantage
of this approach is that the chiral theory can be studied using the
powerful tools of complex analysis since its correlation functions are
analytic functions. The chiral theories also play a crucial r\^ole for
conformal field theories that are defined on manifolds with
boundaries, and that are relevant for the description of open string 
theory.

All known consistent string theories can be obtained by
compactification from a rather small number of theories. These include
the five different supersymmetric string theories in ten dimensions,
as well as a number of non-supersymmetric theories that are defined in
either ten or twenty-six dimensions. The recent advances in string
theory have centered around the idea of {\em duality}, namely that
these theories are further related in the sense that the strong
coupling regime of one theory is described by the weak coupling regime
of another. A crucial element in these developments has been the
realisation that the {\em solitonic} objects that define the relevant
degrees of freedom at strong coupling are {\em Dirichlet-branes} that
have an alternative description in terms of open string theory
\cite{Pol95}. In fact, the effect of a Dirichlet brane is completely
described by adding certain open string sectors (whose end-points
are fixed to lie on the world-volume of the brane) to the theory. 
The possible Dirichlet branes of a given string theory are then
selected by the condition that the resulting theory of open and closed
strings must be consistent. These consistency conditions contain (and
may be equivalent to) the consistency conditions of conformal field
theory on a manifold with a boundary \cite{Car89,Lew92,CarLew91}. Much
of the structure of the theory that we shall explain in this review
article is directly relevant for an analysis of these questions,
although we shall not discuss the actual consistency conditions (and
their solutions) here.  
\bigskip

Any review article of a well-developed subject such as conformal field
theory will miss out important elements of the theory, and this
article is no exception. We have chosen to present one coherent route
through some section of the theory and we shall not discuss in any
detail alternative view points on the subject. The approach  
that we have taken is in essence algebraic (although we shall
touch upon some questions of analysis), and is inspired by the 
work of Goddard \cite{God89} as well as the mathematical theory of
{\em vertex operator algebras} that was developed by Borcherds
\cite{Bor86,Bor92}, Frenkel, Lepowsky \& Meurman \cite{FLM}, Frenkel,
Huang \& Lepowsky \cite{FHL93}, Zhu \cite{Zhu96}, Kac \cite{Kac1} and
others. This algebraic approach will be fairly familiar to many
physicists, but we have tried to give it a somewhat new slant by
emphasising the fundamental r\^ole of the amplitudes. We have also tried
to explain some of the more recent developments in the mathematical
theory of vertex operator algebras that have so far not been widely
appreciated in the physics community, in particular, the work of Zhu. 

There exist in essence two other view points on the subject: a
functional analytic approach in which techniques from algebraic
quantum field theory \cite{Haag} are employed and which has been 
pioneered by Wassermann \cite{Wass95} and Gabbiani and
Fr\"ohlich \cite{GabFro93}; and a geometrical approach that is
inspired by string theory (for example the work of Friedan \& Shenker
\cite{FriShe87}) and that has been put on a solid mathematical
foundation by Segal \cite{Segal} (see also Huang
\cite{Huang92a,Huang92b}). 

We shall also miss out various recent developments of the theory, in
particular the progress in understanding conformal field theories
on higher genus Riemann surfaces
\cite{Ber88a,Ber88b,Zhu94,Gaw95a,Gaw95b}, and 
on surfaces with boundaries
\cite{PSS95a,PSS95b,Run99,BPPZ99,FFFS99a,FFFS99b}. 

Finally, we should mention that a number of treatments of
conformal field theory are by now available, in particular the review
articles of Ginsparg \cite{Gin89} and Gawedzki \cite{Gaw96}, and the
book by Di Francesco, Mathieu and S\'en\'echal \cite{DMS96}. We have
attempted to be somewhat more general, and have put less emphasis on
specific well understood models such as the minimal models or the WZNW
models (although they will be explained in due course). We have also
been more influenced by the mathematical theory of vertex operator
algebras, although we have avoided to phrase the theory in this
language. 
\medskip

The paper is organised as follows. In section~2, we outline the
general structure of the theory, and explain how the various
ingredients that will be subsequently described fit together. Section~3
is devoted to the study of meromorphic conformal field theory; this is
the part of the theory that describes in essence what is sometimes
called the chiral algebra by physicists, or the vertex operator algebra
by mathematicians. We also introduce the most important examples of
conformal field theories, and describe standard constructions such as
the coset and orbifold construction. In section~4 we introduce the
concept of a representation of the meromorphic conformal field theory,
and explain the r\^ole of  Zhu's algebra in classifying (a certain class
of) such representations. Section~5 deals with higher correlation
functions and fusion rules. We explain Verlinde's formula, and give a
brief account of the polynomial relations of Moore \& Seiberg and
their relation to quantum groups. We also describe logarithmic
conformal field theories. We conclude in section~6 with a number of
general open problems that deserve, in our opinion, more
work. Finally, we have included an appendix that contains a brief
summary about the different definitions of rationality.

\section{The General Structure of a Local Conformal Field Theory}

Let us begin by describing somewhat sketchily what the general
structure of a local conformal field theory is, and how the various
structures that will be discussed in detail later fit together.

\subsection{The Space of States}

In essence, a two-dimensional conformal field theory (like any other
field theory) is determined by its space of states and the collection
of its correlation functions. The space of states is a vector space
$\bH$ (that may or may not be a Hilbert space), and the correlation 
functions are defined for collections of vectors in some dense
subspace $\bF$ of $\bH$. These correlation functions are defined on a
two-dimensional space-time, which we shall always assume to be of
Euclidean signature. We shall mainly be interested in the case where
the space-time is a closed compact surface. These surfaces are
classified (topologically) by their genus $g$ which counts the number
of `handles'; the simplest such surface is the sphere with $g=0$,
the surface with $g=1$ is the torus, {\it etc.} In a first step we
shall therefore consider conformal field theories that are defined on
the sphere; as we shall explain later, under certain conditions it is 
possible to associate to such a theory families of theories that are
defined on surfaces of arbitrary genus. This is important in the
context of string theory where the perturbative expansion consists of
a sum over all such theories (where the genus of the surface plays the
r\^ole of the loop order).

One of the special features of conformal field theory is the fact that
the theory is naturally defined on a {\em Riemann surface} (or 
{\em complex curve}), {\it i.e.} on a surface that possesses suitable
complex coordinates. In the case of the sphere, the complex
coordinates can be taken to be those of the complex plane that cover
the sphere except for the point at infinity; complex coordinates
around infinity are defined by means of the coordinate function
$\gamma(z)=1/z$ that maps a neighbourhood of infinity to a
neighbourhood of $0$. With this choice of complex coordinates, the
sphere is usually referred to as the {\em Riemann sphere}, and this
choice of complex coordinates is up to some suitable class of
reparametrisations unique. The correlation functions of a conformal
field theory that is defined on the sphere are thus of the form   
\be
\label{loccorr}
\langle V(\psi_1;z_1,\bz_1) \cdots V(\psi_n;z_n,\bz_n) \rangle \,,
\ee
where $V(\psi,z)$ is the field that is associated to the state $\psi$, 
$\psi_i\in\bF\subset\bH$, and $z_i$ and $\bz_i$ are complex
numbers (or infinity). These correlation functions are assumed to be
{\em local}, \ie\ independent of the order in which the fields appear
in (\ref{loccorr}). 

One of the properties that makes two-dimensional conformal field
theories exactly solvable is the fact that the theory contains a large  
(infinite-dimensional) symmetry algebra with respect to
which the states in $\bH$ fall into representations. This symmetry
algebra is directly related (in a way we shall describe below) to a
certain preferred subspace $\F_0$ of $\bF$ that is characterised by
the property that the correlation functions (\ref{loccorr}) of its
states depend only on the complex parameter $z$, but not on its
complex conjugate $\bz$. More precisely, a state $\psi\in\bF$ is in 
$\F_0$ if for any collection of $\psi_i\in\bF\subset\bH$, the
correlation functions     
\be
\label{corr1}
\langle V(\psi;z,\bz) V(\psi_1;z_1,\bz_1) \cdots V(\psi_n;z_n,\bz_n)
\rangle 
\ee
do not depend on $\bz$. The correlation functions that
involve only states in $\F_0$ are then analytic functions on the
sphere. These correlation functions define the meromorphic (sub)theory
\cite{God89} that will be the main focus of the next
section.\footnote{Our use of the term {\it meromorphic} conformal
field theory is different from that employed by, {\it e.g.},
Schellekens \cite{Schell93}.} 

Similarly, we can consider the subspace of states $\Fbar_0$ 
that consists of those states for which the correlation functions of
the form (\ref{corr1}) do not depend on $z$. These states define
an (anti-)meromorphic conformal field theory which can be analysed by
the same methods as a meromorphic conformal field theory. 
The two meromorphic conformal subtheories encode all the information
about the symmetries of the theory, and for the most interesting
class of theories, the so-called {\em finite} or {\em rational}
theories, the whole theory can be reconstructed from them up to some
finite ambiguity. In essence, this means that the whole
theory is determined by symmetry considerations alone, and this is at
the heart of the solvability of the theory.

The correlation functions of the theory determine the 
{\em operator product expansion} (OPE) of the conformal fields 
which expresses the operator product of two fields in terms of a sum 
of single fields. If $\psi_1$ and $\psi_2$ are two arbitrary states in
$\bF$ then the OPE of $\psi_1$ and $\psi_2$ is an expansion of the
form  
\beq
\label{localope}
\fl V(\psi_1;z_1,\bz_1) V(\psi_2;z_2,\bz_2) \nonumber \\
\lo = \sum_{i} (z_1-z_2)^{\Delta_i} (\bz_1-\bz_2)^{\bar\Delta_i} 
\sum_{r,s\geq 0} V(\phi_{r,s}^{i};z_2,\bz_2) 
(z_1-z_2)^r  (\bz_1-\bz_2)^s\,,
\eeq
where $\Delta_i$ and $\bar\Delta_i$ are real numbers, $r,s\in\Nop$ and
$\phi_{r,s}^{i}\in\bF$. The actual form of this expansion can be read
off from the correlation functions of the theory since the identity
(\ref{localope}) has to hold in {\em all} correlation functions, \ie 
\beq
\label{opeexpansion}
\fl  \Bigl\langle V(\psi_1;z_1,\bz_1) V(\psi_2;z_2,\bz_2) 
V(\phi_1;w_1,\bw_1) \cdots V(\phi_n;w_n,\bw_n) \Bigr\rangle
\nonumber \\ 
\lo = \sum_i (z_1-z_2)^{\Delta_i}
(\bz_1-\bz_2)^{\bar\Delta_i} \sum_{r,s\geq 0} (z_1-z_2)^r
(\bz_1-\bz_2)^s \nonumber \\ 
\qquad \Bigl\langle V(\phi_{r,s}^{i};z_2,\bz_2) 
V(\phi_1;w_1,\bw_1) \cdots V(\phi_n;w_n,\bw_n)
\Bigr\rangle 
\eeq
for all $\phi_j\in\bF$. If both states $\psi_1$ and $\psi_2$ belong to
the meromorphic subtheory $\F_0$, (\ref{opeexpansion}) only depends on
$z_i$, and $\phi_{r,s}^{i}$ also belongs to the meromorphic subtheory
$\F_0$. The OPE therefore defines a certain product on the meromorphic
fields. Since the product involves the complex parameters $z_i$ in a
non-trivial way, it does not directly define an algebra; the 
resulting structure is usually called a {\em vertex (operator)
algebra} in the mathematical literature \cite{Bor86,FLM}, and we shall
adopt this name here as well.    

By virtue of its definition in terms of (\ref{opeexpansion}), the
operator product expansion is {\em associative}, \ie\ 
\be\label{associative}
\fl
\Bigl( V(\psi_1;z_1,\bz_1) V(\psi_2;z_2,\bz_2) \Bigr) 
V(\psi_3;z_3,\bz_3) = V(\psi_1;z_1,\bz_1) \Bigl(V(\psi_2;z_2,\bz_2)
V(\psi_3;z_3,\bz_3) \Bigr) \,,
\ee
where the brackets indicate which OPE is evaluated first. If we
consider the case where both $\psi_1$ and $\psi_2$ are meromorphic
fields (\ie\ in $\F_0$), then the associativity of the OPE implies
that the states in $\bF$ form a {\em representation} of the vertex
operator algebra. The same also holds for the vertex operator algebra
associated to the anti-meromorphic fields, and we can thus decompose 
the whole space $\bF$ (or $\bH$) as 
\be
\label{decomposition}
\bH = \bigoplus_{(j,\bar{\jmath})} \bH_{(j,\bar{\jmath})} \,,
\ee
where each $\bH_{(j,\bar{\jmath})}$ is an (indecomposable)
representation of the two vertex operator algebras. Finite theories
are characterised by the property that only finitely many
indecomposable representations of the two vertex operator algebras
occur in (\ref{decomposition}).

\subsection{Modular Invariance}

The decomposition of the space of states in terms of representations
of the two vertex operator algebras throws considerable light on the
problem of whether the theory is well-defined on higher Riemann
surfaces. One necessary constraint for this (which is believed also to
be sufficient \cite{MooSei89b}) is that the vacuum correlator on the
torus is independent of its parametrisation. Every two-dimensional
torus can be described as the quotient space of $\Rop^2\simeq \Cop$ by
the relations $z\sim z+w_1$ and $z\sim z+w_2$, where $w_1$ and $w_2$
are not parallel. The complex structure of the torus is invariant under
rotations and rescalings of $\Cop$, and therefore every torus is
conformally equivalent to (\ie\ has the same complex structure as) a
torus for which the relations are $z\sim z+1$ and $z\sim z+\tau$,
and $\tau$ is in the upper half plane of $\Cop$. It is also easy to
see that $\tau$, $T(\tau)=\tau+1$ and $S(\tau)=-1/\tau$
describe conformally equivalent tori; the two maps $T$ and $S$
generate the group $\SL(2,\Zop)/ \Zop_2$ that consists of matrices
of the form 
\be
A= \left(\begin{array}{cc} a & b \\ c & d \end{array}\right) \qquad 
\hbox{where} \quad a,b,c,d\in\Zop\,, \quad \quad 
ad - bc = 1 \,,
\ee
and the matrices $A$ and $-A$ have the same action on $\tau$,
\be
\tau \mapsto A\tau = {a\tau + b \over c \tau + d} \,.
\ee
The parameter $\tau$ is sometimes called the modular parameter of the
torus, and the group $\SL(2,\Zop)/\Zop_2$ is called the modular group
(of the torus). 

Given a conformal field theory that is defined on the Riemann sphere,
the vacuum correlator on the torus can be determined as follows. 
First, we cut the torus along one of its non-trivial cycles; the
resulting surface is a cylinder (or an annulus) whose shape depends  
on one complex parameter $q$. Since the annulus is a subset of the
sphere, the conformal field theory on the annulus is determined in
terms of the theory on the sphere. In particular, the states that can
propagate in the annulus are precisely the states of the theory as
defined on the sphere.  

In order to reobtain the torus from the annulus, we have to glue the
two ends of the annulus together; in terms of conformal field theory
this means that we have to sum over a complete set of states. The
vacuum correlator on the torus is therefore described by a trace over
the whole space of states, the {\em partition function} of the theory,
\be
\label{toruscorr}
\sum_{(j,\bar{\jmath})} \Tr_{\bH_{(j,\bar{\jmath})}} \left(
\O(q,\bar{q}) \right) \,,
\ee
where $\O(q,\bar{q})$ is the operator that describes the
propagation of the states along the annulus,
\be
\O(q,\bar{q}) = q^{L_0-{c\over 24}} \,
\bar{q}^{\bar{L}_0 - {\bar{c} \over 24}} \,.
\ee
Here $L_0$ and $\bar{L}_0$ are the scaling operators of the two vertex
operator algebras and $c$ and $\bar{c}$ their central charges; this
will be discussed in more detail in the following section. The
propagator depends on the actual shape of the annulus that is
described in terms of the complex parameter $q$. For a given torus
that is described by $\tau$, there is a natural choice for how to cut
the torus into an annulus, and the complex parameter $q$ that is
associated to this annulus is $q=e^{2\pi i\tau}$. Since the tori that
are described by $\tau$ and $A\tau$ (where $A\in \SL(2,\Zop)$)  
are equivalent, the vacuum correlator is only well-defined provided
that (\ref{toruscorr}) is invariant under this transformation. This
provides strong constraints on the spectrum of the theory. 

For most conformal field theories (although not for all, see for
example \cite{GabKau98}) each of the spaces $\bH_{(j,\bar{\jmath})}$
is a tensor product of an irreducible representation $\H_j$ of the
meromorphic vertex operator algebra and an irreducible representation
$\bar\H_{\bar{\jmath}}$ of the anti-meromorphic vertex operator
algebra. In this case, the vacuum correlator on the torus
(\ref{toruscorr}) takes the form 
\be
\sum_{(j,\bar{\jmath})} \chi_j(q) \,
\bar\chi_{\bar{\jmath}}(\bar{q})\,, 
\ee
where $\chi_j$ is the {\em character} of the representation $\H_j$ of
the meromorphic vertex operator algebra,
\be
\chi_j (\tau) = \Tr_{\H_j} (q^{L_0 - {c\over 24}}) \qquad \hbox{where} 
\quad q=e^{2\pi i \tau}\,,
\ee
and likewise for $\bar\chi_{\bar{\jmath}}$. One of the remarkable
facts about many vertex operator algebras (that has now been proven
for a certain class of them \cite{Zhu96}, see also \cite{Nahm91}) is
the property that the characters transform into one another under
modular transformations,   
\be
\label{modular}
\chi_j(-1/\tau) = \sum_k S_{jk} \,\chi_k(\tau) \qquad \hbox{and}
\qquad  \chi_j(\tau+1) = \sum_k T_{jk} \,\chi_k(\tau) \,,
\ee
where $S$ and $T$ are {\em constant} matrices, \ie\ independent of
$\tau$. In this case, writing 
\be
\bH = \bigoplus_{i,\bar{\jmath}} M_{i \bar{\jmath}} \;
\H_i \otimes \bar\H_{\bar{\jmath}}\,,
\ee
where $M_{i \bar{\jmath}}\in\Nop$ denotes the multiplicity with which
the tensor product $\H_i \otimes \bar\H_{\bar{\jmath}}$ appears in
$\bH$, the torus vacuum correlation function is well defined provided
that  
\be
\sum_{i,\bar{\jmath}} S_{il} M_{i\bar{\jmath}} 
\bar{S}_{\bar{\jmath}\bar{k}}
= \sum_{i,\bar{\jmath}} T_{il} M_{i\bar{\jmath}} 
\bar{T}_{\bar{\jmath}\bar{k}} = M_{l \bar{k}} \,,
\ee
and $\bar{S}$ and $\bar{T}$ are the matrices defined as in
(\ref{modular}) for the representations of the anti-meromorphic vertex 
operator algebra. This provides very powerful constraints 
for the multiplicity matrices $M_{i\bar{\jmath}}$. In particular, in
the case of a finite theory (for which each of the two vertex operator 
algebras has only finitely many irreducible representations) these
conditions typically only allow for a finite number of solutions that
can be classified; this has been done for the case of the so-called
minimal models and the affine theories with group $SU(2)$ by Cappelli,
Itzykson and Zuber \cite{CIZ87a,CIZ87b} (for a modern proof involving
some Galois theory see \cite{Gan99}), and for the affine theories with 
group $SU(3)$ and the $N=2$ superconformal minimal models by Gannon
\cite{Gan96,Gan97}. 
\bigskip

This concludes our brief overview over the general structure of a
local conformal field theory. For the rest of the paper we shall
mainly concentrate on the theory that is defined on the sphere. Let us
begin by analysing the meromorphic conformal subtheory in some detail.

\section{Meromorphic Conformal Field Theory}

In this section we shall describe in detail the structure of a
meromorphic conformal field theory; our exposition follows closely the
work of Goddard \cite{God89} and Gaberdiel \& Goddard \cite{GabGod98},
and we refer the reader for some of the mathematical details (that
shall be ignored in the following) to these papers.

\subsection{Amplitudes and M\"obius Covariance}

As we have explained above, a meromorphic conformal field theory is
determined in terms of its space of states $\H_0$, and the amplitudes
involving arbitrary elements $\psi_i$ in a dense subspace $\F_0$ of
$\H_0$. Indeed, for each state $\psi\in\F_0$, there exists a 
{\em vertex operator} $V(\psi,z)$ that creates the state $\psi$ from
the vacuum (in a sense that will be described in more detail shortly), 
and the amplitudes are the vacuum expectation values of the
corresponding product of vertex operators, 
\be
\label{mero}
\A(\psi_1,\ldots,\psi_n;z_1,\ldots, z_n) = 
\left\langle V(\psi_1,z_1) \cdots V(\psi_n,z_n)\right\rangle \,.
\ee
Each vertex operator $V(\psi,z)$ depends linearly on $\psi$, and the 
amplitudes are {\em mero\-mor\-phic} functions that are defined on
the Riemann sphere $\rs=\Cop\cup\{\infty\}$, \ie\ they are analytic
except for possible poles at $z_i=z_j$, $i\ne j$. The operators are
furthermore assumed to be {\em local} in the sense that for $z\ne\zeta$ 
\be
\label{local}
V(\psi,z) V(\phi,\zeta) = \varepsilon \; V(\phi,\zeta) V(\psi,z) \,,
\ee
where $\varepsilon=-1$ if both $\psi$ and $\phi$ are fermionic, and 
$\varepsilon=+1$ otherwise. In formulating (\ref{local}) we have
assumed that $\psi$ and $\phi$ are states of definite fermion number;
more precisely, this means that $\F_0$ decomposes as
\be
\F_0 = \F_0^B \oplus \F_0^F \,,
\ee
where $\F_0^B$ and $\F_0^F$ is the subspace of bosonic and fermionic
states, respectively, and that both $\psi$ and $\phi$ are either in
$\F_0^B$ or in $\F_0^F$. In the following we shall always only
consider states of definite fermion number.

In terms of the amplitudes, the locality condition (\ref{local}) 
is equivalent to the property that
\beq
\fl \A(\psi_1,\ldots,\psi_i,\psi_{i+1},\ldots,\psi_n;z_1,\ldots, 
z_i,z_{i+1},\ldots,z_n) \nonumber \\
\lo = \varepsilon_{i,i+1}
\A(\psi_1,\ldots,\psi_{i+1},\psi_{i},\ldots,\psi_n;z_1,\ldots,   
z_{i+1},z_{i},\ldots,z_n)\,, 
\label{localamp}
\eeq
and $\varepsilon_{i,i+1}$ is defined as above. As the amplitudes
are essentially independent of the order of the fields, we shall 
sometimes also write them as 
\be
\A(\psi_1,\ldots,\psi_n;z_1,\ldots, z_n) = 
\langle \prod_{i=1}^{n} V(\psi_i,z_i) \rangle \,.
\ee
We may assume that $\F_0$ contains a (bosonic) state $\Omega$ that has
the property that its vertex operator $V(\Omega,z)$ is the identity
operator; in terms of the amplitudes this means that
\be
\langle V(\Omega,z) \prod_{i=1}^{n} V(\psi_i,z_i) \rangle
= \langle \prod_{i=1}^{n} V(\psi_i,z_i) \rangle\,.
\ee
We call $\Omega$ the {\em vacuum (state)} of the theory. Given
$\Omega$, the state $\psi\in\F_0$ that is associated to the vertex
operator $V(\psi,z)$ can then be defined as 
\be\label{statefield}
\psi = V(\psi,0) \Omega \,.
\ee 

In conventional quantum field theory, the states of the theory
transform in a suitable way under the Poincar\'e group, and the
amplitudes are therefore covariant under Poincar\'e
transformations. In the present context, the r\^ole of the Poincar\'e
group is played by the group of M\"obius transformations $\M$, \ie\
the group of (complex) automorphisms of the Riemann sphere. These are 
the transformations of the form
\be
\label{Moeb}
z\mapsto \gamma(z) = {az + b \over cz+d} \,,\qquad \hbox{where}
\quad a,b,c,d\in\Cop\,, \quad ad-bc=1 \,.
\ee
We can associate to each element
\be
A= \left(\begin{array}{cc} a & b \\ c& d \end{array} \right) \in
\SL(2,\Cop) \,,
\ee
the M\"obius transformation (\ref{Moeb}), and since $A\in \SL(2,\Cop)$
and $-A\in \SL(2,\Cop)$ define the same M\"obius transformation, the
group of M\"obius transformations $\M$ is isomorphic to 
$\M\cong \SL(2,\Cop)/\Zop_2$. In the following we shall sometimes use
elements of $\SL(2,\Cop)$ to denote the corresponding elements of $\M$
where no confusion will result.

It is convenient to introduce a set of generators for $\M$ by
\be
\label{Ls}
e^{\lambda \sL_{-1}}(z) = z+\lambda,\qquad
e^{\lambda \sL_0}(z)=e^\lambda z,\qquad
e^{\lambda \sL_1}(z) = {z\over 1-\lambda z} \,,
\ee
where the first transformation is a {\em translation}, the second is a
{\em scaling}, and the last one is usually referred to as a
{\em special conformal transformation}. Every M\"obius transformation
can be obtained as a composition of these transformations, and for
M\"obius transformations with $d\ne 0$, this can be compactly
described as  
\be
\label{gamma}
\gamma = \exp \left[ {b \over d} \sL_{-1} \right] \;
d^{- 2 \sL_0} \exp\left[ - {c \over d} \sL_{1} \right] \,,
\ee
where $\gamma$ is given as in (\ref{Moeb}). In terms of $\SL(2,\Cop)$, 
the three transformations in (\ref{Ls}) are
\be
\fl e^{\lambda \sL_{-1}}= \left(
\begin{array}{cc} 
1&\lambda\\ 
0&1
\end{array} \right)\,,
\qquad
e^{\lambda \sL_0}=\left(
\begin{array}{cc} 
e^{\half\lambda}&0\\
0&e^{-\half\lambda}
\end{array} \right)\,,
\qquad
e^{\lambda \sL_1} = \left(
\begin{array}{cc} 
1&0\\ 
-\lambda&1\end{array} \right)\,.
\ee
The corresponding infinitesimal generators (that are complex 
$2\times 2$ matrices with vanishing trace) are then
\be
\sL_{-1} = \left(
\begin{array}{cc}
0 & 1 \\ 
0 & 0 
\end{array} \right)\,, \qquad
\sL_0 = \left(
\begin{array}{cc}
\half & 0 \\ 
0 & -\half 
\end{array} \right)\,, \qquad
\sL_{1} = \left(
\begin{array}{cc}
0 & 0 \\ 
-1 & 0 \end{array} \right)\,.
\ee
They form a basis for the Lie algebra $sl(2,\Cop)$ of $\SL(2,\Cop)$,
and satisfy the commutation relations 
\be
\label{sl2}
[\sL_m,\sL_n] = (m-n) \sL_{m+n}, \qquad m, n = 0, \pm 1\,.
\ee

As in conventional quantum field theory, the states of the meromorphic
theory form a representation of this algebra which can be decomposed
into irreducible representations. The (irreducible) representations that
are relevant in physics are those that satisfy the condition of 
{\em positive energy}. In the present context, since $L_0$ (the
operator associated to $\sL_0$) can be identified with the energy
operator (up to some constant), these are those representations for
which the spectrum of $L_0$ is bounded from  below. This will follow
from the {\em cluster property} of the amplitudes that will be
discussed below. In a given irreducible highest weight representation,
let us denote by $\psi$ the state for which $L_0$ takes the minimal
value, $h$ say.\footnote{We shall assume here that there is only one
such state; this is always true in irreducible representations.} Using
(\ref{sl2}) we then have 
\be
L_0 L_1 \psi = [L_0,L_1] \psi + h L_1 \psi = (h-1) L_1 \psi \,,
\ee
where $L_n$ denotes the operator corresponding to $\sL_n$. Since
$\psi$ is the state with the minimal value for $L_0$, it
follows that $L_1\psi=0$; states with the property
\be
\label{quasi}
L_1 \psi =0 \qquad L_0 \psi = h \psi 
\ee
are called {\em quasiprimary}, and the real number $h$ is called the
{\em conformal weight} of $\psi$. Every quasiprimary state $\psi$
generates a representation of $sl(2,\Cop)$ that consists of the 
$L_{-1}$-descendants (of $\psi$), \ie\ the states of the form
$L_{-1}^n \psi$ where $n=0,1,\ldots$. This infinite-dimensional
representation is irreducible unless $h$ is a non-positive
half-integer. Indeed,
\beq
L_1 L_{-1}^n \psi & =  
        \sum_{l=0}^{n-1} L_{-1}^l [L_1,L_{-1}] L_{-1}^{n-1-l} \psi \\
     & = 2 \sum_{l=0}^{n-1} (h+n-1-l) L_{-1}^{n-1} \psi \\
     & = 2 n (h+ \half (n-1) ) L_{-1}^{n-1} \psi \,,
\eeq
and thus if $h$ is a non-positive half-integer, the state
$L_{-1}^n\psi$ with $n=1-2h$ and its $L_{-1}$-descendants define a
subrepresentation. In order to obtain an irreducible representation
one has to quotient the space of $L_{-1}$-descendants of $\psi$ by
this subrepresentation; the resulting irreducible representation is
then finite-dimensional.

Since the states of the theory carry a representation of the 
M\"obius group, the amplitudes transform covariantly under M\"obius
transformations. The transformation rule for general states is quite
complicated (we shall give an explicit formula later on), but for
quasiprimary states it can be easily described: let $\psi_i$,
$i=1,\ldots, n$ be $n$ quasiprimary states with conformal weights
$h_i$, $i=1,\ldots, n$, then  
\be
\label{Moebamp}
\langle \prod_{i=1}^{n} V(\psi_i,z_i) \rangle = 
\prod_{i=1}^{n} \left({d\gamma(z_i) \over d z_i} \right)^{h_i}
\langle \prod_{i=1}^{n} V(\psi_i,\gamma(z_i)) \rangle\,,
\ee
where $\gamma$ is a M\"obius transformation as in (\ref{Moeb}).

Let us denote the operators that implement the M\"obius
transformations on the space of states by the same symbols as in
(\ref{Ls}) with $\sL_n$ replaced by $L_n$. Then the transformation
formulae for the vertex operators are given as 
\beq
e^{\lambda L_{-1}} V(\psi,z) e^{-\lambda L_{-1}} & = V(\psi,z+\lambda)
\label{trans} \\ 
x^{L_0}\; V(\psi,z) x^{-L_0} & = x^h \; V(\psi,xz) \label{scale}\\
e^{\mu L_1} V(\psi,z) e^{-\mu L_1} & = (1-\mu z)^{-2h} 
V(\psi, z / (1-\mu z)) \label{special} \,,
\eeq
where $\psi$ is quasiprimary with conformal weight $h$. We also write
more generally 
\be
\label{quasitrans}
D_\gamma V(\psi,z) D_\gamma^{-1} = \Bigl(\gamma'(z)\Bigr)^h
V\Bigl(\psi,\gamma(z)\Bigr)\,,
\ee
where $D_\gamma$ is given by the same formula as in (\ref{gamma}). 
In this notation we then have 
\be
\label{vac}
D_\gamma\Omega=\Omega
\ee
for all $\gamma$; this is equivalent to $L_n\Omega=0$ for $n=0,\pm 1$.  
The transformation formula for the vertex operator associated to
a quasiprimary field $\psi$ is consistent with the identification of
states and fields (\ref{statefield}) and the definition of a
quasiprimary state (\ref{quasi}): indeed, if we apply (\ref{scale})
and (\ref{special}) to the vacuum, use (\ref{vac}) and set $z=0$, we
obtain 
\be
x^{L_0} \psi = x^h \psi \qquad \hbox{and} \qquad
e^{\mu L_{1}} \psi = \psi 
\ee
which implies that $L_1\psi=0$ and $L_0\psi=h\psi$, and is thus in
agreement with (\ref{quasi}). 

The M\"obius symmetry constrains the functional form of all
amplitudes, but in the case of the one-, two- and three-point
functions it actually determines their functional dependence
completely. If $\psi$ is a quasiprimary state with conformal weight
$h$, then $\langle V(\psi,z)\rangle$ is independent of $z$ because of
the translation symmetry, but it follows from (\ref{scale}) that
\be
\langle V(\psi,z) \rangle = \lambda^h \langle V(\psi,\lambda z)
\rangle \,.
\ee
The one-point function can therefore only be non-zero if $h=0$. Under
the assumption of the cluster property to be discussed in the next
subsection, the only state with $h=0$ is the vacuum, $\psi=\Omega$. 

If $\psi$ and $\phi$ are two quasiprimary states with conformal
weights $h_\psi$ and $h_\phi$, respectively, then the translation
symmetry implies that 
\be
\langle V(\psi,z) V(\phi,\zeta) \rangle = 
\langle V(\psi,z-\zeta) V(\phi,0) \rangle = 
F(z-\zeta)\,,
\ee
and the scaling symmetry gives 
\be
\lambda^{h_\psi + h_\phi} F(\lambda x) = F(x) \,,
\ee
so that 
\be
\label{rest}
F(x) = C x^{-h_\psi - h_\phi} \,,
\ee
where $C$ is some constant. On the other hand, the symmetry under
special conformal transformations implies that 
\be
\langle V(\psi,x) V(\phi,0) \rangle = (1-\mu x)^{-2 h_\psi}
\langle V(\psi,x/(1-\mu x)) V(\phi,0) \rangle\,,
\ee
and therefore, upon comparison with (\ref{rest}), the amplitude can
only be non-trivial if $2h_\psi=h_\psi+h_\phi$, \ie\
$h_\psi=h_\phi$. In this case the amplitude is of the form 
\be
\langle V(\psi,z) V(\phi,\zeta) \rangle = C (z-\zeta)^{-2h_\psi}\,.
\ee
If the amplitude is non-trivial for $\psi=\phi$, the locality
condition implies that $h\in\Zop$ if $\psi$ is a bosonic field, and
$h\in\half+\Zop$ if $\psi$ is fermionic. This is the familiar 
{\em Spin-Statistics Theorem}. 

Finally, if $\psi_i$ are quasiprimary fields with conformal weights
$h_i$, $i=1,2,3$, then 
\be
\label{three}
\fl \langle V(\psi_1,z_1) V(\psi_2,z_2) V(\psi_3,z_3) \rangle = 
\prod_{i<j} \left({a_i - a_j \over z_i-z_j} \right)^{h_{ij}}
\langle V(\psi_1,a_1) V(\psi_2,a_2) V(\psi_3,a_3) \rangle\,,
\ee
where $h_{12}=h_1+h_2-h_3$, {\it etc.}, and $a_i$ are three
distinct arbitrary constants. In deriving (\ref{three}) we have used
the fact that every three points can be mapped to any other three
points by means of a M\"obius transformation.

\subsection{The Uniqueness Theorem}

It follows directly from (\ref{trans}), (\ref{vac}) and
(\ref{statefield}) that 
\be
\label{psivac}
V(\psi,z) \Omega = e^{z L_{-1}} V(\psi,0) \Omega = e^{z L_{-1}} \psi
\,.
\ee
If $V(\psi,z)$ is in addition local, \ie\ if it satisfies
(\ref{local}) for every $\phi$, $V(\psi,z)$ is uniquely characterised
by this property; this is the content of the  

\noindent {\bf Uniqueness Theorem} \cite{God89}: If $U_\psi(z)$ is a
local vertex operator that satisfies
\be
\label{Upsi}
U_\psi(z) \Omega = e^{zL_{-1}} \psi
\ee
then
\be
U_\psi(z) = V(\psi,z) 
\ee
on a dense subspace of $\H_0$.

\noindent {\bf Proof:} Let $\chi\in\F_0$ be arbitrary. Then 
\be
U_\psi(z) \chi = U_\psi(z) V(\chi,0)\Omega = \varepsilon_{\chi,\psi}
V(\chi,0) U_\psi(z) \Omega = \varepsilon_{\chi,\psi}
V(\chi,0) e^{zL_{-1}} \psi\,,
\ee
where we have used the locality of $U_\psi(z)$ and (\ref{Upsi}) and
$\varepsilon_{\chi,\psi}$ denotes the sign in (\ref{local}). We
can then use (\ref{psivac}) and the locality of $V(\psi,z)$ to rewrite
this as 
\be
\fl
\varepsilon_{\chi,\psi} V(\chi,0) e^{zL_{-1}} \psi 
= \varepsilon_{\chi,\psi} V(\chi,0) V(\psi,z) \Omega 
= V(\psi,z) V(\chi,0) \Omega = V(\psi,z) \chi \,,
\ee
and thus the action of $U_\psi(z)$ and $V(\psi,z)$ agrees on the dense
subspace $\F_0$. 
\medskip

Given the uniqueness theorem, we can now deduce the transformation
property of a general vertex operator under M\"obius transformations
\be
\label{Moebgen}
D_\gamma V(\psi,z) D_\gamma^{-1} = 
V\left[ \left( {d\gamma \over dz} \right)^{L_0} \;
\exp\left({\gamma''(z)\over 2 \gamma'(z)} L_{1}\right) 
\psi,\gamma(z)\right]\,.
\ee
In the special case where $\psi$ is quasiprimary, 
$\exp(\gamma''(z)/ 2 \gamma'(z) L_{1})\psi=\psi$, and (\ref{Moebgen})
reduces to (\ref{quasitrans}). To prove (\ref{Moebgen}), we observe
that the uniqueness theorem implies that it is sufficient to evaluate
the identity on the vacuum, in which case it becomes 
\be
D_\gamma e^{z L_{-1}} \psi =
e^{\gamma(z) L_{-1}} (cz+d)^{-2 L_0} 
e^{- {c \over cz+d} L_{1}} \psi\,,
\ee
where we have written $\gamma$ as in (\ref{Moeb}). This then follows 
from  
\beq
\fl \left(\begin{array}{cc} a & b \\ c & d \end{array}\right)
\left(\begin{array}{cc} 1 & z \\ 0 & 1 \end{array}\right)
& = \left(\begin{array}{cc} a & az+b \\ c & cz+d \end{array}\right) \\
& = \left(\begin{array}{cc} 1 & {az+b\over cz+d} \\ 
        0 & 1 \end{array}\right)
\left(\begin{array}{cc} (cz+d)^{-1} & 0 \\ 
        0 & (cz+d) \end{array}\right)
\left(\begin{array}{cc} 1 & 0 \\ {c\over cz+d} & 1 \end{array}\right)
\eeq
together with the fact that $\M\cong \SL(2,\Cop)/\Zop_2$. 

We can now also deduce the behaviour under infinitesimal
trans\-formations from (\ref{Moebgen}). For example, if $\gamma$ is an
infinitesimal translation, $\gamma(z)=z+\delta$, then to first order
in $\delta$, (\ref{Moebgen}) becomes 
\be
V(\psi,z) + \delta [L_{-1},V(\psi,z)] = V(\psi,z) 
+ \delta {dV \over dz}(\psi,z) \,,
\ee
from which we deduce that 
\be
\label{inftrans}
[L_{-1},V(\psi,z)] = {dV \over dz}(\psi,z) \,.
\ee
Similarly, we find that 
\be
\label{infscale}
[L_0,V(\psi,z)] =  z {d \over dz} V(\psi,z) + V(L_0\psi,z) \,,
\ee
and
\be
[L_1,V(\psi,z)] =  z^2 {d \over dz} V(\psi,z) + 2 z V(L_0\psi,z)
                      + V(L_1\psi,z) \,.
\ee
If $\psi$ is quasiprimary of conformal weight $h$, the last three
equations can be compactly written as 
\be
\label{quasitranscom}
[L_n,V(\psi,z)] = z^n \left\{ z {d\over dz} + (n+1) h \right\}
V(\psi,z) \qquad \hbox{for $n=0,\pm 1$.}
\ee
Finally, applying (\ref{inftrans}) to the vacuum we have
\be
e^{z L_{-1}} L_{-1} \psi = e^{z L_{-1}} {dV \over dz}(\psi,0)\Omega
\,,
\ee
and this implies, using the uniqueness theorem, that
\be
\label{deri}
{dV \over dz}(\psi,z) = V(L_{-1}\psi,z) \,.
\ee
In particular, it follows that the correlation functions of
$L_{-1}$-descendants of quasi-primary states can be directly deduced
from those that only involve the qua\-si\-pri\-mary states
themselves.

\subsection{Factorisation and the Cluster Property}

As we have explained above, a meromorphic conformal field theory is
determined by its space of states $\H_0$ together with the set of 
amplitudes that are defined for arbitrary elements in a dense subspace
$\F_0$ of $\H_0$. The amplitudes contain all relevant information
about the vertex operators; for example the locality and M\"obius
transformation properties of the vertex operators follow from the
corresponding properties of the amplitudes (\ref{localamp}), 
and (\ref{Moebamp}).

In practice, this is however not a good way to define a conformal
field theory, since $\H_0$ is always infinite-dimensional (unless the
meromorphic conformal field theory consists only of the vacuum), and it
is unwieldy to give the correlation functions for arbitrary
combinations of elements in an infinite-dimensional (dense) subspace
$\F_0$ of $\H_0$. Most (if not all) theories of interest however
possess a finite-dimensional subspace $V\subset\H_0$ that is not
dense in $\H_0$ but that {\em generates} $\H_0$ in the sense that
$\H_0$ and all its amplitudes can be derived from those only involving
states in $V$; this process is called {\em factorisation}.

The basic idea of factorisation is very simple: given the amplitudes
involving states in $V$, we can define the vector space that consists
of linear combinations of states of the form
\be
\Psi= V(\psi_1,z_1) \cdots V(\psi_n,z_n) \Omega \,,
\ee
where $\psi_i\in V$, and $z_i\ne z_j$ for $i\ne j$. We identify two
such states if their difference vanishes in all amplitudes (involving
states in $V$), and denote the resulting vector space by $\hF_0$. We
then say that $V$ generates $\H_0$ if $\hF_0$ is dense in
$\H_0$. Finally we can introduce a vertex operator for $\Psi$ by   
\be
V(\Psi,z) = V(\psi_1,z_1+z) \cdots V(\psi_n,z_n+z) \,,
\ee
and the amplitudes involving arbitrary elements in $\hF_0$ are thus
determined in terms of those that only involve states in $V$. (More
details of this construction can be found in \cite{GabGod98}.) In the 
following, when we shall give examples of meromorphic conformal field
theories, we shall therefore only describe the theory associated to
a suitable generating space $V$.

It is easy to check that the locality and M\"obius transformation
properties of the amplitudes involving only states in $V$ are
sufficient to guarantee the corresponding properties for the
amplitudes involving arbitrary states in $\hF_0$, and therefore for the
conformal field theory that is obtained by factorisation from $V$. 
However, the situation is more complicated with respect to the
condition that the states in $\H_0$ are of positive energy, \ie\ that
the spectrum of $L_0$ is bounded from below, since this clearly does
not follow from the condition that this is so for the states in $V$.
In the case of the meromorphic theory the relevant spectrum condition 
is actually slightly stronger in that it requires that the spectrum of
$L_0$ is non-negative, and that there exists a {\em unique} state, the
vacuum, with $L_0=0$. This stronger condition (which we shall always
assume from now on) is satisfied for the meromorphic theory obtained
by factorisation from $V$ provided the amplitudes in $V$ satisfy the
{\em cluster property};  this states that if we separate the variables
of an amplitude into two sets and scale one set towards a fixed point
({\it e.g.} $0$ or $\infty$) the behaviour of the amplitude is
dominated by the product of two amplitudes, corresponding to the two
sets of variables, multiplied by an appropriate power of the
separation, specifically   
\be
\fl\left\langle\prod_i V(\phi_i,\zeta_i)\prod_j V(\psi_j,\lambda
z_j)\right\rangle
\sim \left\langle\prod_i V(\phi_i,\zeta_i)\right\rangle
\left\langle\prod_j V(\psi_j,z_j)\right\rangle \lambda^{-\Sigma h_j}
\quad \hbox{as $\lambda\rightarrow 0$}\,,
\ee
where $\phi_i, \psi_j\in V$ have conformal weight $h'_i$ and $h_j$,
respectively. (Here $\sim$ means that the two sides of the equation
agree up to terms of lower order in $\lambda$.)  Because of the
M\"obius covariance of the amplitudes this is equivalent to
$$
\left\langle\prod_i V(\phi_i,\lambda\zeta_i)\prod_j V(\psi_j,
z_j)\right\rangle
\sim \left\langle\prod_i V(\phi_i,\zeta_i)\right\rangle
\left\langle\prod_j V(\psi_j,z_j)\right\rangle \lambda^{-\Sigma h_i'}
\quad \hbox{as $\lambda\rightarrow \infty$}\,.
$$

To prove that this implies that the spectrum of $L_0$ is non-negative
and that the vacuum is unique, let us introduce the projection
operators defined by  
\be
P_N = \oint_0 u^{L_0-N-1}du, \qquad\hbox{for } N\in\Zop / 2\,,
\ee
where we have absorbed a factor of $1/2\pi i$ into the definition of
the symbol $\oint$. In particular, we have
\be
P_N \prod_j V(\psi_j,z_j)\Omega = \oint du \; u^{h - N -1}
\prod_j V(\psi_j,uz_j)\; \Omega  \,,
\ee
where $h=\sum_j h_j$. It then follows that the $P_N$ are projection
operators
\be
P_NP_M=0,\hbox{ if } N\ne M, \qquad P_N^2=P_N, \qquad 
\sum_N P_N =1
\ee
onto the eigenspaces of $L_0$,
\be
L_0 P_N=N P_N\,.
\ee
For $N\leq 0$, we then have
\beq
\fl\left\langle\prod_i V(\phi_i,\zeta_i)P_N\prod_j V(\psi_j,z_j)
\right\rangle
&= & \oint_0 u^{\Sigma h_j- N-1}
\left\langle\prod_i V(\phi_i,\zeta_i)\prod_j V(\psi_j,u z_j)
\right\rangle du\nonumber \\
&\sim & \left\langle\prod_i V(\phi_i,\zeta_i)\right\rangle
\left\langle\prod_j V(\psi_j,z_j)\right\rangle \oint_{|u|=\rho}
u^{-N-1}du 
\end{eqnarray}
which, by taking $\rho\rightarrow 0$, is seen to vanish for $N<0$ and,
for $N=0$, to give 
\be
P_0\prod_j V(\psi_j,z_j)\Omega 
=\Omega\left\langle\prod_j V(\psi_j,z_j)\right\rangle\,,
\ee
and so $P_0\Psi = \Omega\,\langle\Psi\rangle$. Thus the cluster 
decomposition property implies that $P_N=0$ for $N<0$, \ie\ that 
the spectrum of $L_0$ is non-negative, and that $\Omega$ is the unique
state with $L_0=0$. The cluster property also implies that the space
of states can be completely decomposed into irreducible
representations of the Lie algebra $sl(2,\Cop)$ that corresponds to
the M\"obius transformations (see Appendix~D of \cite{GabGod98}).

\subsection{The Operator Product Expansion}

One of the most important consequences of the uniqueness theorem is
that it allows for a direct derivation of the {\em duality relation}
which in turn gives rise to the operator product expansion.

\noindent {\bf Duality Theorem} \cite{God89}: Let $\psi$ and $\phi$ be
states in $\F_0$, then 
\be
\label{dual}
V(\psi,z) V(\phi,\zeta) = V \Bigl(V(\psi,z-\zeta) \phi, \zeta \Bigr)
\,.
\ee

\noindent {\bf Proof:} By the uniqueness theorem it is sufficient to
evaluate both sides on the vacuum, in which case (\ref{dual})
becomes 
\beq
V(\psi,z) V(\phi,\zeta) \Omega & = V(\psi,z) e^{\zeta L_{-1}} \phi \\
& = e^{\zeta L_{-1}} V(\psi,z-\zeta) \phi \\
& = V \Bigl(V(\psi,z-\zeta) \phi, \zeta \Bigr) \Omega \,,
\eeq
where we have used (\ref{trans}).
\medskip

For many purposes it is convenient to expand the fields $V(\psi,z)$ in
terms of modes
\be
\label{modes}
V(\psi,z) = \sum_{n\in \Zop-h} V_n(\psi) z^{-n-h}\,,
\ee
where $\psi$ has conformal weight $h$, \ie\ $L_0\psi=h\psi$. The modes
can be defined in terms of a contour integral as
\be
V_n(\psi) = \oint z^{h+n-1} V(\psi,z) dz\,,
\ee
where the contour encircles $z=0$ anticlockwise. In terms of the modes
the identity $V(\psi,0)\Omega=\psi$ implies that 
\be
\label{vacrels}
V_{-h}(\psi)\Omega = \psi \qquad \hbox{and} \qquad
V_{l}(\psi)\Omega = 0 \quad \hbox{for $l>-h$.}
\ee
Furthermore, if $\psi$ is quasiprimary, (\ref{quasitranscom}) becomes 
\be
\label{Lcomm}
[L_m,V_n(\psi)] = \left( m (h-1) - n \right) V_{m+n}(\psi)
\qquad \hbox{$m=0,\pm 1$.}
\ee
Actually, the equations for $m=0,-1$ do not require that $\psi$ is 
quasiprimary as follows from (\ref{inftrans}) and (\ref{infscale});
thus we have that $[L_0,V_n(\psi)]=-n V_n(\psi)$ for all $\psi$, so
that $V_n(\psi)$ lowers the eigenvalue of $L_0$ by $n$. 

Given the modes of the conformal fields, we can introduce the
{\em Fock space} $\widetilde{\F}_0$ that is spanned by eigenstates of
$L_0$ and that forms a dense subspace of the space of states. This
space consists of finite linear combinations of vectors of the form 
\be
\label{Psi}
\Psi = V_{n_1}(\psi_1)V_{n_2}(\psi_2)\cdots
V_{n_N}(\psi_N)\Omega\,,
\ee
where $n_i+h_i\in \Zop$, $h_i$ is the conformal weight of
$\psi_i$, and we may restrict $\psi_i$ to be in the subspace $V$ that 
generates the whole theory by factorisation. Because of (\ref{Lcomm})
$\Psi$ is an eigenvector of $L_0$ with eigenvalue
\be
L_0 \Psi = h \psi \qquad \hbox{where $h= - \sum_{i=1}^{N} n_i$.}
\ee
The Fock space $\widetilde{\F}_0$ is a quotient space of the vector
space $\W_0$ whose basis is given by the states of the form
(\ref{Psi}); the subspace by which $\W_0$ has to be divided consists
of linear combinations of states of the form (\ref{Psi}) that vanish
in all amplitudes. 

We can also introduce a vertex operator for $\Psi$ by the formula
\be
\label{composite}
\fl V(\Psi,z)=\oint_{\C_1}z_1^{h_1+n_1-1}V(\psi_1,z+z_1)dz_1\cdots
\oint_{\C_N}z_N^{h_N+n_N-1}V(\psi_N,z+z_N)dz_N\,,
\ee
where the $\C_j$ are contours about $0$ with $|z_i|>|z_j|$ if
$i<j$. The Fock space $\widetilde{\F}_0$ thus satisfies the conditions
that we have required of the dense subspace $\F_0$, and we may
therefore assume that $\F_0$ {\em is} actually the Fock space of the
theory; from now on we shall always do so.

The duality property of the vertex operators can now be rewritten in
terms of modes as 
\beq
V(\phi,z)V(\psi,\zeta) & = & V(V(\phi,z-\zeta)\psi,\zeta)
\nonumber \\
&= & \sum_{n\leq h_\psi} V(V_n(\phi)\psi,\zeta)
(z-\zeta)^{-n-h_\phi} \label{ope}\,,
\eeq
where $L_0\psi=h_\psi\psi$ and $L_0\phi=h_\phi\phi$, and 
$\psi,\phi\in\F_0$. The sum over $n$ is bounded by $h_\psi$, since 
$L_0 V_n(\phi)\psi = (h_\psi - n) V_n(\phi)\psi$, and the spectrum
condition implies that the theory does not contain any states of
negative conformal weight. The equation (\ref{ope}) is known as the
{\em Operator Product Expansion}. The infinite sum converges provided
that all other meromorphic fields in a given amplitude are further
away from $\zeta$ than $z$. 

We can use (\ref{ope}) to derive a formula for the commutation 
relations of modes as follows.\footnote{To be precise, the following
construction {\it a priori} only defines a Lie bracket for the
quotient space of modes where we identify modes whose action on the
Fock space of the meromorphic theory coincides.} The commutator of two
modes $V_m(\phi)$ and $V_n(\psi)$ is given as
\beq
[V_m(\Phi),\,V_n(\Psi)] & = &
{\oint dz\oint d\zeta}_{\hskip-42pt
{\atop {\atop {\atop \scriptstyle |z|>|\zeta|}}}\hskip16pt}
z^{m+h_\phi-1}
\zeta^{n+h_\psi-1} V(\phi,z) V(\psi,\zeta) \nonumber \\
& & -
{\oint dz\oint d\zeta}_{\hskip-42pt
{\atop {\atop {\atop \scriptstyle |\zeta|>|z|}}}\hskip16pt}
z^{m+h_\phi-1} \zeta^{n+h_\psi-1} V(\phi,z) V(\psi,\zeta) \label{45}
\eeq
where the contours on the right-hand side encircle the origin
anti-clockwise. We can then deform the two contours so as to rewrite 
(\ref{45}) as  
\be
\label{46}
\fl [V_m(\phi),\,V_n(\psi)]=\oint_0 \zeta^{n+h_\psi-1}d\zeta\, 
\oint_\zeta z^{m+h_\phi-1} dz\,
\sum_{l} V(V_l(\phi)\psi,\zeta) (z-\zeta)^{-l-h_\phi} \,,
\ee
where the $z$ contour is a small positive circle about $\zeta$ and the
$\zeta$ contour is a positive circle about the origin. Only terms with  
$l\geq 1-h_\phi$ contribute, and the integral becomes
\be
\label{47}
[V_m(\phi),V_n(\psi)] = \sum_{N=-h_\phi+1}^{h_\psi}
\left( \begin{array}{c}
m+h_\phi-1\\ 
m-N
\end{array} \right) V_{m+n}(V_N(\phi)\psi)\,.
\ee
In particular, if $m\geq-h_\phi+1$, $n\geq-h_\psi+1$, 
then $m-N\geq0$ in the sum, and $m+n\geq N+n\geq N-h_\psi+1$. This
implies that the modes $\left\{V_m(\psi) :m\geq-h_\psi+1\right\}$ 
close as a Lie algebra. The same also holds for the modes
$\left\{V_m(\psi) :m\leq h_\psi-1\right\}$, and therefore for their
intersection 
\be
\sL^0 = \left\{ V_n(\psi) : -h_\psi+1 \leq n \leq h_\psi -1 \right\}
\,.
\ee
This algebra is sometimes called the {\em vacuum-preserving algebra}
since any element in $\sL^0$ annihilates the vacuum. A certain
deformation of $\sL^0$ defines a finite Lie algebra that can be
interpreted as describing the finite $W$-symmetry of the conformal field
theory \cite{BowWat92a}. It is also clear that the subset of all
positive, all negative or all zero modes form closed Lie algebras,
respectively.

\subsection{The Inner Product and Null-vectors}

We can define an (hermitian) inner product on the Fock space $\F_0$ 
provided that the amplitudes are hermitian in the following sense: 
there exists an antilinear involution $\psi \mapsto \psbar$ for each
$\psi\in\F_0$ such that the amplitudes satisfy
\be
\label{herm}
 \left(\langle \prod_{i=1}^n V(\psi_i,z_i)\rangle\right)^\ast =
\langle \prod_{i=1}^n
V(\psbar_i,\bz_i)\rangle\,.
\ee
If this condition is satisfied, we can define an inner product by
\be
\langle \psi,\phi\rangle = \lim_{z\rightarrow 0} 
\left\langle 
V\left( \left(-{1\over \bz^2}\right)^{L_0} 
\exp\left[ - {1\over \bz} L_1 \right] \psbar, 
{1\over \bz} \right) V(\phi,z) \right\rangle \,.
\ee
This inner product is hermitian, \ie
\be
\label{hermtest}
\langle \psi,\phi\rangle^* = \langle \phi,\psi\rangle 
\ee
since (\ref{herm}) implies that the left-hand-side of (\ref{hermtest})
is  
\be
\lim_{z\rightarrow 0} 
\left\langle 
V\left( \left(- {1\over z^2}\right)^{L_0} 
\exp\left[ - {1\over z} L_1 \right] \psi, 
 {1\over z} \right) V(\bar\phi,\bz) \right\rangle \,,
\ee
and the covariance under the M\"obius transformation $\gamma(z)=1/z$
then implies that this equals
\be
\lim_{z\rightarrow 0} 
\left\langle 
V\left(\left( -{1\over \bz^2}\right)^{L_0} 
\exp\left[ - {1\over \bz} L_1 \right] \bar\phi, 
{1\over \bz} \right) V(\psi,z) \right\rangle \,.
\ee
By a similar calculation we find that the adjoint of a vertex operator
is given by 
\be\label{adfield}
\left( V(\psi,\zeta) \right)^\dagger = V\left( 
\left({1\over {\bar\zeta}^2} \right)^{ L_0}
\exp\left[ - {1 \over \bar\zeta} L_1 \right] \bar\psi, 
{1 \over \bar\zeta} \right) \,,
\ee
where the adjoint is defined to satisfy
\be
\langle\chi, V(\psi,\zeta) \phi\rangle = 
\left\langle \left( V(\psi,\zeta) \right)^\dagger \chi,
\phi\right\rangle \,.
\ee
Since $\psi\mapsto\bar\psi$ is an involution, we can choose a basis
of {\em real} states, \ie\ states that satisfy $\bar\psi=\psi$. If
$\psi$ is a quasiprimary real state, then (\ref{adfield}) simplifies to 
\be
\left( V(\psi,\zeta) \right)^\dagger = 
\left(- {1\over {\bar\zeta}^2}\right)^h
V(\psi, 1/ \bar\zeta)\,,
\ee
where $h$ denotes the conformal weight of $\psi$. In this case the
adjoint of the mode $V_n(\psi)$ is
\beq
\left(V_n(\psi)\right)^\dagger & = & 
\oint d \bz \bz^{h+n-1} \left(-{1\over \bz^2}\right)^h 
V(\psi,1/\bz)  \nonumber \\
& = & (-1)^h \oint d \bar\zeta \bar\zeta^{h-n+1} V(\psi,\bar\zeta)
\nonumber \\
& = & (-1)^h V_{-n}(\psi)\,.\label{adjoint}
\eeq
By a similar calculation it also follows that the adjoint of the
M\"obius generators are given as 
\be
\label{Moebadj}
L_{\pm 1}^\dagger = L_{\mp 1} \qquad \qquad L_0^\dagger = L_0 \,.
\ee
All known conformal field theories satisfy (\ref{herm}) and thus
possess a hermitian inner product; from now on we shall therefore
sometimes assume that the theory has such an inner product.

The inner product can be extended to the vector space $\W_0$ whose
basis is given by the states of the form (\ref{Psi}). Typically, the
inner product is degenerate on $\W_0$, \ie\ there exist vectors
$\Nu\in\W_0$ for which 
\be
\label{null}
\langle \psi,\Nu \rangle = 0 \qquad \hbox{for all $\psi\in\W_0$.}
\ee
Every vector with this property is called a {\em null-vector}. Because
of M\"obius covariance, the field corresponding to $\Nu$ vanishes in
all amplitudes, and therefore $\Nu$ is in the subspace by which $\W_0$
has to be divided in order to obtain the Fock space $\F_0$. Since this
is the case for every null-vector of $\W_0$, it follows that the inner 
product is non-degenerate on $\F_0$. 

In general, the inner product may not be positive definite, but there
exist many interesting theories for which it is; in this case the
theory is called {\em unitary}. For unitary theories, the spectrum of
$L_0$ is always bounded by $0$. To see this we observe that if $\psi$
is a quasiprimary state with conformal weight $h$, then
\beq
\langle L_{-1} \psi, L_{-1}\psi \rangle & = &
\langle \psi, L_1 L_{-1} \psi \rangle \nonumber \\
& = & 2 h \langle \psi,\psi \rangle \,,\label{137}
\eeq
where we have used (\ref{Moebadj}). If the theory is unitary then both
sides of (\ref{137}) have to be non-negative, and thus $h\geq 0$.

\subsection{Conformal Structure}

Up to now we have described what could be called `meromorphic field
theory' rather than `meromorphic conformal field theory' (and that is,
in the mathematical literature, sometimes referred to as a vertex
algebra, rather than a vertex operator algebra). Indeed, we have not
yet discussed the conformal symmetry of the correlation functions but
only its M\"obius symmetry. A large part of the structure that we
shall discuss in these notes does not actually rely on the presence of
a conformal structure, but more advanced features of the theory do,
and therefore the conformal structure is an integral part of the
theory. 

A meromorphic field theory is called {\em conformal} if the three
M\"obius generators $L_0$, $L_{\pm 1}$ are the modes of a field
$L$ that is then usually called the {\em stress-energy tensor} or the  
{\em Virasoro field}. Because of (\ref{sl2}), (\ref{Lcomm}) and
(\ref{47}), the field in question must be a quasiprimary field of
conformal weight $2$ that can be expanded as 
\be
L(z) = \sum_{n=-\infty}^{\infty} L_n z^{-n-2} \,.
\ee
If we write $L(z)=V(\psi_L,z)$, the commutator in (\ref{47}) becomes 
\beq
[L_m,L_n] & = & \sum_{N=-1}^{2} 
\left(\begin{array}{c} m+1 \\ m_N \end{array}\right) 
V_{m+n}\left(L_N \psi_L\right) \nonumber \\
& = & {m (m^2-1) \over 6} V_{m+n}(L_2 \psi_L) 
+ {m(m+1) \over 2} V_{m+n}(L_1 \psi_L) \nonumber \\
& & + (m+1) V_{m+n}(L_0 \psi_L) + V_{m+n}(L_{-1} \psi_L)\,.
\label{Lope}
\eeq
All these expressions can be evaluated further
\cite{God89}\footnote{This is also known as the L\"uscher-Mack
Theorem, see \cite{LusMac76,FST89,Mac87}.}: since $L_2\psi_L$ has
conformal weight $h=0$, the uniqueness of the vacuum implies that it
must be proportional to the vacuum vector,  
\be
L_2 \psi_L = L_2 L_{-2} \Omega = {c\over 2} \Omega \,,
\ee
where $c$ is some constant. Also, since the vacuum vector acts as the
identity operator, $V_n(\Omega)=\delta_{n,0}$.  Furthermore,
$L_1\psi_L=0$ since $L$ is quasiprimary, and $L_0\psi_L=2\psi_L$ since
$L$ has conformal weight $2$. Finally, because of (\ref{deri}), 
\be
V(L_{-1}\psi,z) = {d \over dz} \sum_n V_n(\psi) z^{-n-h} 
= - \sum_n (n+h) V_n(\psi) z^{-n-(h+1)} \,,
\ee
and since $L_{-1}\psi$ has conformal weight $h+1$ (if $\psi$ has
conformal weight $h$), 
\be
\label{derimode}
V_{n}(L_{-1}\psi) = - (n+h) V_{n}(\psi) \,.
\ee
Putting all of this together we then find that (\ref{Lope}) becomes
\be
\label{Virasoro}
[L_m,L_n] = (m-n) L_{m+n} 
+ {c \over 12} m (m^2-1) \delta_{m+n,0}\,.
\ee
This algebra is called the {\em Virasoro algebra} \cite{Vir70}, and
the parameter $c$ is called the {\em central charge}. The real algebra
defined by (\ref{Virasoro}) is the Lie algebra of the central
extension of the group of diffeomorphisms of the circle
(see {\it e.g.} \cite{PreSeg}). 

If the theory contains a Virasoro field, the states transform in
representations of the Virasoro algebra (rather than just the Lie
algebra of $sl(2,\Cop)$ that corresponds to the M\"obius
transformations).  Under suitable conditions (for example
if the theory is unitary), the space of states can then
be completely decomposed into irreducible representations of the
Virasoro algebra. Because of the spectrum condition, the relevant
representations are then highest weight representations that are
generated from a {\em primary} state $\psi$, \ie\ a state satisfying  
\be
\label{primary}
L_0 \psi = h \psi \qquad L_n \psi = 0 \quad \hbox{for $n>0$.}
\ee
If $\psi$ is primary, the commutation relation (\ref{Lcomm}) holds for
all $m$, \ie\ 
\be
[L_m,V_n(\psi)] = (m (h-1) - n) V_{m+n}(\psi) \qquad \hbox{for all
$m\in\Zop$} 
\ee
as follows from  (\ref{47}) together with (\ref{derimode}). In this
case the conformal symmetry also leads to an extension of the M\"obius 
transformation formula (\ref{quasitrans}) to arbitrary holomorphic
transformations $f$ that are only locally defined,    
\be
\label{generaltrans}
D_f V(\psi,z) D_f^{-1} = \left(f'(z)\right)^h V(\psi,f(z)) \,,
\ee
where $\psi$ is primary and $D_f$ is a certain product of exponentials
of $L_n$ with coefficients that depend on $f$ \cite{Gab94b}. The
extension of (\ref{generaltrans}) to states that are not primary is
also known (but again much more complicated).

\subsection{Examples}

Let us now give a number of examples that exhibit the structures that
we have described so far.

\subsubsection{The Free Boson}

The simplest conformal field theory is the theory that is associated
to a single free boson. In this case $V$ can be taken to be a
one-dimensional vector space, spanned by a vector $J$ of weight $1$,
in which case we write $J(z)\equiv V(J,z)$. The amplitude of an odd
number of $J$-fields is defined to vanish, and in the case of an even 
number it is given by
\beq
\langle J(z_1) \cdots J(z_{2n}) \rangle & = & {k^n \over 2^n n!} 
\sum_{\pi\in S_{2n}} \,
\prod_{j=1}^{n} {1 \over (z_{\pi(j)} - z_{\pi(j+n)})^2} \,,\label{48a}
\\
& = & k^n \sum_{\pi\in S_{2n}'} \,
\prod_{j=1}^{n} {1 \over (z_{\pi(j)} - z_{\pi(j+n)})^2} \,,
\label{48b}
\eeq
where $k$ is an arbitrary (real) constant and, in (\ref{48a}),
$S_{2n}$ is the permutation group on $2n$ objects, whilst, in
(\ref{48b}), the sum is restricted to the subset $S_{2n}'$ of
permutations $\pi\in S_{2n}$ such that $\pi(i)<\pi(i+n)$ and
$\pi(i)<\pi(j)$ if $1\leq i<j\leq n$. It is clear that these
amplitudes are meromorphic and local, and it is easy to check that
they satisfy the condition of M\"obius invariance with the conformal
weight of $J$ being $1$.   

{}From the amplitudes we can directly read off the operator product
expansion of the field $J$ with itself as 
\be
\label{49}
J(z) J(\z) \sim {k \over (z-\z)^2}  \,,
\ee
where we use the symbol $\sim$ to indicate equality up to terms that
are non-singular at $z=\z$. Comparing this with (\ref{ope}), and using 
(\ref{47}) we then obtain
\be
\label{50}
[J_n,J_m] = n k \delta_{n,-m} \,.
\ee
This defines (a representation of) the affine algebra $\hat{u}(1)$. 
$J$ is also sometimes called a $U(1)$-current. The operator product
expansion (\ref{49}) actually contains all the relevant information
about the theory since one can reconstruct the amplitudes from it; to
this end one defines recursively 
\begin{eqnarray}
\langle \;\rangle & = & 1 \\
\langle J(z) \rangle & = & 0
\end{eqnarray}
and
\be
\left\langle J(z) \prod_{i=1}^{n} J(\zeta_i) \right\rangle = 
\sum_{j=1}^{n} {k \over (z-\zeta_i)^2} \;
\left\langle \prod_{\stackrel{i=1}{i\ne j}}^{n} J(\zeta_i)
\right\rangle 
\,.
\ee
Indeed, the two sets of amplitudes have the same poles, and their
difference describes therefore an entire function; all entire
functions on the sphere are constant and it is not difficult to see
that the constant is actually zero. The equality between the two sets
of amplitudes can also be checked directly. 

This theory is actually conformal since the space of states that is
obtained by factorisation from these amplitudes contains the state
\be
\label{U1Vir}
\psi_L = { 1\over 2k} \, J_{-1}J_{-1}\Omega\,,
\ee
which plays the r\^ole of the stress-energy tensor with central charge
$c=1$. The corresponding field (that is defined by (\ref{composite}))
can actually be given directly as 
\be
\label{U1Virfield}
V(\psi_L,z) = L(z) = {1 \over 2k} \nox J(z) J(z) \nox \,,
\ee
where $\nox \nox$ denotes {\em normal ordering}, which, in this
context, means that the singular part of the OPE of $J$ with itself
has been subtracted. In fact, it follows from (\ref{ope}) that 
\beq
J(w) J(z) & = & {1 \over (w-z)^2} V(J_1 J_{-1}\Omega,z) + 
{1 \over (w-z)} V(J_0 J_{-1}\Omega,z) \\
& & \quad + V(J_{-1} J_{-1}\Omega,z) + O(w-z)\,, 
\eeq
and therefore (\ref{U1Virfield}) implies  (\ref{U1Vir}).

\subsubsection{Affine Theories}

We can generalise this example to the case of an arbitrary
finite-dimensional Lie algebra $g$; the corresponding conformal field 
theory is usually called a {\em Wess-Zumino-Novikov-Witten model}
\cite{WesZum71,Witten84,Nov81,KniZam84,GepWit86}, and the following
explicit construction of the amplitudes is due to Frenkel \& Zhu
\cite{FreZhu92}. Suppose that the matrices $t^a$, 
$1\leq a\leq \dim g$, provide a finite-dimensional 
representation of $g$ so that $[t^a,t^b]={f^{ab}}_ct^c$, where
${f^{ab}}_c$ are the structure constants of $g$. We introduce a field
$J^a(z)$ for each $t^a$, $1\leq a\leq \dim g$. If $K$ is any matrix
which commutes with all the $t^a$, define  
\be
\kappa^{a_1a_2\ldots a_m}
=\hbox{tr}(Kt^{a_1}t^{a_2}\cdots t^{a_m})\,.
\ee
The $\kappa^{a_1a_2\ldots a_m}$ have the properties that
\be
\kappa^{a_1a_2a_3\ldots a_{m-1}a_m}
=\kappa^{a_2a_3\ldots a_{m-1}a_ma_1} 
\ee
and
\be
\kappa^{a_1a_2a_3\ldots a_{m-1}a_m}
-\kappa^{a_2a_1a_3\ldots a_{m-1}a_m}=
{f^{a_1a_2}}_b \kappa^{ba_3\ldots a_{m-1}a_m}\,.
\ee
With a cycle $\sigma=(i_1,i_2,\ldots,i_m)\equiv  (i_2,\ldots,i_m,i_1)$
we associate the function
\be
\label{affu}
\fl f_\sigma^{a_{i_1}a_{i_2}\ldots
a_{i_m}}(z_{i_1},z_{i_2},\ldots,z_{i_m}) 
= {\kappa^{a_{i_1}a_{i_2}\ldots a_{i_m}}
\over (z_{i_1}-z_{i_2})(z_{i_2}-z_{i_3})\cdots (z_{i_{m-1}}-z_{i_m})
(z_{i_m}-z_{i_1})}\,.
\ee
If the permutation $\rho\in S_n$ has no fixed points, it can be
written as the product of cycles of length at least 2,
$\rho=\sigma_1\sigma_2\ldots\sigma_M$. We associate to $\rho$ the
product $f_\rho$ of functions 
$f_{\sigma_1}f_{\sigma_2}\ldots f_{\sigma_M}$ and 
define $\langle J^{a_1}(z_1)J^{a_2}(z_2)\ldots J^{a_n}(z_n)\rangle$ to
be the sum of such functions $f_\rho$ over permutations $\rho\in S_n$
with no fixed point. Graphically, we can construct these amplitudes by
summing over all graphs with $n$ vertices where the vertices carry
labels $a_j$, $1\leq j\leq n$, and each vertex is connected by two
directed lines (propagators) to other vertices, one of the lines at
each vertex pointing towards it and one away. (In the above notation,
the vertex $i$ is connected to $\sigma^{-1}(i)$ and to $\sigma(i)$,
and the line from $\sigma^{-1}(i)$ is directed towards $i$, and from
$i$ to $\sigma(i)$.) Thus, in a given graph, the vertices are divided
into directed loops or cycles, each loop containing at least two
vertices. To each loop, we associate a function as in (\ref{affu}) and
to each graph we associate the product of functions associated to the
loops of which it is composed.  

The resulting amplitudes are evidently local and meromorphic, and one
can verify that they satisfy the M\"obius covariance property with the
weight of $J^a$ being $1$. They determine the operator product
expansion to be of the form\footnote{The terms singular in $(z-w)$
only arise from cycles where the vertices associated to $z$ and $w$
are adjacent. The first term in (\ref{xxx}) comes from the 2-cycle
involving $z$ and $w$. For every larger cycle in which $z$ and $w$ are
adjacent, there exists another cycle where the order of $z$ and $w$ is
reversed; the contributions of these two cycles combine to give the
second term in (\ref{xxx}).} 
\be\label{xxx}
J^a(z) J^b(w) \sim {\kappa^{ab} \over (z-w)^2} 
+ {{f^{ab}}_c J^c(w)\over (z-w)}  \,,
\ee
and the algebra therefore becomes 
\be
[J^a_m,J^b_n] = {f^{ab}}_c J^c_{m+n} + m \kappa^{ab} \delta_{m,-n}\,.
\ee
This is (a representation of) the affine algebra $\hat{g}$
\cite{Kac67,Moo67,Kac}. In the particular case where $g$ is simple, 
$\kappa^{ab}=\hbox{tr}(Kt^at^b)=k\delta^{ab}$ in a suitable basis,
where $k$ is a real number (that is called the {\em level}). The
algebra then becomes
\be
\label{affine}
[J^a_m,J^b_n] = {f^{ab}}_c J^c_{m+n} + m k \delta^{ab}
\delta_{m,-n}\,.
\ee
Again this theory is conformal since it has a stress-energy tensor
given by
\be
\psi_L = {1 \over 2(k+Q)} \sum_a J^a_{-1} J^a_{-1} \Omega \,,
\ee
where $Q$ is the dual Coxeter number of $g$ (\ie\ the value of the
quadratic Casimir in the adjoint representation divided by the length
squared of the longest root). Here the central charge is 
\be
\label{central}
c= {2 k \dim g \over 2k + Q} \,,
\ee
and the corresponding field can be described as 
\be
L(z) = {1 \over 2(k+Q)} \sum_a \nox J^a(z) J^a(z) \nox \,.
\ee
Finally, the modes of $L$ can be expressed in terms of the modes
of $J^a$ as  
\be
\label{Sugawara}
L_n = {1 \over 2(k+Q)} \sum_a \sum_{m} : J^a_m J^a_{n-m} : \,,
\ee
where the colons denote normal ordering
\be
: J^a_m J^b_l : = \left\{\begin{array}{ll}
J^a_m J^b_l & \hbox{if $m<0$} \\
J^b_l J^a_m & \hbox{if $m\geq 0$.}
\end{array} \right.
\ee
The construction of the stress-energy-tensor as a bilinear in the 
{\em currents} $J^a$ is called the {\em Sugawara construction}
\cite{Schu68}. The construction can be generalised directly to the
case where $g$ is semi-simple: in this case, the Sugawara field is the
sum of the Sugawara fields associated to each simple factor, and the
central charge is the sum of the corresponding central charges.  

The conformal field theory associated to the affine algebra $\hat{g}$
is unitary if $k$ is a positive integer \cite{PreSeg,GodOli86}; in
this case $c\geq 1$.

\subsubsection{Virasoro Theories}

Another very simple example of a meromorphic conformal field theory is
the theory where $V$ can be taken to be a one-dimensional vector space
that is spanned by the (conformal) vector $L$ \cite{BPZ84}. Let us
denote the corresponding field by $L(z)=V(L,z)$. Again following
Frenkel and Zhu \cite{FreZhu92}, we can construct the amplitudes
graphically as follows. We sum over all graphs with $n$ vertices,
where the vertices are labelled by the integers $1\leq j\leq n$, and
each vertex is connected by two directed lines (propagators) to other 
vertices, one of the lines at each vertex pointing towards it and one
away. In a given graph, the vertices are now divided into loops, 
each loop containing at least two vertices. To each loop
$\ell=(i_1,i_2,\ldots,i_m)$, we associate a function  
\be
\fl f_\ell(z_{i_1},z_{i_2},\ldots,z_{i_m}) = 
{c/2 \over (z_{i_1}-z_{i_2})^2(z_{i_2}-z_{i_3})^2\cdots 
(z_{i_{m-1}}-z_{i_m})^2 (z_{i_m}-z_{i_1})^2}\,,
\ee
where $c$ is a real number, and, to a graph, the product of the functions
associated to its loops. [Since it corresponds to a factor of the form
$(z_i-z_j)^{-2}$ rather than $(z_i-z_j)^{-1}$, each line or propagator
might appropriately be represented by a double line.] The amplitudes
$\langle L(z_1)L(z_2)\ldots L(z_n)\rangle$ are then obtained by summing the
functions associated with the various graphs with $n$ vertices. [Note
that graphs related by reversing the direction of any loop contribute
equally to this sum.]

These amplitudes determine the operator product expansion to be 
\be
L(z)L(\zeta)\sim {c/2\over (z-\zeta)^4} 
+ {2L(\zeta)\over (z-\zeta)^2} 
+{L'(\zeta)\over z-\zeta} \,,
\ee
which thus agrees with the operator product expansion of the field $L$
as defined in subsection~3.6.

These pure Virasoro models are unitary if either $c\geq 1$ or $c$
belongs to the {\em unitary discrete series} \cite{FQS84a} 
\be
\label{discrete}
c= 1 - { 6 \over m (m+1)} \qquad m=2,3,4,\ldots 
\ee
The necessity of this condition was established in \cite{FQS84a} using
the Kac-determinant formula \cite{Kac79} that was proven by
Feigin \& Fuchs in \cite{FFu82}. The existence of these unitary
representations follows from the coset construction (to be explained
below) \cite{GKO85}.

\subsubsection{Lattice Theories}

Let us recall that a {\em lattice} $\Lambda$ is a subset of an
$n$-dimensional inner product space which has integral coordinates in
some basis, $e_j$, $j=1,\ldots,n$; thus 
$\Lambda=\{\sum m_j e_j : m_j\in\Zop\}$. The lattice is called
{\em Euclidean} if the inner product is positive definite, \ie\ if
$k^2\geq 0$ for each $k\in\Lambda$, and {\em integral} if 
$k\cdot l\in\Zop$ for all $k,l\in\Lambda$. An (integral) lattice is 
{\em even} if $k^2$ is an even integer for every $k\in\Lambda$.

Suppose $\Lambda$ is an even Euclidean lattice with basis $e_j$,
$j=1,\ldots, n$. Let us introduce an algebra consisting of
matrices $\gamma_j$, $1\leq j\leq n$, such that $\gamma_j^2=1$ and 
$\gamma_i\gamma_j = (-1)^{e_i\cdot e_j}\gamma_j\gamma_i$. If we define  
$\gamma_k=\gamma_1^{m_1}\gamma_2^{m_2}\ldots\gamma_n^{m_n}$ for
$k=m_1e_1+m_2e_2+\ldots+m_ne_n$, we can define quantities
$\epsilon(k_1,k_2,\ldots, k_N)$, taking the values $\pm 1$, by 
\be
\gamma_{k_1}\gamma_{k_2}\ldots\gamma_{k_N}=
\epsilon(k_1,k_2,\ldots, k_N)\gamma_{k_1+k_2+\ldots+k_N}\,.
\ee
We define the theory associated to the lattice $\Lambda$ by taking 
a basis for $V$ to consist of the states $k\in\Lambda$. For each $k$,
the corresponding field $V(k,z)$ has conformal weight $\half k^2$, and
the amplitudes are given by 
\be
\fl \langle V(k_1,z_1)V(k_2,z_2)\cdots V(k_N,z_n)\rangle
= \epsilon(k_1,k_2,\ldots, k_N)\prod_{1\leq i<j\leq N}
(z_i-z_j)^{k_i\cdot k_j}
\ee
if $k_1+k_2+\ldots+ k_N=0$ and zero otherwise. The
$\epsilon(k_1,k_2,\ldots, k_N)$ obey the conditions 
\beq
\fl\epsilon(k_1,k_2,\ldots,k_{j-1},k_j,k_{j+1},k_{j+2},\ldots, k_N)\\
\lo= (-1)^{k_j\cdot k_{j+1}}
\epsilon(k_1,k_2,\ldots,k_{j-1},k_{j+1},k_j,k_{j+2},\ldots,k_N)\,,
\eeq
which guarantees the locality of the amplitudes, and 
\beq
\fl \epsilon(k_1,k_2,\ldots, k_j)\epsilon(k_{j+1},\ldots,k_N)\\
\lo=\epsilon(k_1+k_2+\ldots+ k_j,k_{j+1}+\ldots+k_N)
\epsilon(k_1,k_2,\ldots,k_j,k_{j+1},\ldots, k_N)\,,
\eeq
which implies the cluster decomposition property (that guarantees the
uniqueness of the vacuum). It is also easy to check that the
amplitudes satisfy the M\"obius covariance condition. 

This theory is also conformal, but the Virasoro field cannot be easily
described in terms of the fields in $V$. In fact, the theory that is
obtained by factorisation from the above amplitudes contains $n$
fields of conformal weight $1$, $H^i(z)$, $i=1,\ldots, n$, whose
operator product expansion is 
\be
\label{latticcur}
H^i(z) H^j(\z) \sim \delta^{ij} {1 \over (z-\zeta)^2} \,.
\ee
This is of the same form as (\ref{49}), and the corresponding modes
therefore satisfy
\be
[H^i_m,H^j_n] = m \delta^{ij} \delta_{m,-n} \,.
\ee
The Virasoro field is then given as 
\be
L(z) = {1 \over 2} \sum_i \nox H^i(z) H^i(z) \nox \,,
\ee
and the central charge is $c=n$. Details of the construction of this 
unitary meromorphic conformal field theory can be found in
\cite{DGM90a,DGM96}. 

For the case where $\Lambda$ is the root lattice of a simply-laced
finite-dimensional Lie algebra, the lattice theory coincides with the
affine theory of the corresponding affine Lie algebra at level $k=1$;
this is known as the Frenkel-Kac-Segal construction
\cite{FreKac80,Segal81}. 
\medskip

Lattice theories that are associated to even self-dual lattices whose
dimension is a multiple of $24$ provide examples of {\em local
meromorphic conformal field theories}, \ie\ local modular invariant
theories that are meromorphic. Such theories are characterised
by the property that all amplitudes depend only on the chiral
coordinates (\ie\ on $z_i$ but not on $\bar{z}_i$), and that the space
of states of the complete local theory coincides with (rather than
just contains) that of the meromorphic (sub)theory. Recall that for a
given lattice $\Lambda$, the {\em dual lattice} $\Lambda^\ast$ is the
lattice that contains all vectors $y$ for which $x\cdot y\in \Zop$ for
all $x\in\Lambda$. A lattice is integral if
$\Lambda\subset\Lambda^\ast$, and it is {\em self-dual} if
$\Lambda^\ast=\Lambda$. The dimension of an even self-dual lattice has
to be a multiple of $8$.    

It is not difficult to prove that a basis of states for the
meromorphic Fock space $\F_\Lambda$ that is associated to an even
lattice $\Lambda$ can be taken to consist of the states of the form
\be
H^{i_1}_{m_1} H^{i_2}_{m_2} \cdots H^{i_N}_{m_N} |k\rangle \,,
\ee
where $m_1\leq m_2\leq \cdots \leq m_N$, $i_j\in\{1,\ldots,n\}$
and $k\in\Lambda$. Here $H^i_l$ are the modes of the currents
(\ref{latticcur}), and $|k\rangle = V(k,0)\Omega$. The contribution of
the meromorphic subtheory to the partition function (\ref{toruscorr})
is therefore 
\be
\label{latticepart}
\chi_{\F_\Lambda} (\tau) = q^{-{c\over 24}} \tr_{\F_\Lambda}
( q^{L_0}) = \eta(\tau)^{-\dim\Lambda} \Theta_\Lambda (\tau) \,,
\ee
where $\eta(\tau)$ is the famous {\em Dedekind eta-function},
\be
\eta(\tau) = q^{1\over 24} \prod_{n=1}^{\infty} (1- q^n) \,,
\ee
$\Theta_\Lambda$ is the theta function of the lattice,
\be
\Theta_\Lambda(\tau) = \sum_{k\in\Lambda} q^{\half k^2} \,,
\ee
and we have set $q=e^{2\pi i\tau}$. The theta function of a lattice is
related to that of its dual by the Jacobi transformation formula
\be
\Theta_\Lambda (-1/\tau) = (-i\tau)^{\half \dim\Lambda}
|| \Lambda^\ast || \Theta_{\Lambda^\ast}(\tau) \,,
\ee
where $||\Lambda|| = |\det (e_i\cdot e_j)|$. Together with the
transformation formula of the Dedekind function
\be
\eta(-1/\tau) = (-i \tau)^{\half} \eta(\tau) \,,
\ee
this implies that the partition function of the meromorphic theory
transforms under modular transformations as
\be\label{Slattra}
\chi_{\F_\Lambda} (-1/\tau) = || \Lambda^\ast || 
\chi_{\F_{\Lambda^\ast}} (\tau) \,.
\ee
Furthermore, it follows from (\ref{latticepart}) together with the
fact that the spectrum of $L_0$ in $\F_\Lambda$ is integral that
\be
\chi_{\F_\Lambda} (\tau+1) = e^{2\pi i {c\over 24}} 
\chi_{\F_\Lambda} (\tau) \,.
\ee
If $\Lambda$ is a self-dual lattice, $|| \Lambda^\ast ||=1$ and
the partition function is invariant under the transformation in
(\ref{Slattra}). If in addition $c=n=24 k$ where $k\in\Zop$, the
partition function of the meromorphic theory is invariant under the
whole modular group, and the meromorphic theory defines a local
(modular invariant) conformal field theory. For the case where
$n=24$, lattices with these properties have been classified: there
exist precisely 24 such lattices, the $23$ Niemeier lattices and the
Leech lattice (see \cite{ConSlo} for a review). The Leech lattice
plays a central r\^ole in the construction of the Monster conformal
field theory \cite{FLM}.

\subsubsection{More General $W$-Algebras}

In the above examples, closed formulae for all amplitudes of the
generating fields could be given explicitly. There exist however many 
meromorphic conformal field theories for which this is not the
case. These theories are normally defined in terms of the operator 
product expansion of a set of generating fields (that span $V$) from
which the commutation relations of the corresponding modes can be
derived; the resulting algebra is then usually called a 
{\em $W$-algebra}. In general, a $W$-algebra is not a Lie algebra in
the modes of the generating fields since the operator product
expansion (and therefore the associated commutator) of two generating
fields may involve normal ordered products of the generating fields
rather than just the generating fields themselves.  

In principle all amplitudes can be determined from the knowledge of
these commutation relations, but it is often difficult to give closed 
expressions for them. (It is also, {\em a priori}, not clear whether
the power series expansion that can be obtained from these commutation
relations will converge to define meromorphic functions with the
appropriate singularity structure, although this is believed to be the
case for all presently known examples.) The theories that we have
described in detail above are in some sense fundamental in that 
all presently known meromorphic (bosonic) conformal field theories
have an (alternative) description as a {\em coset} or {\em orbifold}
of one of these theories; these constructions will be described in the
next subsection.  

The first example of a $W$-algebra that is not a Lie algebra in the
modes of its generating fields is the so-called $W_3$ algebra
\cite{Zam85,FZ87}; this algebra is generated by the Virasoro algebra
$\{ L_{n} \}$, and the modes $W_{m}$ of a quasiprimary field of
conformal weight $h=3$, subject to (\ref{Virasoro}) and the  relations
\cite{Zam85,FZ87,BowWat92b} 
\beq
\fl [ L_m, W_n ]  = & (2m -n)\, W_{m+n} \,,\nonumber \\
\fl [W_{m},W_{n}] = &
{\displaystyle \frac{1}{48}\, (22+5c)\, \frac{c}{3 \cdot 5!}\; 
(m^{2} - 4)\; (m^{2}-1)\; m\; \delta_{m,-n} + \frac{1}{3} (m-n)\;
\Lambda_{m+n}} \nonumber \\
& {\displaystyle 
+ \frac{1}{48}\, (22+5c)\, \frac{1}{30} \;(m-n)\; 
(2m^{2}-mn+2n^{2}-8)\; L_{m+n}\,,}
\end{eqnarray}
where $\Lambda_{k}$ are the modes of a quasiprimary field of 
conformal weight $h_{\Lambda}=4$. This field is a normal ordered
product of $L$ with itself, and its modes are explicitly given as
\be
\Lambda_{n} = \sum_{k=-1}^{\infty} L_{n-k}\, L_k +
\sum_{k=-\infty}^{-2} L_k \, L_{n-k} 
- \frac{3}{10} \, (n+2) \,(n+3) \, L_n\,.
\ee
One can check that this set of commutators satisfies the
Jacobi-identity. (This is believed to be equivalent to the
associativity of the operator product expansion of the corresponding
fields.) Subsequently, various classes of $W$-algebras have been 
constructed
\cite{FatLyk88,BBSS88a,BBSS88b,Bou88,BilGer88,BilGer89,FigSch90,BFKNRV94}. 
There have also been attempts to construct systematically classes of
$W$-algebras \cite{BFKNRV91,KauWat91} following \cite{Nahm89}. For a
review of these matters see \cite{BouScho93}.

\subsubsection{Superconformal Field Theories}

All examples mentioned up to now have been bosonic theories, \ie\
theories all of whose fields are bosonic (and therefore have integral
conformal weight). The simplest example of a fermionic conformal
field theory is the theory generated by a single free fermion field 
$b(z)$ of conformal weight $h=\half$ with operator product expansion
\be
\label{freef}
b(z) b(\zeta) \sim {1 \over (z-\zeta)} \,.
\ee
The corresponding modes
\be 
b(z) = \sum_{r\in\Zop+\half} b_r z^{-r-\half} 
\ee
satisfy an anti-commutation relation of the form
\be
\label{ffcom}
\{ b_r, b_s \} = \delta_{r,-s} \,.
\ee
The Virasoro field is given in this case by 
\be
L(z) = \half \; \nox {db(z) \over dz}\; b(z) \nox \,,
\ee
and the corresponding modes are
\be
L_n = \half \sum_{r\in\Zop+\half} \left(r-\half\right)  
: b_{n-r} b_r : \,.
\ee
They satisfy the Virasoro algebra with central charge $c=\half$. 

This theory has a {\em supersymmetric} extension whose generating
fields are the free fermion $b(z)$ (\ref{freef}) and the free boson
$J(z)$ defined by (\ref{49}) and (\ref{50}), where we set for
convenience $k=1$. The operator product expansion of $J$ and $b$ is
regular, so that the corresponding commutator vanishes. This theory
then exhibits {\em superconformal  invariance}; in the present context
this means that the theory has in addition to the stress-energy-tensor
$L$, the superpartner field $G$ of conformal weight $3/2$, whose modes
satisfy 
\be
\label{supervir}
\begin{array}{lcl}
{\displaystyle  [L_m,G_r]} & = & {\displaystyle 
\left({m \over 2} - r \right) G_{m+r}} \vspace*{0.1cm} \\
{\displaystyle  \{G_r,G_s\}} & = & {\displaystyle 
2 L_{r+s} + {c \over 3} \left(r^2 - {1\over 4} \right) \delta_{r,-s}}
\,. 
\end{array}
\ee
Together with the Virasoro commutation relations for $L$
(\ref{Virasoro}), this is called the (Neveu Schwarz [NS] sector of
the) {\em $N=1$ superconformal algebra}. In terms of $J$ and $b$, $L$
and $G$ can be written as 
\beq
L(z) & = {1 \over 2} \nox J(z) J(z) \nox 
          + \half \nox {db(z) \over dz} b(z) \nox \nonumber \\
G(z) & = J(z) b(z) \,, 
\eeq
where on the right-hand-side of the second line we have omitted the
normal ordering since the operator product expansion is regular. The
modes of $L$ and $G$ are then given as 
\beq
L_n & = \half \sum_{m\in\Zop} :J_{n-m} J_m : 
+ \half \sum_{r\in\Zop+\half} \left(r-\half\right) : b_{n-r} b_r: 
\nonumber \\
G_r & = \sum_n J_{-n} b_{r+n} \,,
\eeq
and it is easy to check that they satisfy (\ref{Virasoro}) and
(\ref{supervir}) with $c=3/2$. 

The $N=1$ superconformal field theory that is generated by the fields
$L$ and $G$ subject to the commutation relations (\ref{supervir}) is
unitary if \cite{FQS85}  
\be
\label{superdiscrete}
c= \thalf - {12 \over m (m+2)} \qquad m=2,3,4,\ldots
\ee
\medskip

There also exist extended superconformal algebras that contain the
above algebra as a subalgebra. The most important of these is the
{\em $N=2$ superconformal algebra} \cite{ABDD76a,ABDD76b} that
is the symmetry algebra of the world-sheet conformal field theory of
space-time supersymmetric string theories \cite{BDFM88}. In the
so-called NS sector, this algebra is generated by the modes of the
Virasoro algebra (\ref{Virasoro}), the modes of a free boson $J_{n}$
(\ref{50}) (where we set again $k=1$), and the modes of two
supercurrents of conformal dimension $h=\thalf$,  
$\{G^{\pm}_{\alpha}\},\,\alpha\in\Zop+\half$, subject to the relations
\cite{BFK86} 
\beq
{}[L_m, G^{\pm}_r] & = 
\left(\half m - r \right)\; G^{\pm}_{m+r} \nonumber \\
{}[L_m, J_n ] & = - n \; J_{m+n} \nonumber \\
{}[J_m, G^{\pm}_r ] & = \pm G^{\pm}_{m+r}
\label{supervir2} \\ 
\{ G^{\pm}_r, G^{\pm}_s \} & = 0\nonumber \\
\{ G^{+}_r, G^{-}_s \} & = 
2 \;  L_{r + s} + (r - s) \; J_{r + s}
+ {c\over 3} \;\left( r^{2} - \frac{1}{4} \right) 
\delta_{r,-s}\,. \nonumber
\eeq
The representation theory of the $N=2$ superconformal algebra exhibits
many interesting and new phenomena \cite{LVW89,Dor96}.

\subsection{The Coset Construction}

There exists a fairly general construction by means of which a
meromorphic conformal field theory can be associated to a pair of a
meromorphic conformal field theory and a subtheory. In its simplest 
formulation \cite{GKO85,GKO86} (see also \cite{BarHal71,Man73a,Man73b}
for a related construction in a particular case) the pair of theories
are affine theories that are associated to a pair $h\subset g$ of
finite-dimensional simple Lie algebras. Let us denote by $L^g_m$ and
$L^h_m$ the modes of the Sugawara fields (\ref{Sugawara}) of
the affine algebras $\hat{g}$ and $\hat{h}$, respectively. If $J^a_n$
is a generator of $\hat{h}\subset \hat{g}$, then  
\be
[L^g_m,J^a_n ] = - n J^a_{m+n} \qquad \hbox{and} \qquad
[L^h_m,J^a_n ] = - n J^a_{m+n}
\ee
since $J^a_n$ are the modes of a primary field of conformal weight
$h=1$. (This can also be checked directly using (\ref{Sugawara}) and
(\ref{affine}).) It then follows that 
\be
\K_m = L^g_m - L^h_m 
\ee
commutes with every generator $J^a_n$ of $\hat{h}$, and therefore with 
the modes $L^h_m$ (which are bilinear in $J^a_n$). We can thus
write $L^g_m$ as the sum of two commuting terms
\be
L^g_m = L^h_m + \K_m \,,
\ee
and since both $L^g_m$ and $L^h_m$ satisfy the commutation relations
of a Virasoro algebra, it follows that this is also the case for
$\K_m$, where the corresponding central charge is 
\be
c^\K = c^g - c^h \,.
\ee
It is also straightforward to extend the construction to the case
where $g$ (and/or $h$) are semi-simple.

In the unitary case, both $c^g \geq c^h\geq 1$, but $c^\K$ need not be 
greater or equal to one. Indeed, we can consider 
$\hat{g} = \hat{su}(2)_m \oplus \hat{su}(2)_1$ and
$h=\hat{su}(2)_{m+1}$, where the index is the level $k$ of
(\ref{affine}), and the central charge $c^\K$ is then
\be
c^\K = {3m \over m+2} + 1 - {3(m+1) \over (m+3)} =
1 - {6 \over (m+2) (m+3)} \,,
\ee
where $m\in\Zop_+$ and we have used (\ref{central}) together with 
$\dim su(2)=3$ and $Q_{su(2)}=4$. By construction, the subspace that
is generated by $\K_m$ from the vacuum forms a unitary representation of
the Virasoro algebra with $h=0$ and $c=c^\K$, and thus for each value
of $c$ in the unitary discrete series (\ref{discrete}), the vacuum
representation of the Virasoro algebra is indeed unitary
\cite{GKO85}. Similar arguments can also be given for the discrete
series of the $N=1$ superconformal algebra (\ref{superdiscrete})
\cite{GKO86}. 

We can generalise this construction directly to the case where instead
of the affine theory associated to $\hat{g}$, we consider an arbitrary
conformal field theory $\H$ (with stress-energy tensor $L$) that
contains, as a subtheory, the affine theory associated to
$\hat{h}$. Then by the same arguments as above
\be
\K = L - L^h 
\ee
commutes with $\hat{h}$ (and thus with $L^h$), and therefore satisfies
a Virasoro algebra with central charge $c^\K = c - c^h$. By
construction, the Virasoro algebra $\K_m$ leaves the subspace of
states  
\be
\H^h = \left\{ \psi\in\H : J^a_n \psi=0 \; \hbox{for every
$J^a_n\in\hat{h}$ with $n\geq 0$} \right\}
\ee
invariant. Furthermore, the operator product expansion of an
$h$-current $J^a(z)$ and a vertex operator $V(\psi,\z)$ associated to
$\psi\in\H^h$ is regular, and therefore the corresponding modes
commute. This implies that the operator product expansion of two 
states in $\H^h$ only contain states that lie in $\H^h$
\cite{BowGod88}, and thus that we can define a meromorphic 
field theory whose space of states is $\H^h$. Since the commutator of
$L^h$ with any state in $\H^h$ vanishes, it is clear that the Virasoro
field for this theory is $\K$; the resulting meromorphic conformal
field theory is called the {\em coset theory}.

Many $W$-algebras can be constructed as cosets of affine theories
\cite{BBSS88b,GodSch88a,BowGod88,Watts90}. It is also possible to
construct representations of the coset theory from those of $\H$, and
to determine the corresponding modular transformation matrix; details
of these constructions for the case of certain coset theories of
affine theories have been recently worked out in \cite{FSS96}.

\subsection{Orbifolds}

There exists another very important construction that associates to a
given local (modular invariant) conformal field theory another such
theory \cite{DHVW85,DHVW86,Vafa86,NSV87,DVVV89}. This construction is 
possible whenever the theory carries an action of a finite group
$G$. A group $G$ acts on the space of states of a conformal field
theory $\bH$, if each  $g\in G$ defines a linear map
$g:\bH\rightarrow\bH$ (that leaves the dense subspace $\bF$ invariant, 
$g:\bF\rightarrow\bF$), the composition of maps respects the group
structure of $G$, and the amplitudes satisfy
\be
\langle V(g\psi_1;z_1,\bz_1) \cdots  V(g\psi_n;z_n,\bz_n) \rangle = 
\langle V(\psi_1;z_1,\bz_1) \cdots  V(\psi_n;z_n,\bz_n) \rangle \,.
\ee
The space of states of the orbifold theory consists of those states
that are invariant under the action of $G$,
\be
\bH^G = \left\{ \psi\in\bH : g \psi = \psi \quad \hbox{for all 
$g\in G$.} \right\} \,,
\ee
together with additional {\em twisted sectors}, one for each conjugacy
class in $G$. Generically the meromorphic subtheory of the resulting
theory consists of those meromorphic fields of the original theory
that are invariant under $G$, but in general it may also happen that
some of the twisted sectors contain additional meromorphic fields.

The construction of the twisted sectors is somewhat formal in general,
and so is the proof that the resulting local theory is always modular
invariant. There exist however some examples where the construction is
understood in detail, most notably the local meromorphic lattice 
theories that were discussed in section 3.7.4. Let us for simplicity
consider the case where $G=\Zop_2=\{1,\theta\}$, and the action on
$\H_\Lambda$ is determined by 
\be
\label{theta}
\theta H^i(z) \theta = - H^i(z)\,, \qquad
\theta V(x,z) \theta = V(-x,z) \,, \qquad
\theta \Omega=\Omega \,.
\ee
In this case there exists only one twisted sector, $\H'_\Lambda$, and 
it is generated by the operators $c^i_r$, $i=1,\ldots, n$ with
$r\in\Zop+\half$, satisfying the commutation relations 
\be
[c^i_r,c^j_s] = r \delta^{ij} \delta_{r,-s}\,.
\ee
These act on an irreducible representation space $U$ of the algebra
\be
\gamma^i \gamma^j = (-1)^{e_i\cdot e_j} \gamma^j \gamma^i \,,
\ee
where $e_i$, $i=1,\ldots, n$ is a basis of $\Lambda$, and where for
$\chi\in U$, $c^i_r\chi=0$ if $r>0$. The actual orbifold theory
consists then of the states in the untwisted $\H_\Lambda$ and the
twisted sector $\H'_\Lambda$ that are left invariant by $\theta$,
where the action of $\theta$ on $\H_\Lambda$ is given as in 
(\ref{theta}), and on $\H'_\Lambda$ we have
\be
\label{9.19}
\theta c^i_r \theta = - c^i_r \qquad \left.\theta\right|_{U} = \pm 1
\,.
\ee
The generators of the Virasoro algebra act in the twisted sector as 
\be
L_m = \half \sum_{i=1}^{n} \sum_{r\in\Zop+\half} 
: c^i_r c^i_{m-r} : + {n \over 16} \delta_{m,0}\,.
\ee
As in the untwisted sector $L_m$ commutes with $\theta$ and is
therefore well defined in the orbifold theory.

Since the local meromorphic conformal field theory is already modular
invariant, the dimension $n$ of the lattice is a multiple of $24$ and
the orbifold theory is again a meromorphic conformal field
theory. This theory is again bosonic provided the sign in (\ref{9.19})
corresponds to the parity of $\dim\Lambda$ divided by $8$. With this
choice of (\ref{9.19}) the orbifold theory defines another local
meromorphic conformal field theory \cite{DGM90a,DGM90b,DGM96}. 

The most important example of this type is the orbifold theory
associated to the Leech lattice for which the orbifold theory
does not have any states of conformal weight one. This is the famous
Monster conformal field theory whose automorphism group is the Monster
group \cite{CoNNor79,FLM,Bor86,Bor92}, the largest simple sporadic
group. It has been conjectured that this theory is uniquely
characterised by the property to be a local meromorphic conformal
field theory with $c=24$ and without any states of conformal weight
one \cite{FLM}, but this has not been proven so far.

One can also apply the construction systematically to the other $23$ 
Niemeier lattices. Together with the $24$ local meromorphic conformal
field theories that are directly associated to the $24$ self-dual
lattices, this would naively give $48$ conformal field
theories. However, nine of these theories coincide, and therefore
these constructions only produce $39$ different local meromorphic
conformal field theories \cite{DGM90b,DGM96}. If the above conjecture
about the uniqueness of the Monster theory is true, then every local
meromorphic conformal field theory at $c=24$ (other than the Monster
theory) contains states of weight one, and therefore an affine
subtheory \cite{God89}. The theory can then be analysed in terms of
this subtheory, and using arguments of modular invariance, Schellekens
has suggested that at most $71$ local meromorphic conformal field
theories exist for $c=24$ \cite{Schell93}. However this classification
has only been done on the level of the partition functions, and it is
not clear whether more than one conformal field theory may correspond
to a given partition function. Also, none of these additional theories
has been constructed explicitly, and it is not obvious that all $71$
partition functions arise from consistent conformal field theories.

\section{Representations of a Meromorphic Conformal Field
Theory}

For most local conformal field theories the meromorphic fields form a 
proper subspace of the space of states. The additional states of the
theory transform then in representations of the meromorphic (and the
anti-meromorphic) subtheory. Indeed, as we explained in section~2.1,
the space of states is a direct sum of subspaces
$\bH_{(j,\bar{\jmath})}$, each of which forms an indecomposable
representation of the two meromorphic conformal field theories. For
most conformal field theories of interest (although not for all, see
\cite{GabKau98}), each $\bH_{(j,\bar{\jmath})}$ is a tensor product of
an irreducible representation of the meromorphic and the
anti-meromorphic conformal subtheory, respectively 
\be\label{factor}
\bH_{(j,\bar{\jmath})} = \H_j \otimes \bar{\H}_{\bar{\jmath}} \,.
\ee
The local theory is specified in terms of the space of states and the
set of all amplitudes involving arbitrary states in $\bF\subset\bH$. 
The meromorphic subtheory that we analysed above describes the
amplitudes that only involve states in $\F_0$. Similarly, the
anti-meromorphic subtheory describes the amplitudes that only involve
states in $\Fbar_0$. Since the two meromorphic theories commute, a
general amplitude involving states from both $\F_0$ and $\Fbar_0$ is
simply the product of the corresponding meromorphic and
anti-meromorphic amplitude. (Indeed, the product of the meromorphic
and the anti-meromorphic amplitude has the same poles as the original
amplitude.) 

If the theory factorises as in (\ref{factor}), one of the summands in
(\ref{decomposition}) is the completion of $\F_0\otimes\Fbar_0$, and
we denote it by $\bH_{(0,\bar{0})}$. A general amplitude of the
theory contains states from different sectors
$\bH_{(j,\bar{\jmath})}$. Since each $\bH_{(j,\bar{\jmath})}$ is a 
representation of the two vertex operator algebras, we can use 
the operator product expansion (\ref{localope}) to rewrite a given 
amplitude in terms of amplitudes that do not involve states in
$\bH_{(0,\bar{0})}$. It is therefore useful to call an amplitude an
{\em $n$-point function} if it involves $n$ states from sectors other
than $\bH_{(0,\bar{0})}$ and an arbitrary number of states from
$\bH_{(0,\bar{0})}$. In general, each such amplitude can be
expressed as a sum of products of a {\em chiral} amplitude, \ie\ an
amplitude that only depends on the $z_i$, and an {\em anti-chiral}
amplitude, \ie\ one that only depends on the $\bar{z}_i$. However, 
for $n=0,1,2,3$, the sum contains only one term since the functional
form of the relevant chiral (and anti-chiral) amplitudes is uniquely
determined by M\"obius symmetry.

The zero-point functions are simply products of meromorphic
amplitudes, and the one-point functions vanish. The two-point
functions are usually non-trivial, and they define, in essence, the
different representations of the meromorphic and the anti-meromorphic
subtheory that are present in the theory. Since these amplitudes
factorise into chiral and anti-chiral amplitudes, one can analyse them
separately; these chiral amplitudes define then a representation of
the meromorphic subtheory.

\subsection{Highest Weight Representations}

A representation of the meromorphic conformal field theory is defined
by the collection of amplitudes
\be
\label{reprampl}
\Bigl\langle \bar\phi(w) V(\psi_1,z_1)\cdots V(\psi_n,z_n)
\phi(u)\Bigr\rangle \,,
\ee
where $\phi$ and $\bar\phi$ are two fixed fields (that describe the
generating field of a representation and its conjugate), and $\psi_i$
are quasiprimary fields in the meromorphic conformal field theory. The 
amplitudes (\ref{reprampl}) are analytic functions of the variables
and transform covariantly under the M\"obius transformations as in 
(\ref{Moebamp})
\beq
\fl
\Bigl\langle \bar\phi(w) V(\psi_1,z_1)\cdots 
V(\psi_n,z_n) \phi(u)\Bigr\rangle  \nonumber \\
= \left({d\gamma(w) \over dw} \right)^{\bar{h}}
\left({d\gamma(u) \over du} \right)^{h}
\prod_{i=1}^{n} \left({d\gamma(z_i) \over dz_i} \right)^{h_i}
\nonumber \\
\qquad 
\Bigl\langle \bar\phi(\gamma(w)) V(\psi_1,\gamma(z_1))
\cdots V(\psi_n,\gamma(z_n)) \phi(\gamma(u))\Bigr\rangle \,,
\eeq
where $h_i$ is the conformal weight of $\psi_i$, and we call $h$
and $\bar{h}$ the conformal weights of $\phi$ and $\bar\phi$,
respectively. Since $\phi$ and $\bar\phi$ are not meromorphic fields,
$h$ and $\bar{h}$ are in general not half-integer, and the amplitudes
are typically branched about $u=w$. Because of the M\"obius symmetry
we can always map the two points $u$ and $w$ to $0$ and $\infty$,
respectively (for example by considering the M\"obius transformation 
$\gamma(z)={z-u \over z-w}$), and we shall from now always do so. In
this case we shall write $\phi(0)\rangle = |\phi\rangle$. For the
case of $w=\infty$ the situation is slightly more subtle since
(\ref{reprampl}) behaves as $w^{-2\bar{h}}$ for $w\rightarrow\infty$;
we therefore define
\be
\langle\bar\phi| = \lim_{w\rightarrow\infty} 
w^{2\bar{h}} \langle \bar\phi(w)\,. 
\ee
We can then think of the amplitudes as being the expectation value of
the meromorphic fields in the background described by $\phi$ and
$\bar\phi$. 

The main property that distinguishes the amplitudes as 
{\em representations} of the meromorphic conformal field theory is the
condition that the operator product relations of the meromorphic
conformal field theory are preserved by these amplitudes. This is the
requirement that the operator product expansion of the meromorphic
fields (\ref{ope}) also holds in the amplitudes (\ref{reprampl}), \ie\
\beq
\fl \langle \bar\phi| V(\psi_1,z_1) \cdots V(\psi_i,z_i)
V(\psi_{i+1},z_{i+1}) \cdots V(\psi_n,z_n)
|\phi\rangle \nonumber \\
= \sum_{n<h_{i+1}} (z_i-z_{i+1})^{-n-h_i}  \label{operep}\\
\quad \langle \bar\phi| V(\psi_1,z_1) \cdots V(\psi_{i-1},z_{i-1})
V(V_n(\psi_i) \psi_{i+1},z_{i+1}) \cdots V(\psi_n,z_n)
|\phi\rangle\,,\nonumber
\eeq
where $|z_i-z_{i+1}|<|z_j-z_{i+1}|$ for $j\ne i$ and
$|z_i-z_{i+1}|<|z_{i+1}|$. In writing (\ref{operep}) we have also 
implicitly assumed that if $\Nu$ is a null-state of the meromorphic
conformal field theory (\ie\ a linear combination of states of the
form (\ref{Psi}) that vanishes in every meromorphic amplitude) then
any amplitude (\ref{reprampl}) involving $\Nu$ also vanishes; this is
implicit in the above since the operator product expansion of the
meromorphic conformal field theory is only determined up to such
null-fields by the meromorphic amplitudes. 

We call a representation {\em untwisted} if the amplitudes
(\ref{reprampl}) are single-valued as $z_i$ encircles the origin or
infinity; if this is not the case for at least some of the meromorphic 
fields the representation is called {\em twisted}. If the
representation is untwisted, we can expand the meromorphic fields in
terms of their modes as in (\ref{modes}). In this way we can then
define the action of $V_n(\psi)$ on the non-meromorphic state
$|\phi\rangle$, and thus on arbitrary states of the form
\be
\label{Fockgen}
V_{n_1}(\psi_1) \cdots V_{n_N}(\psi_N) |\phi\rangle \,.
\ee
As we explained in section~3.4, the commutation relations of these
modes (\ref{47}) can be derived from the operator product expansion of
the corresponding fields. Since the representation amplitudes
(\ref{reprampl}) preserve these in the sense of (\ref{operep}), it
follows that the action of the modes on the states of the form
(\ref{Fockgen}) also respects (\ref{47}), at least up to null-states
that vanish in all amplitudes. If we thus define  the Fock space $\F$
to be the quotient space of the space of states generated by
(\ref{Fockgen}), where we identify states whose difference vanishes in
all amplitudes, then $\F$ carries a representation of the Lie algebra
of modes (of the meromorphic fields).\footnote{Strictly speaking, the
underlying vector space of this Lie algebra is the vector space of
modes, where we identify two modes if their difference vanishes on the
Fock space of all representations.}

For most of the following we shall only consider untwisted
representations, but there is one important case of a twisted 
representations, the so-called {\em Ramond sector} of a 
fermionic algebra, that should be mentioned here since it can be
analysed by very similar methods. In this case the bosonic fields are
single-valued as $z$ encircles the origin, and the fermionic fields
pick up a minus sign. It is then again possible to expand the
meromorphic fields in modes, where the bosonic fields are treated as
before, and for a fermionic field we now have 
\be
\label{Rsector}
V(\chi,z) = \sum_{r\in\Zop} V_r(\chi) z^{-r-h} \qquad
\hbox{R-sector of fermionic $\chi$.}
\ee
Using the same methods as before, we can deduce the commutation (and
anti-commutation) relations of these modes from the operator product
expansion of the fields in the meromorphic conformal field theory, and
the Fock space of the R-sector representations forms then a
representation of this Lie algebra. Indeed, the actual form of the
commutation and anti-commutation relations is the same except that the
fermionic fields now have integer mode number. For example, the
commutation relations of the R-sector of the $N=1$ superconformal
algebra are 
\be
\label{supervirR}
\begin{array}{lcl}
{\displaystyle  [L_m,L_n]} & = & {\displaystyle (m-n) L_{m+n} + 
{c \over 12} m (m^2-1) \delta_{m,-n}} \vspace*{0.1cm} \\
{\displaystyle  [L_m,G_r]} & = & {\displaystyle 
\left({m \over 2} - r \right) G_{m+r}} \vspace*{0.1cm} \\
{\displaystyle  \{G_r,G_s\}} & = & {\displaystyle 
2 L_{r+s} + {c \over 3} \left(r^2 - {1\over 4} \right) \delta_{r,-s}}
\,,
\end{array}
\ee
where now $r,s\in\Zop$, and this agrees formally with
(\ref{supervir}). 

The physically relevant representations satisfy again the condition
that the spectrum of $L_0$ (the zero mode of the conformal
stress-energy-tensor) is bounded from below. This implies that there
exists a state $\phi_0$ in $\F$ that is annihilated by all modes
$V_n(\psi)$ with $n>0$ (since $V_n(\psi)$ lowers the $L_0$ eigenvalue
by $n$ as follows from (\ref{Lcomm})); such a state is called a 
{\em (Virasoro) highest weight state}. If the representation is
irreducible, \ie\ if it does not contain any proper
subrepresentations, then the representation generated from $\phi_0$ by
the action of the modes reproduces the whole representation space, and
we may therefore assume that $\phi$ (and $\bar\phi$) are highest weight
states. Using the mode expansion of the meromorphic field $V(\psi,z)$,
the highest weight property of $\phi$ can be rewritten as the
condition that the pole in $z$ of an amplitude involving
$V(\psi,z)|\phi\rangle$ is at most of order $h$, where $h$ is the
conformal weight of $\psi$. Because of the M\"obius covariance 
of the amplitudes, this is  then also equivalent to the condition that
the order of the pole in $(z_i-u)$ in (\ref{reprampl}) is bounded by
the conformal weight of $\psi_i$, $h_i$. 

If $\phi_0$ is a Virasoro highest weight state, then so is  
$V_0(\psi)\phi_0$ for any $\psi$. The space of Virasoro highest weight
states therefore forms a representation of the zero modes of the
meromorphic fields. Conversely, given a representation $R$ of the zero
modes we can consider the space of states that is generated from a
state in $R$ by the action of the negative modes; this is called the
{\em Verma module}. More precisely, the Verma module $\V$ is the
vector space that is spanned by the states of the form
\be
V_{n_1}(\psi_1) \cdots V_{n_N}(\psi_N) \phi \qquad \phi\in R \,,
\ee
where $n_i < 0 $, modulo the relations
\beq
\fl
V_{l_1}(\psi_1) \cdots V_{l_r}(\psi_r)
\Bigl(V_n(\psi) V_m(\chi) - V_m(\chi) V_n(\psi) \Bigr)
V_{l_{r+1}}(\psi_{r+1}) \cdots 
V_{l_N}(\psi_N) \phi \nonumber \\
= V_{l_1}(\psi_1) \cdots V_{l_r}(\psi_r) 
\Bigl[V_n(\psi),V_m(\chi)\Bigr] V_{l_{r+1}}(\psi_{r+1}) \cdots 
V_{l_N}(\psi_N) \phi \,,
\eeq
where $\phi\in R$, and $[V_n(\psi),V_m(\chi)]$ stands for the
right-hand-side of (\ref{47}). If the me\-ro\-mor\-phic conformal field
theory is generated by a finite set of fields, $W^1(z),\ldots W^s(z)$,
one can show \cite{Watts90} that the Verma module is spanned by the
so-called {\em Poincar\'e-Birkhoff-Witt basis} that consists of
vectors of the form 
\be
W^{i_1}_{-m_1} \cdots W^{i_l}_{-m_l} \phi \,,
\ee
where $\phi\in R$, $m_j>0$, $1 \leq i_{j+1}\leq i_{j}\leq s$ and 
$m_j\geq m_{j+1}$ if $i_j=i_{j+1}$. The actual Fock space of the
representation $\F$ is again a certain quotient of the Verma module,
where we set to zero all states that vanish identically in all
amplitudes; these states are again called {\em null-vectors}.

\subsection{An Illustrative Example}

It should be stressed at this stage that the condition to be a
representation of the meromorphic conformal field theory is usually  
{\em stronger} than that of being a representation of the Lie algebra
(or W-algebra) of modes of the generating fields. For example, in the
case of the affine theories introduced in section~3.7.2, the latter
condition means that the representation space has to be a
representation of the affine algebra (\ref{affine}), and for any value
of $k$, there exist infinitely many (non-integral) representations. On
the other hand, if $k$ is a positive integer, the meromorphic theory
possesses null-vectors, and only those representations of the affine
algebra are representations of the meromorphic conformal field theory
for which the null-fields act trivially on the representation space;
this selects a finite number of representations. For example, if
$g=su(2)$, the affine algebra can be written (in the Cartan-Weyl
basis) as \cite{GodOli86}  
\beq
{}[H_m,H_n] & =  \half k m \delta_{m,-n} \nonumber \\
{}[H_m,J^\pm_n] & =  \pm J^{\pm}_{m+n} \label{su2} \\
{}[J^+_m,J^-_n] & =  2 H_{m+n} +  k m \delta_{m,-n}\,, \nonumber
\eeq
and if $k$ is a positive integer, the vector
\be
\Nu = \left(J^+_{-1}\right)^{k+1} \Omega
\ee
is a {\em singular} vector, \ie\ it is a descendant of the highest
weight vector that is annihilated by all modes $V_n(\psi)$ with
$n>0$. This follows from the fact that the positive modes of the
generating fields annihilate $\Nu$ which is obvious for  
$H_n\Nu=J^+_n\Nu=0$, and for $J^-_n\Nu$ is a consequence of  
$L_n\Nu=0$ for $n>0$ together with
\beq
J^-_1 \Nu & = \sum_{l=0}^{k} (J^+_{-1})^l [ J^-_1,J^+_{-1} ] 
(J^+_{-1})^{k-l} \Omega \nonumber \\
& = \left[ k (k+1) - 2 \sum_{l=0}^{k} (k-l) \right](J^+_{-1})^k \Omega 
= 0 \,.
\eeq
Every singular vector is a null-vector as follows from
(\ref{adjoint}). In the above example, $\Nu$ actually generates
the null-space in the sense that every null-vector of the meromorphic
conformal field theory can be obtained by the action of the modes from
$\Nu$ \cite{Kac74} (see also \cite{Kac}). 

The zero modes of the affine algebra $\hat{su}(2)$ form the finite Lie
algebra of $su(2)$. For every (finite-dimensional) representation $R$
of $su(2)$, \ie\ for every spin $j\in\Zop/2$, we can construct a Verma 
module for $\hat{su}(2)$ whose Virasoro highest weight space
transforms as $R$. This (and any irreducible quotient space thereof)
defines a representation of the affine algebra $\hat{su}(2)$.
On the other hand, the zero mode of the null-vector $\Nu$ acts on
any Virasoro highest weight state $\phi$ as   
\be
\label{nulleq}
V_0(\Nu) \phi = \left(J^+_0\right)^{k+1} \phi \,,
\ee
and in order for the Verma module to define a representation of the
meromorphic conformal field theory, (\ref{nulleq}) must vanish. This
implies that $j$ can only take the values $j=0,1/2,\ldots,k/2$. Since
$\Nu$ generates all other null-fields \cite{FreZhu92}, one may suspect
that this is the only additional condition, and this is indeed correct.

Incidentally, in the case at hand the meromorphic theory is actually
unitary, and the allowed representations are precisely those
representations of the affine algebra that are unitary with respect to
an inner product for which
\be
\left( J^\pm_n \right)^\dagger = J^\mp_{-n} \qquad
\left( H_m \right)^\dagger = H_{-m} \,.
\ee
Indeed, if $|j,j\rangle$ is a Virasoro highest weight state with
$J^+_0|j,j\rangle=0$ and  $H_0 |j,j\rangle=j |j,j\rangle$, then  
\beq
\Bigl( J^+_{-1} |j,j\rangle, J^+_{-1}|j,j\rangle \Bigr) & = 
\Bigl( |j,j\rangle, J^-_1 J^+_{-1} |j,j\rangle \Bigr) \nonumber \\
& = (k-2j) \Bigl( |j,j\rangle, |j,j\rangle \Bigr) \,,
\eeq
and if the representation is unitary, this requires that $(k-2j)\geq 0$, 
and thus that $j=0,1/2,\ldots, k/2$. As it turns out, this is also
sufficient to guarantee unitarity. In general, however, the
constraints that select the representations of the meromorphic
conformal field theory from those of the Lie algebra of modes cannot
be understood in terms of unitarity.

\subsection{Zhu's Algebra and the Classification of Representations}

The above analysis suggests that to each representation of the zero
modes of the meromorphic fields for which the zero modes of the
null-fields vanish, a highest weight representation of the meromorphic
conformal field theory can be associated, and that all highest weight
representations of a meromorphic conformal field theory can be
obtained in this way \cite{EFHHNV92}. This idea has been made precise
by Zhu \cite{Zhu96} who constructed an algebra, now commonly referred
to as {\em Zhu's algebra}, that describes the algebra of zero modes 
modulo zero modes of null-vectors, and whose representations are in
one-to-one correspondence with those of the meromorphic conformal
field theory. The following explanation of Zhu's work follows closely 
\cite{GabGod98}. 

In a first step we determine the subspace of states whose zero modes
always vanish on Virasoro highest weight states. This subspace
certainly contains the states of the form $(L_{-1}+L_0)\psi$, where
$\psi\in\F_0$ is arbitrary, since (\ref{derimode}) implies that 
\be
\label{null1}
V_0 \left( (L_{-1} + L_0) \psi \right) = V_0\left(L_{-1}\psi\right)
+ h V_0\left(\psi\right) = 0 \,.
\ee
Furthermore, the subspace must also contain every state whose zero
mode is of the form $V_0(\chi) V_0((L_{-1}+L_0)\psi)$ or
$V_0((L_{-1}+L_0)\psi) V_0(\chi)$. In order to describe states of this
form more explicitly, it is useful to observe that if both $\phi$ and
$\bar\phi$ are Virasoro highest weight states 
\beq
\langle \bar\phi | V(\psi,1) |\phi\rangle & = 
\sum_l \langle \bar\phi | V_l(\psi) |\phi\rangle \nonumber \\
& = \langle \bar\phi| V_0(\psi) | \phi\rangle \,, \label{zhu1}
\eeq
since the highest weight property implies that
$V_l(\psi)|\phi\rangle=0$ for $l>0$ and similarly, using
(\ref{adjoint}), $\langle\bar\phi| V_l(\psi)=0$ for $l<0$. Because of
the translation symmetry of the amplitudes, this can then be rewritten
as  
\be
\label{zhu2}
\langle \bar\phi | \phi(-1)\; \psi\rangle = 
\langle \bar\phi| V_0(\psi)| \phi\rangle \,.
\ee
Let us introduce the operators
\be
\label{zhu3}
V^{(N)}(\psi) = \oint_{0} {d w \over w^{N+1}} 
V\left( \left(w+1 \right)^{L_0} \psi, w \right) \,,
\ee
where $N$ is an arbitrary integer, and the contour is a small circle
that encircles $w=0$ but not $w=-1$. Then if both $\phi$ and
$\bar\phi$ are arbitrary Virasoro highest weight states and $N>0$, we
have that 
\be
\label{zhu4}
\langle \bar\phi | \phi(-1) \;V^{(N)}(\psi)\chi\rangle = 0 \,,
\ee
since the integrand in (\ref{zhu4}) does not have any poles at $w=-1$
or $w=\infty$. Because of (\ref{zhu2}) this implies that the zero mode
of the corresponding state vanishes on an arbitrary highest weight
state (since the amplitude with any other highest weight state
vanishes). Let us denote by $O(\F_0)$ the subspace of $\F_0$ that is 
generated by states of the form $V^{(N)}(\psi)\chi$ with $N>0$, and
define the quotient space $\A(\F_0)=\F_0 / O(\F_0)$. The above then
implies that we can associate a zero mode (acting on a highest weight
state) to each state in $\A(\F_0)$. We can write (\ref{zhu3}) in terms
of modes as  
\be
\label{zhumode}
V^{(N)}(\psi) = \sum_{n=0}^{h} {h\choose n} V_{-n-N}(\psi) \,,
\ee
where $\psi$ has conformal weight $h$, and it therefore follows that
\be
V^{(1)}(\psi)\Omega = V_{-h-1}(\psi)\Omega + h V_{-h}(\psi)\Omega 
                  = (L_{-1}+L_0) \psi \,.
\ee
Thus $O(\F_0)$ contains the states in (\ref{null1}). Furthermore, 
\beq
V^{(N)}(L_{-1}\psi)  & = 
\oint_{0} {d w \over w^{N+1}} \left( w +1 \right)^{h_\psi+1}
{dV( \psi, w ) \over dw} \nonumber \\
& = -\oint_{0} dw{d\over dw}\left(
{( w +1)^{h_\psi+1} \over w^{N+1}}\right)
V( \psi, w ) \nonumber \\
& =  (N-h_\psi)V^{(N)}(\psi) + (N+1)V^{(N+1)}(\psi)  \,,
\eeq
and this implies, for $N\ne -1$, 
\be
\label{recursive}
V^{(N+1)}(\psi) = {1\over N+1} V^{(N)}(L_{-1}\psi) 
- {N-h_\psi\over N+1}V^{(N)}(\psi)\,.
\ee 
Thus $O(\F_0)$ is actually generated by the states of the form
$V^{(1)}(\psi)\chi$, where $\psi$ and $\chi$ are arbitrary states in
$\F_0$. 

As we shall show momentarily, the vector space $\A(\F_0)$ actually has
the structure of an associative algebra, where the product is defined
by  
\be
\label{prodd}
\psi \ast_L \chi \equiv V^{(0)}(\psi) \chi \,,
\ee
and $V^{(0)}(\psi)$ is given as in (\ref{zhu3}) or (\ref{zhumode});
this algebra is called Zhu's algebra. The analogue of (\ref{zhu4}) is
then 
\beq
\label{zhuaction}
\langle \bar\phi | \phi(-1) \; V^{(0)}(\psi)\chi\rangle & = 
(-1)^{h_\psi} 
\langle V_0(\psi) \bar\phi | \phi(-1) \; \chi\rangle \nonumber \\
& = (-1)^{h_\psi} 
\langle V_0(\psi) \bar\phi |V_0(\chi) | \phi \rangle \\
& = \langle \bar\phi | V_0(\psi) V_0(\chi) | \phi \rangle\,,
\eeq
and thus the product in $\A(\F_0)$ corresponds indeed to the action of
the zero modes. 

In order to exhibit the structure of this algebra it is useful to
introduce a second set of modes by  
\be
\label{zhumodec}
V^{(N)}_c(\psi) 
= (-1)^N \oint {d w \over w} {1 \over (w+1)} 
\left({w+1 \over w}\right)^N V\left( (w+1)^{L_0} \psi, w \right)\,.
\ee
These modes are characterised by the property that 
\beq
\langle \bar\phi | \phi(-1) \;V^{(N)}_c(\psi) \; \chi\rangle & = 0
\qquad \hbox{for $N>0$,} \\
\langle \bar\phi | \phi(-1) \;V^{(0)}_c(\psi) \; \chi\rangle & = 
\langle \bar\phi | \left( V_0(\psi) \phi\right)(-1) \; \chi\rangle\,.
\label{zhuaction1}
\eeq
It is obvious from (\ref{zhumode}) and (\ref{zhumodec}) that 
$V^{(1)}(\psi)=V^{(1)}_c(\psi)$, and the analogue of (\ref{recursive})
is 
\be
\label{recursive1}
V^{(N+1)}_c(\psi) = - {1 \over N+1} V^{(N)}_c(L_{-1}\psi)
-  {N+h_\psi\over N+1} V^{(N)}_c(\psi) \,.
\ee
The space $O(\F_0)$ is therefore also generated by the states of the
form $V^{(1)}_c(\psi)\chi$. Let us introduce, following
(\ref{zhuaction}) and (\ref{zhuaction1}), the notation   
$$
V_L(\psi) \equiv V^{(0)}(\psi) \qquad 
V_R(\psi) \equiv V^{(0)}_c(\psi) \qquad
N(\psi) = V^{(1)}(\psi)=V^{(1)}_c(\psi) \,.
$$
We also denote by $N(\F_0)$ the vector space of operators that are
spanned by $N(\psi)$ for $\psi\in\F_0$; then $O(\F_0)=N(\F_0) \F_0$. 
Finally it follows from (\ref{vacrels}) that 
$V_L(\psi)\Omega =V_R(\psi)\Omega=\psi$.

The equations (\ref{zhuaction}) and (\ref{zhuaction1}) suggest that
the modes $V_L(\psi)$ and $V_R(\chi)$ commute up to an operator in
$N(\F_0)$. In order to prove this it is sufficient to consider the
case where $\psi$ and $\chi$ are both eigenvectors of $L_0$ with
eigenvalues $h_\psi$ and $h_\chi$, respectively. Then we have 
\begin{eqnarray}
\label{commutator}
\fl {}[V_L(\psi),V_R(\chi)] & = &
\oint\oint_{|\zeta|>|w|} {d\zeta \over \zeta}(\zeta+1)^{h_\psi}
{dw \over w } (w+1)^{h_\chi-1}
V(\psi,\zeta) V(\chi,w) \nonumber \\
& & \qquad 
- \oint\oint_{|w|>|\zeta|} {dw \over w} 
(w+1)^{h_\chi -1} {d\zeta \over \zeta} 
(\zeta+1)^{h_\psi} V(\chi,w) V(\psi,\zeta) \nonumber \\
& = & 
\oint_0 \left\{ \oint_w {d\zeta \over \zeta}
(\zeta+1)^{h_\psi}  V(\psi,\zeta) V(\chi,w) \right\}
{dw \over w} (w+1)^{h_\chi-1}  \nonumber \\
& = & \sum_{n}
\oint_0 \left\{ \oint_w {d\zeta \over \zeta}
(\zeta+1)^{h_\psi} V(V_n(\psi)\chi,w) (\zeta-w)^{-n-h_\psi} \right\}
{dw \over w} (w+1)^{h_\chi-1} \nonumber \\
& = &
\sum_{h_\chi\geq n \geq 0} \sum_{l=0}^{n+h_\psi - 1}
(-1)^l \, {h_\psi \choose l+1-n} \nonumber \\
&  &  \qquad \qquad 
\oint_0 {dw \over w (w+1)} \left( {w+1 \over w}\right)^{l+1}
(w+1)^{h_\chi-n} V(V_n(\psi)\chi,w) \nonumber \\
& \in & N(\F_0)\,.
\end{eqnarray}
Because of (\ref{recursive1}), every element in $N(\F_0)$ can be written
as $V_R(\phi)$ for a suitable $\phi$, and (\ref{commutator}) thus
implies that $[V_L(\psi),N(\chi)] \in N(\F_0)$; hence
$V_L(\psi)$ defines an endomorphism of $\A(\F_0)$. 

For two endomorphisms, $\Phi_1, \Phi_2$, of $\F_0$, which leave
$O(\F_0)$ invariant (so that they induce endomorphisms of $\A(\F_0))$, 
we shall write $\Phi_1\approx\Phi_2$ if they agree as endomorphisms of 
$\A(\F_0)$, \ie\ if $(\Phi_1-\Phi_2)\F_0\subset O(\F_0)$. Similarly 
we write $\phi\approx 0$ if $\phi\in O(\F_0)$.

In the same way in which the action of $V(\psi,z)$ is uniquely
characterised by locality and (\ref{psivac}), we can now prove the
following  

\noindent {\bf Uniqueness Theorem for Zhu modes} \cite{GabGod98}:
Suppose $\Phi$ is an endomorphism of $\F_0$ that leaves $O(\F_0)$
invariant and satisfies  
\beq
\Phi\Omega =\psi \nonumber \\
{}[\Phi,V_R(\chi)]\in N(\F_0) \quad \hbox{for all $\chi\in\F_0$.}
\nonumber 
\eeq
Then $\Phi\approx V_L(\psi)$. 

\noindent {\bf Proof}: This follows from 
$$
\Phi \, \chi=\Phi \, V_R(\chi)\Omega \approx V_R(\chi)\, \Phi\, \Omega=
V_R(\chi) \psi = V_R(\chi) \, V_L(\psi) \, \Omega \approx 
V_L(\psi)\chi \,,
$$
where we have used that $V_L(\psi)\Omega=V_R(\psi)\Omega=\psi$.
\medskip

It is then an immediate consequence that
\be
\label{A}
V_L(V_L(\psi)\chi)\approx V_L(\psi)V_L(\chi) \,,
\ee
and a particular case of this (using again the fact that every element
in $N(\F_0)$ can be written as $V_L(\phi)$ for some suitable $\phi$) 
is that 
\be
V_L(N(\psi)\chi)\approx N(\psi) V_L(\chi)\,. 
\ee
In particular this implies that the product (\ref{prodd})
$\phi\ast_L\psi$  is well-defined for both
$\phi,\psi\in\A(\F_0)$. Furthermore, (\ref{A}) shows that this product
is associative, and thus $\A(\F_0)$ has the structure of an algebra. 

We can also define a product by $\phi\ast_R\psi=V_R(\phi)\psi$. Since 
$$
\phi\ast_L\psi= V_L(\phi)\psi= V_L(\phi)V_R(\psi)\Omega\approx
V_R(\psi)V_L(\phi)\Omega=V_R(\psi)\phi=\psi\ast_R\phi
$$
this defines the reverse ring (or algebra) structure. 

As we have explained before, this algebra plays the r\^ole of the
algebra of zero modes. Since it has been constructed in terms of
the space of states of the meromorphic theory, all null-relations have
been taken into account, and one may therefore expect that its
irreducible representations are in one-to-one correspondence with the
irreducible representation of the meromorphic conformal field
theory. This is indeed true \cite{Zhu96}, although the proof is rather
non-trivial.

\subsection{Finite (or Rational) Theories}

Since Zhu's algebra plays a central r\^ole in characterising the
structure of a conformal field theory, one may expect that the 
theories for which it is finite-dimensional are particularly simple
and tractable. In the physics literature these theories are sometimes 
called {\em rational} \cite{MooSei88}, although it may seem more 
appropriate to call them {\em finite}, and we shall from now on do
so. The name rational originates from the observation that the
conformal weights of all states as well as the central charge are
rational numbers in these theories
\cite{Vafa88a,AndMoor88}. Unfortunately, there is no uniform
definition of rationality, and indeed, the notion is used somewhat
differently in mathematics and physics; a survey of the most common
definitions is given in the appendix. In this paper we shall adopt the
convention that a theory is called {\em finite} if Zhu's algebra is
finite-dimensional, and it is called {\em rational} if it satisfies
the conditions of Zhu's definition together with the $C_2$ criterion
(see the appendix). 

The determination of Zhu's algebra is usually rather difficult since
the modes $N(\psi)$ that generate the space $O(\F_0)$ are not
homogeneous with respect to $L_0$. It would therefore be interesting
to find an equivalent condition for the finiteness of a conformal
field theory that is easier to analyse in practice. One such condition
that implies (and may be equivalent to) the finiteness of a 
meromorphic conformal field theory is the so-called $C_2$ condition of
Zhu \cite{Zhu96}: this is the condition that the quotient space 
\be
\label{homo}
\A_{(1)}(\F_0) = \H / O_{(1)}(\F_0) 
\ee
is finite-dimensional, where $O_{(1)}(\F_0)$ is spanned by the states
of the form $V_{-l}(\psi)\chi$ where $l\geq h$, the conformal weight
of $\psi$.\footnote{Incidentally, $\A_{(1)}(\F_0)$ also has the 
structure of an abelian algebra; the significance of this algebra
is however not clear at present.} It is not difficult to show that the
dimension of Zhu's algebra is bounded by that of the above quotient
space \cite{Zhu96}, \ie\ $\dim(\A(\F_0))\leq \dim(\A_{(1)}(\F_0))$. In
many cases the two dimensions are actually the same, but this is not
true in general; the simplest counter example is the theory associated
to the affine algebra for $g=e_8$ at level $k=1$. As we have mentioned
before, this theory can equivalently be described as the meromorphic
conformal field theory that is associated to the self-dual root
lattice of $e_8$, and it is well known that its only representation is
the meromorphic conformal field theory itself \cite{Dong93}; the
highest weight space of the vacuum representation is one-dimensional,
and Zhu's algebra is therefore also one-dimensional. On the other
hand, it is clear that the dimension of $\A_{(1)}(\F_0)$ is at least
$249$ since the vacuum state and the $248$ vectors of the form
$J^a_{-1} \Omega$ (where $a$ runs over a basis of the
$248$-dimensional adjoint representation of $e_8$) are linearly
independent in $\A_{(1)}(\F_0)$.   

Many examples of finite conformal field theories are known. Of the
examples we mentioned in section~3.7, the theories associated to
even Euclidean lattices (3.7.4) are always finite (and unitary)
\cite{Dong93,DLM96b}, the affine theories (3.7.2) are finite if the
level $k$ is a positive integer \cite{KniZam84,GepWit86,FreZhu92} (in
which case the theory is also unitary), and the Virasoro models
(3.7.3) are finite if they belong to the so-called minimal series
\cite{BPZ84,Wang93}. This is the case provided the central charge $c$
is of the form 
\be\label{minimal}
c_{p,q} = 1 - {6 (p-q)^2 \over pq} \,,
\ee
where $p,q \geq 2$ are coprime integers. In this case there exist only
finitely many irreducible representations of the meromorphic conformal
field theory. Each such representation is the irreducible quotient
space of a Verma module generated from a highest weight state with
conformal weight $h$, and the allowed values for $h$ are
\be\label{hminimal}
h_{(r,s)} = {(rp - qs)^2 - (p-q)^2 \over 4 pq}  \,,
\ee
where $1\leq r \leq q-1$ and $1\leq s \leq p-1$, and $(r,s)$ defines
the same value as $(q-r,p-s)$. Each of the corresponding Verma modules
has two null-vectors at conformal weights $h+rs$ and $h+(p-r)(q-s)$,
respectively, and the actual Fock space is the quotient space of the
Verma module by the subspace generated by these two null-vectors
\cite{Kac79,FFu82,FFu84}.  

There are therefore $(p-1)(q-1)/2$ inequivalent irreducible highest
weight representations, and Zhu's algebra has dimension
$(p-1)(q-1)/2$, and is of the form  
\be
\Cop[t] / \prod_{(r,s)} (t - h_{(r,s)}) \,.
\ee
In this case the dimension of Zhu's algebra actually agrees with 
the dimension of the homogeneous quotient space (\ref{homo}). Indeed,
we can choose a basis for (\ref{homo}) to consist of the states of the
form $L_{-2}^l\Omega$, where $l=0,1,\ldots$. For $c=c_{p,q}$ the
meromorphic Verma module has a null-vector at level $(p-1)(q-1)$
(since it corresponds to $r=s=1$), and since the coefficient of
$L_{-2}^{(p-1)(q-1)/2}\Omega$ in the null-vector does not vanish
\cite{FFu84,FNO92,Wang93}, this allows us to express
$L_{-2}^{(p-1)(q-1)/2}\Omega$ in terms of states in $O_{(1)}(\F_0)$,
and thus shows that the dimension of $\A_{(1)}(\F_0)$ is indeed
$(p-1)(q-1)/2$. 

The minimal models include the unitary discrete series
(\ref{discrete}) for  which we choose $p=m$ and $q=m+1$, but they also 
include non-unitary finite theories. The theory with $(p,q)=(2,3)$ is
trivial since $c=0$, and the simplest (non-trivial) unitary theory is
the so called {\em Ising model} for which $(p,q)=(3,4)$ \cite{BPZ84}:
this is the theory with $c=\half$, and its allowed representations
have conformal weight 
\be
\label{Ising}
h=0 \quad \hbox{(vacuum)} \qquad h=\half \quad \hbox{(energy)}
\qquad  h={1\over 16} \quad \hbox{(spin)}\,.
\ee
The simplest non-unitary finite theory is the {\em Yang-Lee edge}
theory with $(p,q)=(2,5)$ \cite{Car85} for which 
$c=-{22\over 5}$. This theory has only two allowed representations, 
the vacuum representation with $h=0$, and the representation with
$h=-1/5$. It has also been observed that the theory with 
$(p,q)=(4,5)$ can be identified with the {\em tricritical Ising model}
\cite{FQS84b}.

\section{Fusion Rules, Correlation Functions and Verlinde's Formula}

Upto now we have analysed in detail the meromorphic subtheory and
its re\-pre\-sen\-ta\-tions, \ie\ the zero- and two-point functions of
the theory. In order to understand the structure of the theory further
we need to analyse next the amplitudes that involve more than two 
non-trivial representations of the meromorphic subtheory.  In a first
step we shall consider the three-point functions that describe the
allowed {\em couplings} between the different subspaces of $\bH$. We
shall then also consider higher correlation functions; their structure
is in essence already determined in terms of the three-point
functions.

\subsection{Fusion Rules and the Comultiplication Formula}
\setcounter{footnote}{0}

As we have explained before, the three-point amplitudes factorise into
chiral and anti-chiral functions. We can therefore restrict ourselves
to discussing the corresponding chiral amplitudes of the meromorphic
theory, say. Their functional form is uniquely determined, and one of
the essential pieces of information is therefore whether the
corresponding amplitudes can be non-trivial or not; this is encoded in
the so-called {\em fusion rules}. 

The definition of the fusion rule is actually slightly more
complicated since there can also be non-trivial multiplicities. In
fact, the problem is rather analogous to that of decomposing a tensor
product representation (of a compact group, say) into
irreducibles. Because of the M\"obius covariance of the amplitudes, it
is sufficient to consider the amplitudes of the form
\be
\langle \phi_k(\infty) \, V(\psi_1,z_1) \cdots V(\psi_l,z_l) 
\, \phi_i(u_1) \phi_j(u_2) \rangle \,,
\ee
where the three non-meromorphic fields are $\phi_i$, $\phi_j$ and
$\phi_k$, and we could set $u_1=1$ and $u_2=0$, for example. The
amplitude defines, in essence, an action of the meromorphic fields on
the product $\phi_i(u_1) \phi_j(u_2)$, and the amplitude can only be
non-trivial if this product representation contains the representation
that is conjugate to $\phi_k$. Furthermore, if this representation is
contained a finite number of times in the product representation  
$\phi_i(u_1) \phi_j(u_2)$, there is a finite ambiguity in defining the
amplitude. We therefore define, more precisely, the fusion rule
$N_{ij}^{k}$ to be the multiplicity with which the representation
conjugate to $\phi_k$ appears in $\phi_i(u_1) \phi_j(u_2)$.

The action of the meromorphic fields (or rather their modes) on the
product of the two fields can actually be described rather explicitly
using the comultiplication formula
\cite{MooSei89b,Gab93,Gab94a}:\footnote{An alternative (more
mathematical) definition of this tensor product was developed by Huang
\& Lepowsky \cite{HuaLep93b,HuaLep93c,HuaLep95,Hua95b}.} 
let us denote by $\A$ the algebra of modes of the meromorphic
fields. A comultiplication is a homomorphism 
\be
\Delta : \A \rightarrow \A\otimes\A \,, \qquad
V_n(\psi) \mapsto \sum  \Delta^{(1)}(V_n(\psi))\otimes 
\Delta^{(2)}(V_n(\psi))\,,
\ee
and it defines an action on the product of two fields as
\be\label{copp}
\fl
V_n(\psi) \Bigl( \phi_i(u_1) \phi_j(u_2) \Bigr) = 
\sum \left(\Delta^{(1)}(V_n(\psi)) \phi_i\right)(u_1) 
\left(\Delta^{(2)}(V_n(\psi)) \phi_j\right)(u_2) \,.
\ee
Here the action of the modes of the meromorphic fields on $\phi_i$ or
$\phi_j$ is defined as in (\ref{Fockgen}). The comultiplication
depends on $u_1$ and $u_2$, and for the modes of a field $\psi$ of
conformal weight $h$, it is explicitly given as
\begin{eqnarray}
{\displaystyle \Delta_{u_1,u_2}(V_{n}(\psi))} & = &
{\displaystyle \sum_{m=1-h}^{n} \left( \begin{array}{c} n+h-1 \\ m+h-1
\end{array} \right)
u_1^{n-m} \left(V_{m}(\psi) \otimes \bbbone\right)  }
\nonumber \\
\label{chir1}
&&  \qquad{\displaystyle +\, \varepsilon_{1}
\sum_{l=1-h}^{n} \left( \begin{array}{c} n+h-1 \\ l+h-1
\end{array} \right)
u_2^{n-l} \left(\bbbone \otimes V_{l}(\psi) \right)}  \\
{\displaystyle \Delta_{u_1,u_2}(V_{-n}(\psi))} & = &
{\displaystyle \sum_{m=1-h}^{\infty} \left( \begin{array}{c} n+m-1 \\ n-h
\end{array} \right) (-1)^{m+h-1}
u_1^{-(n+m)} \left(V_{m}(\psi) \otimes \bbbone \right) }
\nonumber \\
\label{chir2}
& &  \qquad{\displaystyle +\, \varepsilon_{1}
\sum_{l=n}^{\infty} \left( \begin{array}{c} l-h \\ n-h
\end{array} \right)
(-u_2)^{l-n} \left(\bbbone \otimes V_{-l}(\psi)  \right)\,,}
\end{eqnarray}
where in (\ref{chir1}) $n\geq 1-h$, in (\ref{chir2}) $n\geq h$ and
$\varepsilon_{1}$ is $\pm 1$ according to whether the left hand vector
in the tensor product and the meromorphic field $\psi$ are both
fermionic or not.\footnote{It is {\it a priori} ambiguous whether a
given vector in a representation space is fermionic or not. However,
once a convention has been chosen for one element, the fermion number
of any element that can be obtained from it by the action of the modes
of the meromorphic fields is well defined.}  (In
(\ref{chir1},\,\ref{chir2}) $m$ and $l$ are in $\Zop - h$.) This
formula holds in every amplitude, 
\ie\  
\beq\label{comult}
\fl
\left\langle \prod_j V(\chi_j,\zeta_j)  V_n(\psi) \Bigl( \phi_i(u_1)
\phi_j(u_2) \Bigr) \right\rangle \nonumber \\
= \sum \left\langle \prod_j V(\chi_j,\zeta_j) 
\left(\Delta^{(1)}(V_n(\psi)) \phi_i\right)(u_1) 
\left(\Delta^{(2)}(V_n(\psi)) \phi_j\right)(u_2) \right\rangle \,,
\eeq
where each $\chi_j$ can be a meromorphic or a non-meromorphic
field and we have used the notation of (\ref{copp}). In fact, the
comultiplication formula can be derived from (\ref{comult}) using the
fact that the amplitude $\langle \prod_j V(\chi_j,\zeta_j) V(\psi,z)
\phi_i(u_1) \phi_j(u_2)\rangle$ from which the above expression can be
obtained by integration has only poles (as a function of $z$)
for $z=\zeta_j$ and $z=u_i$ \cite{Gab93,Gab94a}. 

The above formula is not symmetric under the exchange of $\phi_i$ and
$\phi_j$. In fact, it is manifest from the derivation that
(\ref{comult}) must also hold if the comultiplication formulae
(\ref{chir1}) and (\ref{chir2}) are replaced by
\begin{eqnarray}
\fl{\displaystyle \widetilde{\Delta}_{u_1,u_2}(V_{n}(\psi))} & = & 
{\displaystyle 
\sum_{m=1-h}^{n} \left( \begin{array}{c} n+h-1 \\ m+h-1
\end{array} \right)
u_1^{n-m} \left(V_{m}(\psi) \otimes \bbbone\right) }
\nonumber \\
\label{chir1'}
&& \qquad {\displaystyle + \varepsilon_1
\sum_{l=1-h}^{n} \left( \begin{array}{c} n+h-1 \\ l+h-1
\end{array} \right)
u_2^{n-l} \left(\bbbone \otimes V_{l}(\psi) \right)\,,}
\end{eqnarray}
for $n\geq 1-h$, and 
\begin{eqnarray}
\label{chir2'}
\fl{\displaystyle \widetilde{\Delta}_{u_1,u_2}(V_{-n}(\psi))} & = &
{\displaystyle 
\sum_{m=n}^{\infty} \left( \begin{array}{c} m-h \\ n-h
\end{array} \right) 
(-u_1)^{m-n} \left(V_{-m}(\psi) \otimes \bbbone \right) }
\nonumber \\
&& \qquad {\displaystyle + \varepsilon_1
\sum_{l=1-h}^{\infty} \left( \begin{array}{c} n+l-1 \\ n-h
\end{array} \right) (-1)^{l+h-1}
u_2^{-(n+l)} \left(\bbbone \otimes V_{l}(\psi)  \right)\,,}
\end{eqnarray}
for $n\geq h$. Since the two formulae agree in every amplitude, 
the product space is therefore the {\em ring-like} tensor product,
\ie\ the quotient of the direct product by the relations that
guarantee that $\Delta=\widetilde\Delta$; this construction is based
on an idea of Richard Borcherds, {\it unpublished} (see \cite{Gab93}).   

A priori it is not clear whether the actual product space may not be
even smaller. However, the fusion rules for a number of models have
been calculated with this definition \cite{Gab94a,Gab97}, and the
results coincide with those obtained by other methods. Indeed, fusion
rules were first determined, for the case of the minimal models, by
considering the implications for the amplitudes of the differential
equations that follow from the condition that a null-vector of a
representation must vanish in all amplitudes \cite{BPZ84}: if the
central charge $c$ is given in terms of $(p,q)$ as in (\ref{minimal}),
the highest weight representations are labelled by $(r,s)$, where $h$
is defined by (\ref{hminimal}) and $1\leq r \leq q-1$ and  $1\leq s
\leq p-1$; the fusion rules are then 
given as    
\be\label{minfusion}
(r_1,s_1) \otimes (r_2,s_2) 
\bigoplus_{r=|r_1-r_2|+1}^{\min(r_1+r_2-1,2q-1-r_1-r_2)} \;\;
\bigoplus_{s=|s_1-s_2|+1}^{\min(s_1+s_2-1,2p-1-s_1-s_2)} 
\; (r,s)\,,
\ee
where $r$ and $s$ attain only every other value, \ie\ $r$ ($s$) is
even if $r_1+r_2-1$ $(s_1+s_2-1$) is even, and odd otherwise.

The analysis of \cite{BPZ84} was adapted for the
Wess-Zumino-Novikov-Witten models in \cite{GepWit86}. For a general
affine algebra $\hat{g}$, the fusion rules can be determined from the
so-called {\em depth rule}; in the specific case of $g=su(2)$ at level
$k$, this leads to  
\be
j_1 \otimes j_2 = \bigoplus_{j=|j_1-j_2|}^{\min(j_1+j_2,k-j_1-j_2)}\;
j\,,
\ee
where $j$ is integer if $j_1+j_2$ is integer, and half-integer
otherwise, and the highest weight representations are labelled by
$j=0,1/2,\ldots,k/2$. A closed expression for the fusion rules in the
general case is provided by the {\em Kac-Walton formula}
\cite{Wal90a,Kac,FGP90a,FucDri90}. 

Similarly, the fusion rules have been determined for the $W_3$ algebra
in \cite{BPT92}, the $N=1$ superconformal minimal models in
\cite{SotSta86} and the $N=2$ superconformal minimal models in
\cite{MSS89b,MSS89a}. For finite theories the fusion rules can also be
obtained by performing the analogue of Zhu's construction in each
representation space; this was first done (in a slightly different
language) by Feigin \& Fuchs for the minimal models \cite{FFu88}, and
later by Frenkel \& Zhu for general vertex operator algebras
\cite{FreZhu92}. (As was pointed out by Li \cite{Li94}, the
analysis of Frenkel \& Zhu only holds under additional assumptions,
for example in the rational case.) 

One of the advantages of the approach that we have adopted here is the
fact that structural properties of fusion can be derived in this
framework \cite{Nahm94}. For each representation $\H_j$, let us define
the subspace $\F_j^{-}$ of the Fock space $\F_j$ to be the space that
is spanned by the vectors of the form 
\be
V_{-n}(\psi) \Phi \qquad \hbox{where $\Phi\in\F_j$ and 
$n\geq h(\psi)$.} 
\ee
We call a representation {\em quasi-rational} provided that the
quotient space 
\be
\F_j / \F_j^{-}
\ee
is finite-dimensional. This quotient space (or rather a realisation of
it as a subspace of $\F_j$) is usually called the {\em special
subspace}. 

It was shown by Nahm \cite{Nahm94} that the fusion product of a
quasi-rational re\-pre\-sen\-tation and a highest weight representation
contains only finitely many highest weight representations. He
also showed that the special subspace of the fusion product of two
quasi-rational representations is finite-dimensional. In fact, 
if we denote by $d^s_j$ the dimension of the special subspace, we have 
\be\label{quasiineq}
\sum_k N^k_{ij} d^s_k \leq d^s_i d^s_j  \,.
\ee
In particular, these results imply that the set of quasi-rational
representations of a meromorphic conformal field theory is closed
under the operation of taking fusion products. It is believed that
every representation of a finite meromorphic conformal field theory is
quasi-rational \cite{Nahm94}, but quasi-rationality is a weaker
condition and there also exist quasi-rational representations of
meromorphic theories that are not finite. The simplest example is the
Virasoro theory for which $c$ is given by (\ref{minimal}), but $p$ and
$q$ are not (coprime) integers greater than one. This theory is not
finite, but every highest weight representation with $h=h_{r,s}$ as in 
(\ref{hminimal}) and $r,s$ positive integers is quasi-rational. (This
is a consequence of the fact that such a representation has a
null-vector with conformal weight $h+rs$, whose coefficient of 
$L_{-1}^{rs}$ is non-zero.) In fact, the collection of all of these
representations is closed under fusion, and forms a `quasi-rational'
chiral conformal field theory.

\subsection{Indecomposable Fusion Products and Logarithmic Theories}

In much of the above discussion we have implicitly assumed that the 
fusion product of any two irreducible representations of the chiral
conformal field theory can be completely decomposed into irreducible
representations. Whilst this is indeed correct for most theories of
interest, there exist a few models where this is not the case. These
theories are usually called {\em logarithmic} theories since, as we
shall explain, some of their correlation functions contain
logarithms. In this subsection we shall give a brief account of this 
class of theories; since the general theory has only been developed
for theories for which this problem is absent, the present subsection
is something of an interlude and not crucial for the rest of this
article. 

The simplest example of a logarithmic theory is the (quasi-rational)
Virasoro model with $p=2,q=1$ whose conformal charge  is $c=-2$
\cite{Gur93}. As we have explained before, the 
quasi-rational (irreducible) representations of this theory have a
highest weight vector with conformal weight $h=h_{r,s}$
(\ref{hminimal}), where $r$ and $s$ are positive integers. Since the
formula for $h_{r,s}$ has the symmetry (for $p=2,q=1$)
\be
h_{r,s}=h_{1-r,2-s}=h_{r-1,s-2} \,,
\ee
we can restrict ourselves to the values $(r,s)$ with $s=1,2$.

The vacuum representation is $(r,s)=(1,1)$ with conformal weight
$h=0$; the null-vector at level $rs=1$ is $L_{-1}\Omega$. The simplest
non-trivial representation is $(r,s)=(1,2)$ with $h=-1/8$; it has a
null-vector at level $rs=2$. As we have alluded to before, a 
null-vector of a representation gives rise to a differential equation
for the corresponding amplitude \cite{BPZ84}. In the present case, if
we denote by $\mu$ the highest weight state with conformal weight
$h=-1/8$, the $4$-point function involving four times $\mu$ has the
form
\be
\langle \mu(z_1) \mu(z_2) \mu(z_3) \mu(z_4) \rangle = 
(z_1 - z_3)^{1 \over 4} (z_2 - z_4)^{1 \over 4} 
\left( x (1-x) \right)^{1\over 4} F(x) \,,
\ee
where we have used the M\"obius symmetry, and $x$ denotes the 
{\em cross-ratio}
\be
x = {(z_1 - z_2) (z_3 - z_4) \over (z_1 - z_3) (z_2 - z_4)} \,.
\ee
The null-vector for $\mu$ gives then rise to a differential equation
for $F$ which, in the present case, is given by
\be\label{diff}
x (1-x) F''(x) + (1 - 2 x) F'(x) - {1 \over 4} F(x) = 0 \,.
\ee
We can make an ansatz for $F$ as 
\be\label{ansatz}
F(x) = x^s \left( a_0 + a_1 x + a_2 x^2 + \cdots \right) \,,
\ee
where $a_0\ne 0$. The differential equation (\ref{diff}) then
determines the $a_i$ recursively provided we can solve the 
{\em indicial} equation, \ie\ the equation that comes from the
coefficient of $x^{s-1}$,
\be\label{indicial}
s(s-1) + s = s^2 = 0 \,.
\ee
Generically, the indicial equation has two distinct roots (that do not
differ by an integer), and for each solution of the indicial equation
there is a solution of the original differential equation that is of
the form (\ref{ansatz}). However, if the two roots coincide (as in our
case), only one solution of the differential equation is of the form
(\ref{ansatz}), and the general solution to (\ref{diff}) is
\be
F(x) = A G(x) + B \left[ G(x) \log(x) + H(x) \right] \,,
\ee
where $G$ and $H$ are regular at $x=0$ (since $s=0$ solves
(\ref{indicial})), and $A$ and $B$ are constants. In fact,
\be
G(x) = \int_{0}^{\pi\over 2} {d \varphi \over \sqrt{1-x\sin^2\varphi}}
\,,
\ee
and 
\be\label{monodromy}
G(x) \log(x) + H(x)  = G(1-x) \,.
\ee
This implies that the 4-point function necessarily has a logarithmic
branch cut: if $F$ is regular at $x=0$, \ie\ if we choose $B=0$, then 
because of (\ref{monodromy}), $F$ has a logarithmic branch cut at
$x=1$. 

In terms of the representation theory this logarithmic behaviour is
related to the property of the fusion product of $\mu$ with itself not
to be completely reducible: by considering a suitable limit of
$z_1,z_2\rightarrow \infty$ in the above $4$-point function we can
obtain a state $\Omega'$ satisfying 
\be
\Bigl\langle \Omega'(\infty)\; \mu(z) \mu(0) \Bigr\rangle 
= z^{1 \over 4} \Bigl(A + B \log (z)\Bigr) \,,
\ee
where $A$ and $B$ are constants (that depend now on $\Omega'$). We can
therefore write   
\be\label{logope}
\mu(z) \mu(0) \sim z^{1\over 4} \Bigl( \omega(0) 
+ \log(z) \Omega (0) \Bigr) \,, 
\ee
where $\langle \Omega'(\infty) \omega(0)\rangle = A$ and 
$\langle \Omega'(\infty) \Omega(0) \rangle = B$. Next we consider the
transformation of this amplitude under a rotation by $2\pi$; this is 
implemented by the M\"obius transformation $\exp(2\pi i L_0)$,
\begin{eqnarray}
\Bigl\langle \Omega'(\infty) \; e^{2\pi i L_0} 
\Bigl(\mu(z) \mu(0)\Bigr) \Bigr\rangle 
& = & e^{-{2\pi i \over 4}} 
\Bigl\langle \Omega'(\infty) \; \mu(e^{2\pi i}z)  \mu(0) \Bigr\rangle 
\nonumber \\
& = & z^{1\over 4} \Bigl(A + B \log(z) + 2\pi i B \Bigr) \,, \label{rech1}
\end{eqnarray}
where we have used that the transformation property of vertex
operators (\ref{scale}) also holds for non-meromorphic fields. On the
other hand, because of (\ref{logope}) we can rewrite 
\be\label{rech2}
\fl
\Bigl\langle \Omega'(\infty) \; e^{2\pi i L_0} 
\Bigl(\mu(z) \mu(0) \Bigr) \Bigr\rangle 
= z^{1\over 4} \Bigl\langle \Omega'(\infty) \; e^{2\pi i L_0}
\Bigl(\omega(0) + \log(z) \Omega(0) \Bigr) \Bigr\rangle \,.
\ee
Comparing (\ref{rech1}) with (\ref{rech2}) we then find that 
\begin{eqnarray}
e^{2 \pi i L_0} \Omega & = & \Omega \\
e^{2 \pi i L_0} \omega & = & \omega + 2\pi i  \Omega \,,
\end{eqnarray}
\ie\ $L_0 \Omega = 0$, $L_0 \omega = \Omega$. Thus we find that
the scaling operator $L_0$ is not diagonalisable, but that it acts as
a {\em Jordan block}  
\be
\left(
\begin{array}{cc} 
0 & 1\\ 
0& 0
\end{array} \right)
\ee
on the space spanned by $\Omega$ and $\omega$. Since
$L_0$ {\em is} diagonalisable in every irreducible representation,
it follows that the fusion product is necessarily not completely
de\-com\-po\-sable. This conclusion holds actually more generally
whenever any correlation function contains a logarithm.

One can analyse the fusion product of $\mu$ with itself using the
comultiplication formula, and this allows one to determine the
structure of the resulting representation ${\cal R}_{1,1}$ in detail 
\cite{GabKau96a}: the representation is generated from a highest
weight state $\omega$ satisfying  
\be
L_0 \omega = \Omega\,, \qquad L_0 \Omega = 0\,, \qquad
L_n \omega = 0 \quad \hbox{for $n>0$}
\ee
by the action of the Virasoro algebra. The state $L_{-1}\Omega$ is a
null-state of ${\cal R}_{1,1}$, but $L_{-1} \omega$ is not null since  
$L_1 L_{-1} \omega = [L_1,L_{-1}] \omega = 2 L_0 \omega = 2 \Omega$. 
Schematically the representation can therefore be described as 
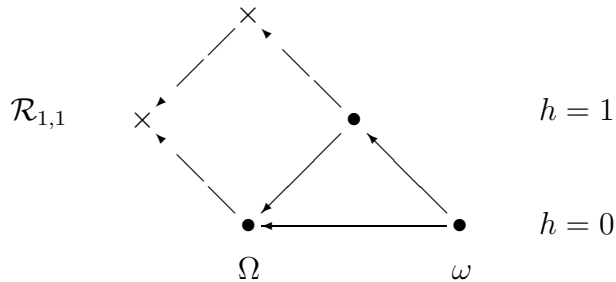
\begin{center}
  \begin{picture}(180,110)(-10,20)
    \multiput(40,40)(80,0){2}{\vbox to 0pt
      {\vss\hbox to 0pt{\hss$\bullet$\hss}\vss}}
    \put(80,80){\vbox to 0pt
      {\vss\hbox to 0pt{\hss$\bullet$\hss}\vss}}
    \put(0,80){\vbox to 0pt
      {\vss\hbox to 0pt{\hss$\times$\hss}\vss}}
    \put(40,120){\vbox to 0pt
      {\vss\hbox to 0pt{\hss$\times$\hss}\vss}}

    \put(115,40){\vector(-1,0){70}}
    \put(115,45){\vector(-1,1){30}}
    \put(75,75){\vector(-1,-1){30}}

    \multiput(35,45)(-12,12){2}{\line(-1,1){10}}
    \put(11,69){\vector(-1,1){6}}
    \multiput(75,85)(-12,12){2}{\line(-1,1){10}}
    \put(51,109){\vector(-1,1){6}}
    \multiput(35,115)(-12,-12){2}{\line(-1,-1){10}}
    \put(11,91){\vector(-1,-1){6}}

    \put(150,37){$h=0$}       
    \put(150,80){$h=1$}
    \put(36,20){$\Omega$}
    \put(117,20){$\omega$}
    \put(-50,80){${\cal R}_{1,1}$}
  \end{picture}
\end{center}
Here each vertex $\bullet$ denotes a state of the representation
space, and the vertices $\times$ correspond to null-vectors. An arrow
$A\longrightarrow B$ indicates that the vertex $B$ is in the image of
$A$ under the action of the Virasoro algebra. The representation
${\cal R}_{1,1}$ is not irreducible since the states that are obtained
by the action of the Virasoro algebra from $\Omega$ form the
subrepresentation $\H_0 $ of ${\cal R}_{1,1}$ (that is actually
isomorphic to the vacuum representation). On the other hand, 
${\cal R}_{1,1}$ is not completely reducible since we cannot find a
complementary subspace to $\H_0$ that is a representation by itself;
${\cal R}_{1,1}$ is therefore called an {\em indecomposable} (but
reducible) representation. 

Actually, ${\cal R}_{1,1}$ is the simplest example of a whole class of
indecomposable representations that appear in fusion products of the
irreducible quasi-rational re\-pre\-sen\-tations; these
indecomposable representations are labelled by $(m,n)$ where now
$n=1$, and their structure is schematically described as 
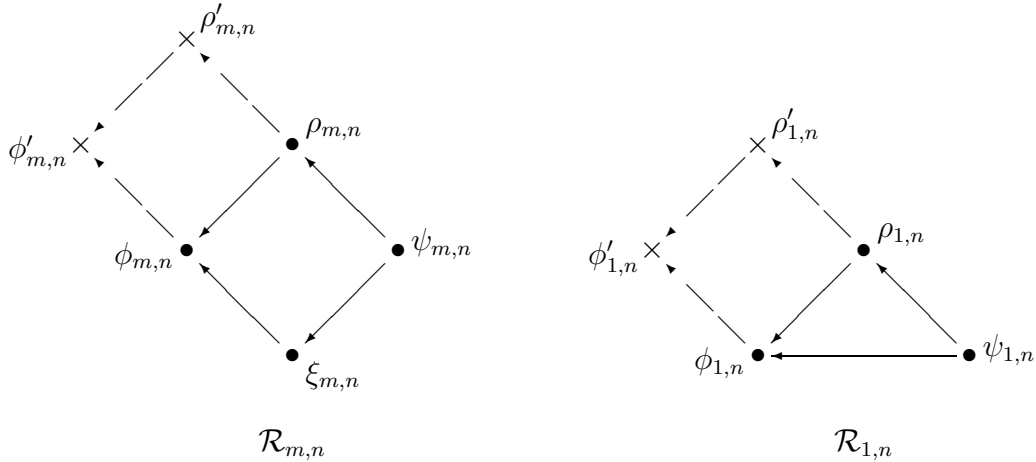
\begin{center}
  \begin{tabular}{c@{\hskip0.5in}c}
  \begin{picture}(180,180)(-10,-20)
    \put(80,0){\vbox to 0pt
      {\vss\hbox to 0pt{\hss$\bullet$\hss}\vss}}
    \multiput(40,40)(80,0){2}{\vbox to 0pt
      {\vss\hbox to 0pt{\hss$\bullet$\hss}\vss}}
    \put(80,80){\vbox to 0pt
      {\vss\hbox to 0pt{\hss$\bullet$\hss}\vss}}
    \put(0,80){\vbox to 0pt
      {\vss\hbox to 0pt{\hss$\times$\hss}\vss}}
    \put(40,120){\vbox to 0pt
      {\vss\hbox to 0pt{\hss$\times$\hss}\vss}}

    \put(75,5){\vector(-1,1){30}}
    \put(115,35){\vector(-1,-1){30}}
    \put(115,45){\vector(-1,1){30}}
    \put(75,75){\vector(-1,-1){30}}

    \multiput(35,45)(-12,12){2}{\line(-1,1){10}}
    \put(11,69){\vector(-1,1){6}}
    \multiput(75,85)(-12,12){2}{\line(-1,1){10}}
    \put(51,109){\vector(-1,1){6}}
    \multiput(35,115)(-12,-12){2}{\line(-1,-1){10}}
    \put(11,91){\vector(-1,-1){6}}

    \put(85,-10){$\xi_{m,n}$}
    \put(125,40){$\psi_{m,n}$}
    \put(85,85){$\rho_{m,n}$}
    \put(35,35){\hbox to 0pt{\hss$\phi_{m,n}$}}
    \put(-5,75){\hbox to 0pt{\hss$\phi'_{m,n}$}}
    \put(45,125){$\rho'_{m,n}$}
  \end{picture}
&
  \begin{picture}(180,120)(-10,20)
    \multiput(40,40)(80,0){2}{\vbox to 0pt
      {\vss\hbox to 0pt{\hss$\bullet$\hss}\vss}}
    \put(80,80){\vbox to 0pt
      {\vss\hbox to 0pt{\hss$\bullet$\hss}\vss}}
    \put(0,80){\vbox to 0pt
      {\vss\hbox to 0pt{\hss$\times$\hss}\vss}}
    \put(40,120){\vbox to 0pt
      {\vss\hbox to 0pt{\hss$\times$\hss}\vss}}

    \put(115,40){\vector(-1,0){70}}
    \put(115,45){\vector(-1,1){30}}
    \put(75,75){\vector(-1,-1){30}}

    \multiput(35,45)(-12,12){2}{\line(-1,1){10}}
    \put(11,69){\vector(-1,1){6}}
    \multiput(75,85)(-12,12){2}{\line(-1,1){10}}
    \put(51,109){\vector(-1,1){6}}
    \multiput(35,115)(-12,-12){2}{\line(-1,-1){10}}
    \put(11,91){\vector(-1,-1){6}}

    \put(125,40){$\psi_{1,n}$}
    \put(85,85){$\rho_{1,n}$}
    \put(35,35){\hbox to 0pt{\hss$\phi_{1,n}$}}
    \put(-5,75){\hbox to 0pt{\hss$\phi'_{1,n}$}}
    \put(45,125){$\rho'_{1,n}$}
  \end{picture}
\\
  ${\cal R}_{m,n}$ & ${\cal R}_{1,n}$
  \end{tabular}
\end{center}
The representation ${\cal R}_{m,n}$ is generated from the vector
$\psi_{m,n}$ by the action of the Virasoro algebra, where
\begin{eqnarray} 
L_0 \psi_{m,n} &=& h_{(m,n)} \psi_{m,n} + \phi_{m,n}\,, \\ 
L_0 \phi_{m,n} &=& h_{(m,n)} \phi_{m,n}\,, \\
L_k \psi_{m,n} &=& 0 \qquad \hbox{for $k\geq 2$}\,.
\end{eqnarray}
If $m=1$ we have in addition $L_1\psi_{m,n}=0$, whereas if $m\geq 2$, 
$L_1\psi_{m,n}\ne 0$, and 
\be
L_1^{(m-1)(2-n)} \psi_{m,n} = \xi_{m,n} \,,
\ee
where $\xi_{m,n}$ is a Virasoro highest weight vector of conformal
weight $h=h_{(m-1,2-n)}$. The Verma module generated by $\xi_{m,n}$
has a  singular vector of conformal weight 
\begin{eqnarray}
\fl
h_{(m-1,2-n)} + (m-1)(2-n) 
& = & {(2(m-1) - (2-n))^2 + 8 (m-1)(2-n) - 1 \over 8} \\
& = & {(2m-n)^2 - 1 \over 8} = h_{(m,n)} \,,
\end{eqnarray}
and this vector is proportional to $\phi_{m,n}$; this singular vector
is however not a null-vector in ${\cal R}_{m,n}$ since it does not
vanish in an amplitude with $\psi_{m,n}$. It was shown in
\cite{GabKau96a} that the set of representations that consists of all
quasi-rational irreducible representations and the above
indecomposable representations closes under fusion, \ie\ any fusion
product of two such representations can be decomposed as a direct sum
of these representations.  

This model is not an isolated example; the same structure is also
present for the Virasoro models with $q=1$ where $p$ is any positive
integer \cite{Flohr95,Kausch95,GabKau96a}. Furthermore, the WZNW model
on the supergroup $GL(1,1)$ \cite{RSal92} and gravitationally dressed
conformal field theories \cite{Schm93,BKog95} are also known to define
logarithmic theories. It was conjectured by  Dong \& Mason
\cite{DonMas96} (using a different language) that logarithms can only
occur if the theory is not finite, \ie\ if the number of irreducible
representations is infinite. However, this does not seem to be correct
since the triplet algebra \cite{Kausch91} has only finitely many
irreducible representations, but contains indecomposable
representations in their fusion products that lead to logarithmic
correlation functions \cite{GabKau96b}. Logarithmic conformal field
theories are not actually pathological; as was shown in
\cite{GabKau98} a consistent local conformal field theory that
satisfies all conditions of a local theory (including modular
invariance of the partition function) can be associated to this
triplet algebra. The space of states of this local theory is then a
certain quotient of the direct sum of tensor products of
indecomposable representations of the two chiral algebras.

\subsection{Verlinde's Formula}

If all chiral representations $\H_j$ as well as their fusion products
are completely reducible into irreducibles, we can define for each
irreducible representation $\H_j$, its {\em conjugate} representation
$\H_{j^\vee}$. The conjugate representation has the property
that at least one two-point function involving a state from $\H_j$ and 
a state from $\H_{j^\vee}$ is non-trivial. Because of Schur's
lemma the conjugation map is uniquely defined, and by construction it
is clearly an involution, \ie\ $(j^\vee)^\vee=j$. 

If conjugation is defined, the condition that the fusion product of
$\H_i$ and $\H_j$ contains the representation $\H_k$ is equivalent to
the condition that the three-point function involving a state from
$\H_i$, one from $\H_j$ and one from $\H_{k^\vee}$ is non-trivial;
thus we can identify $N_{ij}^{k}$ with the number of different
three-point functions of suitable $\phi_i\in\H_i$, $\phi_j\in\H_j$ and
$\phi_{k^\vee}\in\H_{k^\vee}$. It is then natural to define
\be\label{symmetric}
N_{ijk} \equiv N_{ij}^{k^\vee} \,,
\ee
which is manifestly symmetric under the exchange of $i$, $j$ and $k$.  

The fusion product is also associative, 
\be\label{asso1}
\sum_{k} N_{ij}^k N_{kl}^m = \sum_k N_{ik}^m  N_{jl}^k \,.
\ee
If we define ${\bf N}_i$ to be the matrix with matrix elements
\be
\left({\bf N}_i\right)_j^k \equiv N_{ij}^{k} \,,
\ee
then (\ref{asso1}) can be rewritten as 
\be
\sum_k \left({\bf N}_i\right)_j^k \left({\bf N}_l\right)_k^m 
= \sum_k \left({\bf N}_l\right)_j^k \left({\bf N}_i\right)_k^m \,,
\ee
where we have used (\ref{symmetric}). Thus the matrices ${\bf N}_i$
commute with each other.

Because of (\ref{symmetric}) the matrices ${\bf N}_i$ are normal, \ie\
they commute with their adjoint (or transpose) since 
${\bf N}_i^\dag= {\bf N}_{i^\vee}$. This implies that each ${\bf N}_i$
can be diagonalised, and since the different ${\bf N}_i$
commute, there exists a common matrix $S$ that diagonalises all 
${\bf N}_i$ simultaneously. If we denote the 
different eigenvalues of ${\bf N}_i$ by $\lambda_i^{(l)}$, we
therefore have that 
\be\label{diag}
N_{ij}^{k} 
= \sum_{lm} S_j^l \lambda_i^{(l)} \delta_l^m \left(S^{-1}\right)_m^k 
= \sum_l S_j^l \lambda_i^{(l)} \left(S^{-1}\right)_l^k\,.
\ee
If $j$ is the vacuum representation, \ie\ $j=0$, then
$N_{i0}^k=\delta_i^k$ if all representations labelled by $i$ are
irreducible.  In this case, multiplying all three expressions of
(\ref{diag}) by $S_k^n$ from the right (and summing over $k$), we find 
\be
S_i^n = \sum_{lm} S_0^l \lambda_i^{(l)} \delta_l^n 
      = S_0^n \lambda_i^{(n)} \,.
\ee
Hence $\lambda_i^{(n)} = S_i^n / S_0^n$, and we can rewrite
(\ref{diag}) as 
\be\label{Verlinde}
N_{ij}^{k} = \sum_l 
{S_j^l S_i^{l} \left(S^{-1}\right)_l^k \over S_0^l} \,.
\ee
What we have done so far has been a rather trivial
manipulation. However, the deep conjecture is now that the matrix $S$
that diagonalises the fusion rules coincides precisely with the matrix
$S$ (\ref{modular}) that describes the modular transformation 
properties of the characters associated to the irreducible
representations \cite{Ver88}. Thus (\ref{Verlinde}) provides an
expression for the fusion rules in terms of the modular properties of
the corresponding characters; with this interpretation,
(\ref{Verlinde}) is called the  {\em Verlinde formula}. This is a
remarkable formula, not least because it is not obvious in general why
the right-hand-side of (\ref{Verlinde}) should define a non-negative
integer. In fact, using techniques from Galois theory, one can show
that this property implies severe constraints on the matrix elements
of $S$ \cite{BoeGoe91,CosGan94}.

The Verlinde formula has been tested for many conformal field
theories, and whenever it makes sense (\ie\ whenever the theory is
rational), it is indeed correct. It has also been proven for the case
of the WZNW models of the classical groups at integer level
\cite{Fal94}, and it follows from the polynomial equations of Moore \&
Seiberg \cite{MooSei88} to be discussed below.

\subsection{Higher Correlation Functions and the Polynomial Relations
of Moore \& Seiberg} 

The higher correlation functions of the local theory do not directly
factorise into products of chiral and anti-chiral functions, but they
can always be written as sums of such products. It is therefore useful
to analyse the functional form of these chiral functions. The actual
(local) amplitudes (that are linear combinations of products of the
chiral and anti-chiral amplitudes) can then be determined from these 
by the conditions that (i) they have to be local, and (ii) the
operator product expansion (\ref{associative}) is indeed
associative. These constraints define the so-called {\em bootstrap
equations} \cite{BPZ84}; in practice they are rather difficult to
solve, and explicit solutions are only known for a relatively small
number of examples
\cite{DotFat84,DotFat85b,ChrFlu87,Pet89,FFK89b,FucKle89,
FGP90b,Dot91,GabKau98,Tesch99}.  

The chiral $n$-point functions are largely determined in terms of the
three-point functions of the theory. In particular, the number of
different solutions for a given set of non-meromorphic fields can be
deduced from the fusion rules of the theory.\footnote{Since the
functional form of the amplitudes is no longer determined by 
M\"obius symmetry, it is possible (and indeed usually
the case) that there are more than one amplitude for a given set of
irreducible highest weight representations; see for example the
$4$-point function that we considered in section~5.2.} Let us
consider, as an example, the case of a $4$-point function, where the
four non-meromorphic fields $\phi_i \in \H_{m_i}$, $i=1,\ldots, 4$ are 
inserted at $u_1,\ldots, u_4$. (It will become apparent from the
following discussion how this generalises to arbitrary higher
correlation functions.) In the limit in which $u_2\rightarrow u_1$
(with $u_3$ and $u_4$ far away), the $4$-point function can be thought
of as a three-point function whose non-meromorphic field at
$u_1\approx u_2$ is the fusion product of $\phi_1$ and $\phi_2$; we
can therefore write every $4$-point function involving
$\phi_1,\ldots,\phi_4$ as 
\be\label{expan}
\fl
\left\langle\phi_1(u_1)\phi_2(u_2)\phi_3(u_3)\phi_4(u_4)\right\rangle
= \sum_k \sum_{i=1}^{N_{m_1 m_2}^{k}}
\alpha_{k,i} \left\langle\Phi_{12}^{k,i}(u_1,u_2) \;
                  \phi_3(u_3)\phi_4(u_4)\right\rangle\,,
\ee
where $\Phi_{12}^{k,i}(u_1,u_2) \in \H_k$, $\alpha_{k,i}$ are
arbitrary constants, and the sum extends over those $k$ for which 
$N_{m_1 m_2}^{k}\geq 1$. The number of different three-point functions
involving  $\Phi_{12}^k(u_1,u_2) \in \H_k$, $\phi_3$ and $\phi_4$, is  
given by $N_{m_3 m_4}^{k^\vee}$, and the number of different
solutions is therefore altogether
\be\label{number1}
\sum_k N_{m_1 m_2}^k N_{m_3 m_4}^{k^\vee} \,.
\ee

The space of chiral $4$-point functions is a vector space (since any
linear combination of $4$-point functions is again a $4$-point
function), and in the above we have selected a specific {\em basis}
for this space; in fact the different basis vectors (\ie\ the
solutions in terms of which we have expanded (\ref{expan})) are
characterised by the condition that they can be approximated by a
product of three-point functions as $u_2\rightarrow u_1$. In the
notation of Moore \& Seiberg \cite{MooSei89b}, these solutions are
described by 
\be\label{basis1}
\langle \phi_1 | {m_2 \choose m_1 \; k}_{u_2;a} (\phi_2)\;
{m_3 \choose k \; m_4}_{u_3;b} (\phi_3) |\phi_4\rangle\,,
\ee
where we have used the M\"obius invariance to set, without loss of
generality,  $u_1=\infty$ and $u_4=0$. Here, 
\be
{i \choose j \; k}_{u;a} (\phi) : \H_k \rightarrow \H_j
\ee
describes the so-called {\em chiral vertex operator} that is
associated to $\phi\in\H_i$; it is the restriction of $\phi(u)$ to
$\H_k$, where the image is projected onto $\H_j$ and $a$ labels the
different such projections (if $N_{ik}^j\geq 2$). This definition has 
to be treated with some care since $\phi(u)$ is strictly speaking 
{\em not} a well-defined operator on the direct sum of the chiral
spaces, $\oplus_i \H_i$ --- indeed, if it were, there would only be one
$4$-point function!\footnote{It is possible to give an operator
description for the chiral theory (at least for the case of the
WZNW-models) by considering, instead of $\H_i$, the tensor product of
$\H_i$ with a finite-dimensional vector space that is a certain
truncation of the corresponding anti-chiral representation
$\overline{\H}_{\bar{\imath}}$ \cite{MooRes89,Gab95}. This also
provides a  natural interpretation for the quantum group symmetry to
be discussed below.} 

In the above we have expanded the $4$-point functions in terms of
a {\em basis} of functions each of which approximates a product of
three-point functions as $u_2\rightarrow u_1$. We could equally 
consider the basis of functions to consist of those functions that
approximate products of three-point functions as $u_2\rightarrow u_3$;
in the notation of Moore \& Seiberg \cite{MooSei89b} these are
described by 
\be\label{basis2}
\langle \phi_1 | {k \choose m_1 \; m_4}_{u_3;c} (\chi)\; |\phi_4\rangle
\cdot \langle \chi| {m_2 \choose k \; m_3}_{u_2-u_3;d} (\phi_2) 
| \phi_3\rangle
\,.
\ee
Since both sets of functions form a basis for the same vector space,
their number must be equal; there are (\ref{number1}) elements in the
first set of basis vectors, and the number of basis elements of the
form (\ref{basis2}) is 
\be\label{number2}
\sum_k N_{m_1 m_4}^k N_{m_2 m_3}^{k^\vee} \,.
\ee
The two expressions are indeed equal, as follows from
(\ref{asso1})  upon setting $m_1=j$, $m_2=i$, $m_3=\bar{m}$, $m_4=l$, 
and using (\ref{symmetric}). We can furthermore express the two sets
of basis vectors in terms of each other; this is achieved by the
so-called {\em fusing matrix} of Moore \& Seiberg \cite{MooSei89b}, 
\be\label{fusion}
\fl
{m_2 \choose m_1 \; p}_{u_2;a} {m_3 \choose p \; m_4}_{u_3;b}
= \sum_{q;c,d} F_{pq} \left[
\begin{array}{cc}
m_2 & m_3 \\
m_1 & m_4
\end{array}
\right]_{ab}^{cd} \;
{q \choose m_1 \; m_4}_{u_3;c} {m_2 \choose q \; m_3}_{u_2-u_3;d}
\,.
\ee

We can also consider the basis of functions that are approximated by
products of three-point functions as $u_3\rightarrow u_1$ (rather than 
$u_2\rightarrow u_1$). These basis functions are described, in the 
notation of Moore \& Seiberg \cite{MooSei89b}, by
\be\label{basis3}
\langle \phi_1 | {m_3 \choose m_1 \; k}_{u_3;c} (\phi_3)\;
{m_2 \choose k \; m_4}_{u_2;d} (\phi_2) |\phi_4\rangle\,.
\ee
By similar arguments to the above, it is easy to see that the 
number of such basis vectors is the same as (\ref{number1}) or
(\ref{number2}). Furthermore, we can express the basis vectors 
in (\ref{basis1}) in terms of the new basis vectors (\ref{basis3}) as  
\be\label{braiding}
\fl
{m_2 \choose m_1 \; p}_{u_2;a} {m_3 \choose p \; m_4}_{u_3;b}
= \sum_{q;c,d} B(\pm)_{pq} \left[
\begin{array}{cc}
m_2 & m_3 \\
m_1 & m_4
\end{array}
\right]_{ab}^{cd} \;
{m_3 \choose m_1 \; q}_{u_3;c} {m_2 \choose q \; m_4}_{u_2;d}
\,,
\ee
where $B$ is the so-called {\em braiding matrix}. Since the
correlation functions are not single-valued, the braiding matrix
depends on the equivalence class of paths along which the 
configuration $u_2\approx u_1$ (with $u_4$ far away) is analytically 
continued to the configuration $u_3\approx u_1$. In fact, there are
two such equivalence classes which differ by a path along which $u_3$
encircles $u_2$ once; we distinguish the corresponding braiding
matrices by $B(\pm)$.

The two matrices (\ref{fusion}) and (\ref{braiding}) have been derived
in the context of $4$-point functions, but the notation we have used
suggests that the corresponding identities should hold more generally,
namely for products of chiral vertex operators in any correlation
function. As we can always consider the limit in which the remaining
coordinates coalesce (so that the amplitude approximates a $4$-point
function), this must be true in every consistent conformal field
theory. On the other hand, the identities (\ref{fusion}) and
(\ref{braiding}) {\em can} only be true in general provided that the
matrices $F$ and $B$ satisfy a number of consistency conditions; these
are usually called the  {\em polynomial} equations \cite{MooSei88}.  

The simplest relation is that which allows to describe the braiding
matrix $B$ in terms of the fusing matrix $F$, and the diagonalisable
matrix $\Omega$. The latter is defined by 
\be
{k \choose l \; m}_{z;a} = \left(\Omega(\pm)^{k}_{lm} \right)_{a}^{b}
{k \choose m \; l}_{z;b}\,, 
\ee
where again the sign $\pm$ distinguishes between clockwise (or
anti-clockwise) analytic continuation of the field in $\H_m$ around
that in $\H_l$. Since all three representations are irreducible,
$\Omega(\pm)$ is just a phase,
\be\label{Omega}
\left(\Omega(\pm)^{k}_{lm}\right)_a^b = s_a 
e^{\pm i\pi (h_k - h_l - h_m)} \delta_{a}^{b}\,,
\ee
where $s_a=\pm 1$, and $h_i$ is the conformal weight of the highest
weight state in $\H_i$. In order to describe now $B$ in terms of $F$,
we apply the  fusing matrix to obtain the right-hand-side of
(\ref{fusion}); braiding now corresponds to $\Omega$ (applied to the
second and third representation), and in order to recover the
right-hand-side of (\ref{braiding}), we have to apply the inverse of
$F$ again. Thus we find \cite{MooSei89b}
\be
B(\epsilon) = F^{-1} \left(\bbbone\otimes \Omega(-\epsilon)\right) F
\,.
\ee
The consistency conditions that have to be satisfied by $B$ and $F$
can therefore be formulated in terms of $F$ and $\Omega$. In essence,
there are two non-trivial identities, the {\em pentagon} identity, and
the {\em hexagon} identity. The former can be obtained by considering 
sequences of fusing identities in a $5$-point function, and is
explictly given as \cite{MooSei89b}
\be\label{pentagon}
F_{23} F_{12} F_{23} = P_{23} F_{13} F_{12} \,,
\ee
where $F_{12}$ acts on the first two representation spaces, {\it etc.},
and $P_{23}$ is the permutation matrix that exchanges the second and
third representation space. The hexagon identity can be derived by
considering a sequence of transformations involving $F$ and $\Omega$
in a $4$-point function \cite{MooSei89b}
\be\label{hexagon}
F (\Omega(\epsilon)\otimes\bbbone) F =
(\bbbone\otimes\Omega(\epsilon)) \, F \,
(\bbbone\otimes\Omega(\epsilon)) \,.
\ee
It was shown by Moore \& Seiberg \cite{MooSei89b} using category
theory that all relations that arise from comparing different
expansions of an arbitrary $n$-point function on the sphere are
a consequence of the pentagon and hexagon identity. This is a deep
result which allows us, at least in principle (and ignoring problems
of convergence, {\it etc.}), to construct all $n$-point functions of
the theory from the three-point functions. Indeed, the three-point
functions determine in essence the chiral vertex operators, and by
composing these operators as above, we can construct a basis for an
arbitrary $n$-point function. If $F$ and $\Omega$  satisfy the
pentagon and hexagon identities, the resulting space will be
independent of the particular expansion we used. 

Actually, Moore \& Seiberg also solved the problem for the case of
arbitrary $n$-point functions on an arbitrary surface of genus
$g$. (The proof in \cite{MooSei89b} is not quite complete; see
however \cite{Tur94}.) In this case there are three additional
consistency conditions that originate from considering correlation
functions on the torus and involve the modular transformation matrix
$S$ (see \cite{MooSei89b} for more details). As was also shown in
\cite{MooSei89b} this extended set of relations implies Verlinde's
formula.

\subsection{Quantum Groups}

It was observed in \cite{MooSei89b} that every (compact) group $G$
gives rise to matrices $F$ and $B$ (or $\Omega$) that satisfy the
polynomial equations: let us denote by $\{R_i\}$ the set of
irreducible representations of $G$. Every tensor product of two
irreducible representations can be decomposed into irreducibles,
\be\label{tensor}
R_i \otimes R_j = \oplus_{k} V_{ij}^k \otimes R_k \,,
\ee
where the vector space $V_{ij}^k$ can be identified with the space of
intertwining operators,
\be
{k \choose i \; j} : R_i \otimes R_j \rightarrow R_k\,,
\ee
and $\dim(V_{ij}^k)$ is the multiplicity with which $R_k$ appears in 
the tensor product of $R_i$ and $R_j$. There exist natural
isomorphisms between representations,
\be\label{aa1}
\hat\Omega: R_i \otimes R_j  \cong  R_j \otimes R_i 
\ee
\be\label{aa2}
\hat{F}: (R_i \otimes R_j) \otimes R_k  \cong  R_i \otimes (R_j \otimes R_k)
\,,
\ee
and they induce isomorphisms on the space of intertwining operators,
\begin{eqnarray}
\Omega & : & V_{ij}^{k}  \cong  V_{ji}^{k} \label{bb1}\\
F & : & \oplus_r V_{ij}^{r} \otimes V_{rk}^{l}  \cong  
   \oplus_s V_{is}^{l} \otimes V_{jk}^{s} \,.\label{bb2}
\end{eqnarray}
The pentagon commutative diagram  
$$
\begin{array}{ccccc}
R_1 \otimes (R_2\otimes(R_3\otimes R_4)) & \stackrel{\hat{F}}{\rightarrow} &
(R_1 \otimes R_2) \otimes (R_3 \otimes R_4) & 
\stackrel{\hat{F}}{\rightarrow} & ((R_1\otimes R_2)\otimes R_3)\otimes R_4
\\
\downarrow {\scriptstyle (1\otimes \hat{F})} & & & & 
\downarrow {\scriptstyle (\hat{F}\otimes 1)} \\
R_1 \otimes((R_2\otimes R_3)\otimes R_4) & &
\stackrel{\hat{F}}{\longrightarrow} & & 
(R_1 \otimes(R_2\otimes R_3))\otimes R_4 
\end{array}
$$
then implies the pentagon identity for $F$ (\ref{pentagon}), while the  
hexagon identity follows from 
$$
\begin{array}{ccccc}
R_1 \otimes (R_2\otimes R_3) & \stackrel{\hat{F}}{\rightarrow} &
(R_1 \otimes R_2) \otimes R_3 & 
\stackrel{\hat\Omega}{\rightarrow} & R_3 \otimes (R_1\otimes R_2)
\\
\downarrow {\scriptstyle (1\otimes \hat\Omega)} & & & & 
\downarrow {\scriptstyle \hat{F}} \\
R_1 \otimes(R_3\otimes R_2) & \stackrel{\hat{F}}{\rightarrow} & 
(R_1 \otimes R_3) \otimes R_2 & 
\stackrel{\hat\Omega\otimes 1}{\rightarrow} & 
(R_3 \otimes R_1) \otimes R_2 \,.
\end{array}
$$
Thus the representation ring of a compact group gives rise to a
solution of the polynomial relations; the fusion rules are then
identified as $N_{ij}^k = \dim(V_{ij}^k)$.

The $F$ and $B$ matrices that are associated to a chiral conformal
field theory are, however, usually not of this form. Indeed, for
compact groups we have $\Omega^2=1$ since the tensor product is
symmetric, but because of (\ref{Omega}) this would require that the
conformal weights of all highest weight states are half-integer which
is not the case for most conformal field theories of interest. For a
general conformal field theory the relation $\Omega^2=1$ is replaced
by $\Omega(+) \Omega(-) = 1$; this is a manifestation of the fact that
the fields of a (two-dimensional) conformal field theory  obey braid
group statistics rather than permutation group statistics
\cite{RehSch89,Fro87}.

On the other hand, many conformal field theories possess a `classical
limit' \cite{MooSei89b} in which the conformal dimensions tend to
zero, and in this limit the $F$ and $B$ matrices come from compact
groups. This suggests \cite{MooSei89b,AGS89a} that the actual $F$ and
$B$ matrices of a chiral conformal field theory can be thought 
of as being associated to the representation theory of a {\em quantum
group} \cite{Drin85,Drin86,Jim85,FRT90}, a certain deformation of a
group (for reviews on quantum groups see
\cite{ChaPre94,Maj96}). Indeed, the chiral conformal field theory of
the WZNW model associated to the affine algebra $\hat{su(2)}$ at level
$k$ has the same $F$ and $B$ matrices \cite{TsuKan87,TsuKan88} as the
quantum group $U_q(sl(2))$ \cite{KirRes88} at $q=e^{i\pi / (k+2)}$
\cite{AGS89b}. Similar relations have also been found for the 
WZNW models associated to the other groups \cite{Fin96}, and the
minimal models \cite{AGS89a,FFK89b}. 

These observations suggest that chiral conformal field theories may
have a hidden quantum group symmetry. Various attempts have been made
to realise the relevant quantum group generators in terms of the
chiral conformal field theory \cite{GomSie90,GomSie91,RRR90,RRR91} but
no clear picture has emerged so far. A different proposal has been put
forward in \cite{Gab95} following \cite{MooRes89} (see also
\cite{Rec94,ARS97}). According to this idea the quantum group symmetry
acts naturally on a certain (finite-dimensional) truncation of the
anti-chiral representation space, namely the special subspace; these
anti-chiral degrees of freedom arise naturally in an operator
formulation of the chiral theory. 
\smallskip

The actual quantum symmetries that arise for rational theories are
typically quantum groups at roots of unity; tensor products of
certain representations of such quantum groups are then not completely
reducible \cite{FroKer92}, and in order to obtain a structure as in
(\ref{tensor}), it is necessary to truncate the tensor products in a
suitable way. The resulting symmetry structure is then more correctly
described as a quasi-Hopf algebra \cite{Drin90b,MacSch92}. In fact,
the underlying structure of a chiral conformal field theory {\em must}  
be a quasi-Hopf algebra (rather than a normal quantum group) whenever
the {\em quantum dimensions} are not integers: to each representation
$\H_i$  we can associate (because of the Perron-Frobenius theorem) a
unique positive real number $d_i$, the quantum dimension, so that
\be\label{qdim}
d_i d_j = \sum_k N_{ij}^k d_k \,,
\ee
where $N_{ij}^k$ are the fusion rules; if the symmetry structure of
the conformal field theory is described by a quantum group, the choice 
$d_i=\dim(R_i)$ satisfies (\ref{qdim}), and thus each quantum
dimension must be a positive integer. It follows from (\ref{Verlinde}) 
that 
\be
d_i = {S_{i0} \over S_{00}} 
\ee
satisfies (\ref{qdim}); for unitary theories this expression is a
positive number, and thus coincides with the quantum dimension. For
most theories of interest this number is not an integer for all $i$.

It has been shown in \cite{Scho95} (see also \cite{Maj91a}) that for
every rational chiral conformal field theory, a weak quasi-triangular
quasi-Hopf algebra exists that reproduces the fusing and the braiding
matrices of the conformal field theory. This quasi-Hopf algebra is
however not unique; for every choice of positive integers $D_i$
satisfying  
\be\label{inequality}
D_i D_j \geq \sum_k N_{ij}^k D_k
\ee
such a quasi-Hopf algebra can be constructed. As we have seen above,
the dimensions of the special subspaces $d_i^s$ satisfy this
inequality (\ref{quasiineq}) provided that they are finite; this gives
rise to one preferred such quasi-Hopf algebra in this case
\cite{Gab95,ARS97}. 

The quantum groups at roots of unity also play a central r\^ole in the
various knot invariants that have been constructed starting with the
work of Jones \cite{Jon83,Jon85a,ResTur90}. These have also a direct
relation to the braiding matrices of conformal field theories
\cite{RehSch89,Fro87} and can be interpreted in terms of $2+1$
dimensional Chern-Simons theory \cite{Witten89,FroKin89}.

\section{Conclusions and Outlook}

Let us conclude this review with a summary of general problems in
conformal field theory that deserve, in our opinion, further work.
\smallskip

\noindent {\it 1. The local theory}: It is generally believed that to
every modular invariant partition function of tensor products of
representations of a chiral algebra, a consistent local theory can be
defined. Unfortunately, only very few local theories have been
constructed in detail, and there are virtually no general results (see
however \cite{MooSei89c}). 

Recently it has been realised that the operator product expansion
coefficients in a boundary conformal field theory can be expressed
in terms of certain elements of the fusing matrix $F$
\cite{Run99,BPPZ99,FFFS99a}. Since there exists a close relation
between the operator product expansion coefficients of the boundary
theory and those of the bulk theory, this may open the way for a
general construction of local conformal field theories.
\smallskip

\noindent {\it 2. Algebraic formulae for fusing and braiding
matrices}: Essentially all of the structure of conformal field theory
can be described in terms of the representation theory of certain
algebraic structures. However, in order to obtain the fusing and
braiding matrices that we discussed above, it is necessary to analyse
the {\em analytical} properties of correlation functions, in
particular their monodromy matrices. If it is indeed true that the
whole structure is determined by the algebraic data of the theory,
a direct (representation theoretic) expression should exist for
these matrices as well.
\smallskip

\noindent {\it 3. Finite versus rational}: As we have explained in this
article (and as is indeed illustrated by the appendix), there are
different conditions that guarantee that different aspects of the
theory are well-behaved in some sense. Unfortunately, it is not clear
at the moment what the precise logical relation between the different
conditions are, and which of them is crucial in distinguishing between
theories that are tractable, and those that are less so. In this
context it would also be very interesting to understand under which
conditions the correlation functions (of representations of the
meromorphic conformal field theory) do not contain logarithmic branch
cuts. 
\smallskip

\noindent {\it 4. Existence of higher correlation functions}: It is
generally believed that the higher correlation functions of
representations of a finite conformal field theory define analytic
functions that have appropriate singularities and branch cuts. This is
actually a crucial assumption in the definition of the fusing and
braiding matrices, and therefore in the derivation of the polynomial
relations of Moore \& Seiberg (from which Verlinde's formula can be
derived). It would be interesting to prove this in general.
\smallskip

\noindent {\it 5. Higher genus}: Despite recent advances in our
understanding of the theory on higher genus Riemann surfaces
\cite{Zhu94,Gaw95a,Gaw95b}, a completely satisfactory treatment for
the case of a general conformal field theory is not available at
present.

\appendix

\section{Definitions of Rationality}

\subsection{Zhu's Definition}

According to \cite{Zhu96}, a meromorphic conformal field theory is
{\em rational} if it has only finitely many irreducible highest weight
representations. The Fock space of each of these representations has
finite-dimensional weight spaces, \ie\ for each eigenvalue of $L_0$,
the corresponding eigenspace is finite-dimensional. Furthermore, each
finitely generated representation is a direct sum of these
irreducible representations. 

If a meromorphic conformal field theory is rational in this sense,
Zhu's algebra is a semi-simple complex algebra, and therefore
finite-dimensional \cite{Zhu96,DLM95}. Zhu has also conjectured that
every such theory satisfies the $C_2$ criterion, \ie\ the condition
that the quotient space (\ref{homo}) is finite-dimensional. 

If a meromorphic conformal field theory is rational in this sense and
satisfies the $C_2$ condition then the characters of its
representations define a representation of the modular group
$SL(2,\Zop)$ \cite{Zhu96}.

\subsection{The DLM Definitions}

Dong, Li \& Mason call a representation {\em admissible} if it
satisfies the representation criterion (\ie\ if the corresponding
amplitudes satisfy the condition (\ref{operep})), and if it possesses
a decomposition of the form  $\oplus_{n=0}^\infty M_{n+\lambda}$,
where $\lambda$ is fixed and $V_n(\psi)M_{\mu}\subset M_{\mu-n}$ for 
$\mu=\lambda+m$ for some $m$. A meromorphic conformal field theory is
then called {\em rational} if every admissible representation can be
decomposed into irreducible admissible representations. 

If a meromorphic conformal field theory is rational in this sense,
then Zhu's algebra is a semi-simple complex algebra (and hence
finite-dimensional), every irreducible admissible representation is an
irreducible representation for which each $M_{\mu}$ is
finite-dimensional and an eigenspace of $L_0$, and the number of  
irreducible representations is finite \cite{DLM95}. 

This definition of rationality therefore implies Zhu's notion of
rationality, but it is not clear whether the converse is true.

It has been conjectured by Dong \& Mason \cite{DonMas96} that
the finite-dimensionality of Zhu's algebra implies rationality (in
either sense). This is not true as has been demonstrated by the
counterexample of Gaberdiel \& Kausch \cite{GabKau96b}.

\subsection{Physicists Definition}

Physicists call a meromorphic conformal field theory rational
if it has finitely many irreducible highest weight representations. 
Each of these has a Fock space with finite-dimensional $L_0$
eigenspaces, and the characters of these representations form a
representation of the modular group $SL(2,\Zop)$. (Sometimes this last
condition is not imposed.)

If a meromorphic conformal field theory is rational in the sense of
Zhu and satisfies the $C_2$ condition, then it is rational in the
above sense. 

The notion of rationality is also sometimes applied to the whole
conformal field theory: a conformal field theory is called rational if
its meromorphic and anti-meromorphic conformal subtheories are
rational.

\ack

I am indebted to Peter Goddard for explaining to me much of the
material that is presented here, and for a very enjoyable
collaboration. I also thank Matthias D\"orrzapf, Terry Gannon, Horst
Kausch, Andy Neitzke, Andreas Recknagel, Sakura Sch\"afer-Nameki and
Volker Schomerus for a careful reading of the manuscript and many
useful suggestions.

\section*{References}
\bibliography{biblio1}

\end{document}